\RequirePackage{fix-cm}
\documentclass[smallextended]{svjour3}       
\smartqed  
\usepackage{graphicx}
\usepackage{xcolor}
\usepackage{amsmath}
\usepackage{hyperref}

\usepackage{natbib}
\usepackage[version=4]{mhchem}
\begin{document}


\title{The long-term evolution of the atmosphere of Venus: processes and feedback mechanisms.



}
\subtitle{Interior-exterior exchanges}


\titlerunning{Long-Term Atmosphere Interior Evolution of Venus}        

\author{Cedric Gillmann \and
        M.J. Way \and Guillaume Avice \and Doris Breuer \and Gregor J. Golabek \and Dennis H\"oning \and Joshua Krissansen-Totton \and Helmut Lammer \and Joseph G. O'Rourke \and Moa Persson \and Ana-Catalina Plesa \and Arnaud Salvador \and Manuel Scherf \and Mikhail Yu. Zolotov 
}

\institute{C. Gillmann \at
              Rice University, Department of Earth, Environmental and Planetary Sciences, Houston, TX 77005, USA \\
              Tel.: +1-346-4289057\\
              \email{cedric.gillmann@rice.edu}
           \and
           M.J. Way \at
              NASA Goddard Institute for Space Studies,
              2880 Broadway, New York, NY 10025\\
              Theoretical Astrophysics,
              Department of Physics and Astronomy,
              Uppsala University, Uppsala, Sweden\\
              \email{Michael.J.Way@nasa.gov}
           \and
            G. Avice \at
              Université Paris Cité, Institut de physique du globe de Paris, CNRS, 75005 Paris, France \\
              \email{avice@ipgp.fr}
           \and
            D. Breuer \at
              DLR, Institute of Planetary Research,
              12489 Berlin, Germany  \\
              \email{doris.breuer@dlr.de}
            \and
            G. J. Golabek \at
              Bayerisches Geoinstitut, University of Bayreuth, 95440 Bayreuth, Germany \\
              \email{gregor.golabek@uni-bayreuth.de}
            \and
            D. H\"oning \at
              Potsdam Institute for Climate Impact Research, Potsdam, Germany \\
              Department of Earth Sciences, VU Amsterdam, The Netherlands \\
              \email{dennis.hoening@pik-potsdam.de}   
            \and  
             J. Krissansen-Totton \at
              Department of Astronomy and Astrophysics,
              University of California, Santa Cruz,
              CA, USA\\
              \email{jkt@ucsc.edu}
            \and
            H. Lammer \at
              Space Research Institute,
              Austrian Academy of Sciences,
              Graz, Austria\\
              \email{helmut.lammer@oeaw.ac.at}  
              \and  
           Joseph G. O'Rourke \at
              School of Earth and Space Exploration, Arizona State University, Tempe, USA            
              \and
           M. Persson \at
              Institut de Recherche en Astrophysique et Planétologie, Centre National de la Recherche Scientifique, Université Paul Sabatier - Toulouse III, Centre National d’Etudes Spatiales, Toulouse, France  \\
              \email{moa.persson@irap.omp.eu}
              \and  
           A.-C. Plesa \at
               Institute of Planetary Research, DLR, Berlin, Germany
              \and
           A. Salvador \at
              Department of Astronomy and Planetary Science, Northern Arizona University, Box 6010, Flagstaff, AZ 86011, USA\\
              Habitability, Atmospheres, and Biosignatures Laboratory, University of Arizona, Tucson, AZ, USA\\
	   Lunar and Planetary Laboratory, University of Arizona, Tucson, AZ, USA\\
              \email{arnaudsalvador@arizona.edu}
              \and
           M. Scherf \at
              Space Research Institute,
              Austrian Academy of Sciences\\
              Institute of Physics, University of Graz\\
              Institute for Geodesy, Technical University,
              Graz, Austria\\
              \email{manuel.scherf@oeaw.ac.at}       
              \and
           M. Yu. Zolotov \at
              School of Earth and Space Exploration, Arizona State University, Tempe, USA
              \email{zolotov@asu.edu}
}

\date{Received: date / Accepted: date}

\maketitle

\begin{abstract}

This work reviews the long-term evolution of the atmosphere of Venus, and modulation of its composition by interior/exterior cycling. The formation and evolution of Venus’s atmosphere, leading to contemporary surface conditions, remain hotly debated topics, and involve questions that tie into many disciplines. We explore these various inter-related mechanisms which shaped the evolution of the atmosphere, starting with the volatile sources and sinks. Going from the deep interior to the top of the atmosphere, we describe volcanic outgassing, surface-atmosphere interactions, and atmosphere escape.
Furthermore, we address more complex aspects of the history of Venus, including the role of Late Accretion impacts, how magnetic field generation is tied into long-term evolution, and the implications of geochemical and geodynamical feedback cycles for atmospheric evolution. We highlight plausible end-member evolutionary pathways that Venus could have followed, from accretion to its present-day state, based on modeling and observations. In a first scenario, the planet was desiccated by atmospheric escape during the magma ocean phase. In a second scenario, Venus could have harbored surface liquid water for long periods of time, until its temperate climate was destabilized and it entered a runaway greenhouse phase. In a third scenario, Venus's inefficient outgassing could have kept water inside the planet, where hydrogen was trapped in the core and the mantle was oxidized. We discuss existing evidence and future observations/missions required to refine our understanding of the planet's history and of the complex feedback cycles between the interior, surface, and atmosphere that have been operating in the past, present or future of Venus.

\end{abstract}

\section{Introduction and Overview}\label{sec:introduction}

\begin{figure}
\centering
\includegraphics[width=12cm]{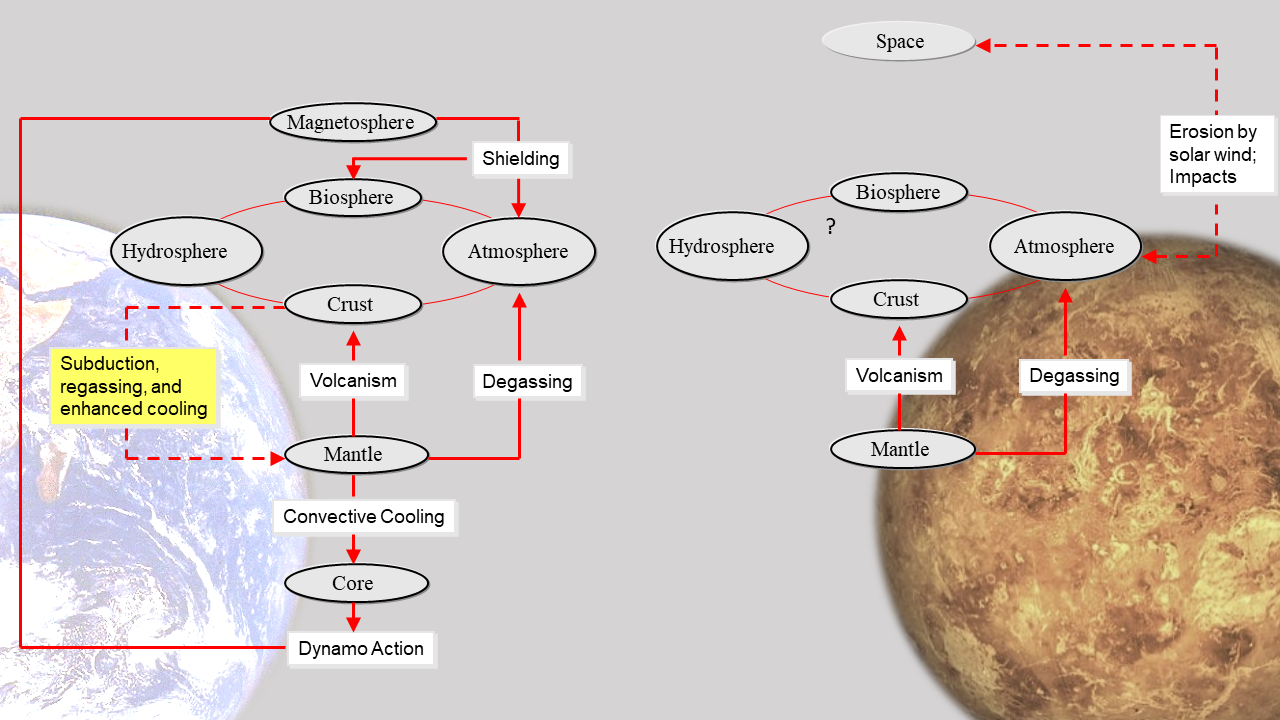}
\caption{Sketch of a plate tectonic planet like Earth (left) and a stagnant lid or single plate planet like present Venus (right) showing the interactions between the different reservoirs. Venus shares characteristics with stagnant lid planets but also shows evidence of deformation and horizontal motion. Major difference between plate tectonic planets and stagnant lid planets is the lack of subduction and regassing of volatiles from the surface (yellow rectangle) for the latter. In the plate tectonic scenario, subduction of the cold surface layers (i.e., oceanic crust and lithosphere) leads to efficient cooling of the mantle and core. This can facilitate the generation of a strong magnetic field in the core, which in turn affects atmosphere loss mechanisms caused by the solar wind (see section \ref{magneticfields} for details). 
The volatile exchanges in a stagnant lid or a single plate planet are therefore reduced and shows only a one-way path from the mantle into the atmosphere. The reduced volatile circulation may prevent the formation of a biosphere and a hydrosphere over long periods of time. Photos of Earth and Venus in the background, courtesy of NASA/JPL}
\label{Fig:Venus_tectonics}
\end{figure}

\emph{In situ} and remote studies of Venus, numerical modelling and experimental work, as well as comparison with the Earth, inform us about the current state of Venus and can provide clues on its past (Fig.~\ref{Fig:Venus_tectonics}). Despite past missions and advances in computational models and experimental techniques, many uncertainties remain when it comes to which exact evolutionary pathway Venus has followed. The observation of Venus has challenged many of our assumptions on the evolution and present-day state of terrestrial planets. For example, before observations revealed its surface environment, Venus was imagined to be warm but temperate, while its surface proved to be dry and hot enough to melt lead \citep[see][]{ORourke2022}. The planet was thought to be geologically moribund or even dead, but data for the atmosphere and surface from numerous missions indicate that Venus might instead be an active world. 

 These advances, however, highlight the many gaps in our understanding of Venus. This is even more evident when it comes to Venus' past and its long-term evolution. Whether Venus could have sustained temperate surface conditions at any time during its evolution and for how long this could have lasted, is not known. Whether Venus could have had a magnetic field in the past, and when and why the core dynamo stopped operating is also unknown. What convection regime is most representative for the interior dynamics of Venus, if and how it changed over the planet's history, and what became of Venus's water, are difficult to constrain with current data \citep[see][for a detailed review]{Rolfetal2022}. Even Venus' present-day mantle convection regime is peculiar, lacking the subduction zones and the clear dichotomy between continental and oceanic crust we see on Earth, it also displays more traces of horizontal motion and deformation than the clear stagnant lids operating on Mars and Mercury. For these reasons we label it as single plate planet in this manuscript.
In fact, one of the main challenges for reconstructing Venus' evolution is that many observations defy straightforward interpretation. Instead, present-day measurements depend on the cumulative effects of many mechanisms. A good example of this is the high D/H ratio measured for the atmosphere of Venus, indicative of strong fractionation relative to the Earth \citep{Donahue1982,DeBergh1991}. This is traditionally interpreted as an indication of strong water loss during the evolution of the planet. However, it is actually the result of hydrogen loss combined with at least a possible volcanic source and an external source (impact delivery), both of unknown isotopic composition \citep{Grinspoon1993}. Therefore, timing, mechanism and quantity of water involved remain uncertain. 

Recently the selection of new Venus missions (DAVINCI, VERITAS, ENVISION and Shukrayaan-1) recognized both the importance of Venus in our understanding of terrestrial planets and the limited current state of our knowledge. We can only make sense of Venus if we understand how it became the planet we now observe. Despite being Earth's closest neighbor, exploration of Venus by spacecraft has been neglected relative to Mars; we need more data at every level, from its interior structure to its composition, from its surface to its upper atmosphere.

A host of mechanisms have contributed to shape Venus over billions of years into the planet we observe today. Venus, as a whole, is a complex system of interacting processes. While it is important to understand each of them by itself, the evolution of the planet is the result of their joined actions and possible feedback loops. A major goal of this review is to assess the consequences of a wide range of mechanisms through changing planetary conditions. While \citep{Rolfetal2022} deals with the interior of Venus, this work is focused on its atmospheric evolution, and how the atmosphere changes over time under the influence of various processes and how, in turn, it affects all other parts of the planet. We first describe the mechanisms involved in volatile evolution that affect atmosphere composition and thickness: sources and sinks of volatiles. Then we discuss peripheral and external mechanisms such as magnetic field generation and the effects of impacts. Finally, we describe the possible evolutionary pathways from the formation of Venus to its present-day state and future, and how the many processes fit into that broad picture. We conclude with highlights of what observations would be needed to distinguish between various evolution scenarios described herein.

\section{\bf Volatiles exchanges}

Despite the atmosphere being the most readily observable part of Venus at present-day, its evolution is another matter. It is suspected that the thick $\approx$ 90 bar CO$_2$-rich atmosphere has been in place since before the current basalt-dominated surface was created, based on the lack of small craters (smaller meteoroids are destroyed in the current thick atmosphere during entry).

For comparison, despite remaining uncertainties, the evolution of the atmospheres of both Earth and Mars is known to a much finer degree. For Mars, there are estimates for the palaeopressure $\approx$3.6 Gyr ago based on the size distribution of ancient craters \citep{Kite2014}. For Earth, hypotheses exist for the mechanisms that sculpt atmospheric evolution, and these are supported by a variety of geological proxies. For example, despite uncertainties, especially in the Archean and Hadean, we have an idea of the range of pressures at Earth's surface in its distant past, and know about variations of its atmospheric composition \citep[e.g., the rise of O$_2$,][]{Lyons2014}. Earth has remained habitable for at least the last $\approx$4.3 Gyr \citep{mojzsis2001} with probably moderate variations of its atmospheric pressure \citep{Som2012,Som2016,marty_nitrogen_2013,avice_evolution_2018}. In the case of Venus we have no direct constraints on atmospheric pressure through time, although a number of scenarios have been proposed. One end member suggests that the present-day atmosphere was produced in the distant past of the evolution of Venus and was generated by ancient processes \citep{Head2021}, while on the other extreme, a possible geologically more recent origin for the 90 bar worth of CO$_2$ \citep{way2020venusian} has also been proposed. 

Additional calculations related to magma ocean lifetime by \cite{Hamano2013} and \cite{Lebrun2013} concern the fate of water, a related crucial question. These works advocate that the divergence between Earth and Venus could have occurred at a primitive epoch due to magma ocean duration, with one planet (Earth) cooling fast and remaining moist, while the other (Venus) would have dried up and stayed hot. Those studies still do not provide a clear way to discriminate between the scenarios, although a 3-D global circulation model (GCM) by \cite{Turbet2021} points to an early and persistent hot-house Venus, based on cloud behaviour. With the lack of definitive answers from currently available data, it is useful to examine mechanisms and try to understand how they interact to produce Venus’ modern day atmosphere based on observational clues and the modelling of their respective effects. 

Volatile exchanges are at the heart of the interactions between the interior and exterior of Venus. This involves a balance between sources and sinks that governs what the atmosphere looks like, from surface conditions to the edge of space. As such, they define the ``face" of the planet: the present-day atmosphere is the outcome of volatile exchange and loss mechanisms throughout its history. Conversely, that also means that it can inform us about what has occurred during the planet’s evolution, both inside its mantle and outside. This goes both ways: understanding the mechanisms is necessary for interpreting observations, and data is the basis for our understanding of the mechanisms at play. 

Many of these mechanisms of volatile exchange from within and without potentially affect atmosphere composition, as evidenced on Earth. The most straightforward processes can clearly be classified as sources (outgassing from the interior) and sinks (atmospheric escape, chemical interaction with the surface) and are described herein. More complex interactions are left for the next section of this work.
\par

\subsection{\bf The source: Outgassing}\label{sec:source}

\par

A main source of volatile elements on Venus after the initial magma ocean stage and the accretion phase is thought to be volcanic outgassing from the mantle. \cite{Avice2022} offers insights into the initial conditions for the main evolution phase lasting from 4+ Ga to present-day. The outgassing process is complex: to produce outgassing, volatile elements and compounds from the solid mantle must be first (i) extracted during partial melting of the mantle (partitioning into the silicate melt), ascend with the melt to reach the vicinity of the surface and then (ii) be released from magma chambers, volcanic conduits and the lava flows into the atmosphere.
 
\subsubsection{General principles}
\label{outgass}
$\textbf{Volatile extraction from the mantle}$

In a first stage, volatile extraction from the solid mantle depends on melting conditions (pressure, temperature) and mantle composition. For outgassing to occur, the mantle must partially melt: melt is produced when mantle material temperature reaches its solidus temperature (usually due to decompression). Local conditions in the mantle (temperature, pressure, mantle composition) govern how much melt is generated. Melt fraction $F$ is defined as the ratio between the volume of melt that has been generated locally by the partial melting of the mantle and the original volume of mantle affected by said set of physical conditions.  

When melt is produced, chemical species can be extracted and enter the liquid phase. 
However, they distribute unevenly between the melt and the solid phase. The amount of volatiles in the melt depends on their solubility as well as the partitioning behaviour between the melt and solid phase; the smaller value of the two limits the amount of volatiles. For water, the partitioning behaviour is the limiting factor and is represented by the distribution coefficient 
$D_i$= $X_i^{sol}$/$X_i^{liq}$, where $X_i$ denotes the concentration of species $i$ in a given phase at equilibrium. A species with a $D<<$1 is said to be incompatible and will transfer easily into the melt, such as water \citep{Katz2003}, while compatible elements will remain in the solid phase. The concentration of a species in the melt will depend on its concentration in the mantle material, on the melt fraction of the material, and on the distribution coefficient for the specific species. The exact relation between these parameters will depend on the type of melting considered: whether it is equilibrium (batch) melting or fractional melting. 

The type of melting mechanism is related to the mode of the ascent/extraction of the magma. A plume rising adiabatically would be best described by equilibrium melting: melt and solid stay intermingled as a closed system until melting ceases. Fractional melting rather describes fast-ascending melt in a dike, where melt is extracted as soon as it forms and is only in equilibrium with the solid at the instant it forms before accumulating in a magmatic chamber outside the system.

The amount of H$_2$O and C in the melt depend on the partition coefficient, the degree of partial melting and their solubility in the melt. Water is much more soluble than CO$_2$. However the amount of water in magmas (and the mantle) on present-day Venus is still unknown, with no direct measurement, but could be low. Indirect estimates, based on Magellan data (maps of volcanic features) and old calculations of escape rates, or modeling of crustal reservoirs, suggest low water abundances in the lava \citep[ about 50 ppm, see][and references therein]{BullockGrinspoon2001}, but are to be taken with caution (see below for the limitation of outgassing by surface pressure).

Solubility of carbon in silicate melts affects the amount of C extracted to the melt in mantle source regions and, ultimately, quantities of C-bearing gases delivered to the atmosphere with volcanism. The solubility depends on oxygen fugacity, melt composition, temperature, water content, and pressure \citep[e.g.]{Holloway1992,Ni2013}. The solubility of C decreases with decreasing fO$_2$. Reducing mantle conditions are suggested for planets with stagnant lids, but also for the early Earth \citep[e.g.,][]{Wadhwa2008}, where substantial recycling of oxidizing agents such as ferric iron, water, or carbonates is/was lacking. At fO$_2$ roughly below conditions of the iron-wüstite (IW) buffer, C is stable in graphite rather than in melt and C partitions inefficiently in the melt in Fe(CO)$_5$ and minor CH$_4$ complexes \citep[e.g.][]{Wetzel2013,Stanley2014,Armstrong2015}. CO gas will ultimately form through decompression of such melts via decomposition of Fe(CO)$_5$ and/or low-pressure oxidation of graphite grains \citep[e.g.][]{Fogel1995}. 
Venus’s present-day mantle fO$_2$ is estimated to be between those on Earth and Mars \citep{Schaefer_Fegley2017,Wadhwa2008}. Then, C in melts is likely present in carbonaceous melt complexes (e.g., CO$_3^2-$) and CO$_2$ dominates over CO in volcanic gases. C solubility varies with the melt composition \citep[e.g.,][and references therein]{dixon1997}. Although the SiO$_2$ content in melts has a minor and unclear effect on the solubility of CO$_2$, solubility increases with increasing alkali metal content. Tholeiitic basaltic magmas, such as suggested for landing sites of Venera 9, 10, 14, and Vega 2 \citep{Surkov1984,Surkov1986,kargel1993}, display the lowest solubility among common silicate melts. Alkali-rich mafic melts, that could have formed Venera 13 and possibly Venera 8 rocks \citep{Surkov1986,kargel1993}, show a C solubility that is 2-3 times higher than in tholeiitic melts. Note that formation of K-rich Venera 13 type magmas could be related to the partial melting of carbonated mantle source regions \citep{kargel1993,Filiberto2014}. In other words, the presence of K-rich mafic igneous rock may suggest a C-rich mantle that could indicate an heterogeneous mantle, past burial of C-bearing rocks and future CO$_2$ degassing. For mafic melts, temperature has no definite effect on CO$_2$ solubility, though solubility in felsic and alkaline melts could be lower or higher at higher temperature \citep{Ni2013}. Solubility of CO$_2$ commonly decreases with increasing dissolved water content in magma \citep[e.g.][for a review]{Ni2013}. The effect of water is not linear and is minor in water-poor melts that could characterize tholeiitic melts on Venus. For all silicate melts, solubility of CO$_2$ strongly decreases with decreasing pressure \citep{Holloway1994,Ni2013}.

\medskip
$\textbf{Volatile release into the atmosphere}$

 In the second stage, when melt has reached the surface, volatile species can then be released into the atmosphere. The total amount of volatiles released is affected by how much magma reaches the surface (intrusive/extrusive volcanism), and by pre-erupting sub-surface conditions \citep{Berlo2011, Edmonds2018}. The outgassing process is further affected by surface conditions, such as atmospheric pressure \citep{Gaillard2014}, since solubility depends strongly on pressure \citep[e.g.,][]{Sparks1994}.
 
 For Venus, it is currently unknown how much of the rising magma was extruded at the surface, although modelling attempts have suggested that large crustal magma reservoirs \citep{Head_Wilson1992} with dikes \citep{Parfitt_Head1993} could be favoured. Earth shows a wide range of possible extrusive efficiencies based on the volcanic environment. The ratio of intrusive to extrusive magma for Earth ranges from 12:1 (in continental environments) to 5:1 (in oceanic environments) \citep{Crisp1984,Cawood2013}. It is uncertain whether comparable values are valid for the case of Venus. Direct observation remains limited to estimates of the composition of Venus lava (more basaltic overall; e.g., \citealt{Surkov1984, Surkov1986}), and surface morphology \citep[indicating the basalt-like morphology of lava in the plains and volcanic centers such as Maat Mons]{head1992,barsukov1986}. Additionally, studies of the physics of recent Venus' volcanism and volcanic-tectonic settings indicate no global, or limited subduction-related features. Hot-spots and flood basaltic volcanism dominate recent history \citep{wilson1983,Head1986,Head_Wilson1992,Smrekar2022, Herrick2022}. In the absence of observational data for Venus, Earth-like ranges of the ratio of intrusive to extrusive magma have been assumed. Volcanic outgassing models for Venus usually use a fixed relative extrusion rate in the range of 10:1 \citep{Noack2012,Gillmann2014}, as a simplification of a very complex mechanism. Self-consistent time- and conditions-dependent calculations have not been achieved yet, and a more precise modeling of melt extraction variations will be needed in the future. Since little observational constraint exist for Venus, it is impossible to completely rule out higher values. Higher surface temperatures on Venus might imply a more ductile deformation that translates into relatively low eruption to intrusion ratios. For this reason, it has been suggested that intrusion process may be common on Venus \citep{Gerya2014}. It can have strong implications for the tectonic regime: when hot magma is emplaced in the crust, it can favour a so-called ``plutonic squishy lid" regime \citep[a regime where the thin strong lithosphere is divided by warm weak regions generated by plutonism][]{Rozel2017}, which could be favoured on Venus \citep{Lourenco2020} as it tends to form both a thinner crust and a warmer and thinner lithosphere \citep{Rozel2017,Lourenco2018} that could be more in line with more recent lithosphere estimates for Venus \citep{Anderson_Smrekar2006, James2013}. However, the consequences of plutonic squishy lid convection on outgassing are yet not fully understood.
 
 One should note that knowing the extrusive magma flux is not sufficient to estimate the total fraction of volatiles reaching the atmosphere. It is possible that intrusive magmatism also leads to subsequent outgassing as volatiles find their way through the crust over time. For example, it has been suggested that on Mars 40$\%$ of the total volatiles produced by partial melting would ultimately be outgassed into the atmosphere, assuming a fractured upper crust \citep{Grott2011}. The latter number takes into account the mechanical state of the lithosphere, porosity of the crust and calculates the depth at which the pores/fractures would close due to plastic deformation. This number could act as an upper limit for present-day Venus, considering the higher surface temperatures. 

An additional uncertainty regarding degassing is a potential change in the tectonic mode associated with resurfacing events \citep[][]{Herrick2022}: the mantle convection mode can change during the history of the planet. Such changes can be as straightforward as a switch from mobile lid regime to stagnant lid, or involve a more complicated pattern \citep{Gillmann2014,Weller2020}. Changes of the convection regime directly affect mantle conditions (for example temperatures), the amount and location of partial melting, the ability of melt to reach the surface and thus outgassing. However, the exact consequences of those changes on outgassing scenarios are uncertain. For stagnant lid planets partial melting occurs below the lid within a region, where the temperature exceeds the solidus temperature \citep{Grott2011}. On planets with plate tectonics, in addition to volcanism related to hotspots and subduction zones, decompression melting below mid-ocean ridges occurs closer to the surface \citep{kite2009,kruijver2021} which can affect the total amount of melt. As a consequence, if plate tectonics operated on early Venus the degassing rate may have been much higher than today, depending on surface conditions (see section \ref{surfcond}). Issues still remain as the volcanic evolution of Venus is poorly constrained, with no solid data about its ancient history available to date.

\subsubsection{Water and the role of surface conditions.} \label{surfcond}
Looking at the present-day composition of Venus' atmosphere, it has long been assumed that outgassing of water was limited \citep[e.g.,][]{Grinspoon1993}, and that the reason could have been that the planet started off relatively dry \citep[e.g.,][]{Namiki_Solomon1998}. A scenario explaining this possible characteristic relied on desiccation of the planet by thermal escape during the slow cooling and solidification of the magma ocean \citep{Hamano2013,Hamano2015}. It was further supported by the observation that the topography of Venus was positively correlated with its geoid \citep{Simons1994, Johnson2003}, unlike Earth. This could be explained by Venus having a stiffer interior than Earth, lacking an asthenosphere and possibly being drier, at least in the upper mantle \citep{Kaula1994, Simons1994, Kucinskas1996, Solomatov1996}. 

However, even a relatively wet Venus mantle could fail to produce significant water outgassing \citep{Holloway1992}. The amount and species of volatiles released can change drastically depending on specific conditions. While surface temperature does not substantially affect the solubility of gaseous species in silicate melt, it can modify some gas speciations \citep[carbon and sulphur being the most sensitive elements;][]{Gaillard2021}. However, lithostatic and surface pressures have been found to be a dominant parameter governing the amount of outgassed species due to their differential effect on volatile solubility. Provided all species are present in sufficient quantities in the mantle that undergoes melting, high surface pressure (100-1000 bar) produces CO$_2$-rich, N$_2$-rich, and H$_2$O-poor volcanic gases, for a high oxyen fugacity (e.g., Venus;
\citealt{Holloway1992,Zolotov2002}), while low pressure ($<$10$^{-4}$ bar) promotes sulfur-rich gases (e.g., Io). Earth's intermediate surface pressure (1 bar) favors H$_2$O-rich outgassing. In the case of Venus, with an atmospheric pressure just below 100 bar (and oxygen fugacities above IW+1), outgassing of CO$_2$ would be favored, even if water degassing may not be completely suppressed ($\sim$ 1\% of the initial lava content, \cite{Gaillard2021,Ortenzi2020}). 

However, the effect of surface pressure should be taken with care given the fact that before reaching and erupting at the surface, volatile exsolution primary occurs during magma ascent and associated lithostatic decompression. Then, magmas are stored within crustal reservoirs and magmatic chambers where volatile exsolution depends on the ambient pressure. The lithostatic pressure of ascending magmas and the pressure within the magmatic chamber is likely higher than the atmospheric pressure and volatile melt-gas partitioning is first ruled by them. Thus, some volatiles exsolution, responsible for the increase of pressure within the chamber, and therefore for the erupting process, already occurred before the magma reaches the surface \citep[e.g.,][]{Sparks1994, Berlo2011, Edmonds2018}. Melt composition is also determinant in the process. Thus, an important fraction of volatile species are already outgassed and form gas bubbles before magma erupts so that melt-gas volatile partitioning occurring at depth - and in the magmatic chamber - remains a key aspect of volcanic outgassing \citep[e.g.,][]{Gonnermann2015, Wallace2015} and needs to be accounted for to reconstruct a consistent outgassing sequence. All outgassed compositions are conditional on the presence of various species in the mantle melt zone. The mantle may be well mixed but not homogeneous \citep[see][]{salvador2022}, or species may have been depleted earlier during the evolution. For example, it is possible that a large part of N$_2$ was degassed very early in the evolution of Venus, leaving little N$_2$ in magmatic gases throughout geological history, especially if no N$_2$ recycling occurred on Venus (in the case of a full stagnant lid/single plate history, for example). Indeed, the processes affecting the large-scale degassing of the early magma ocean, and likely responsible for most of the atmospheric mass and composition, may significantly differ from the local-scale, volcanic outgassing of the solid-state mantle.

The effect of surface pressure could contribute to understanding the state of the current atmosphere of Venus, but its complete evolution needs to be taken into account. As long as the surface pressure was low, water could have effectively outgassed, but as soon as the surface pressure rose above around 20 bar, mainly CO$_2$ would have accumulated in the atmosphere. It follows that a CO$_2$ sink is needed early in the evolution of terrestrial planets to ensure the stability of the low pressure of about 1 bar under which the Earth has been for much of its history. On Earth, this sink is part of the silicate carbon cycle, while recent Venus is unlikely to have been able to sustain it. Conversely, it is likely difficult to return to low pressure conditions that favour water outgassing once a dense and dry CO$_2$ atmosphere has been established, even through catastrophic scenarios that remove a substantial part of the atmosphere like very large impactors \citep{Schlichting2018}.

\subsection{Outgassing of N$_2$ and SO$_2$}

Nitrogen is the second most abundant constituent in the atmosphere of Venus after CO$_2$. Data from the Venera 11, Venera 12 and the Pioneer Venus atmospheric probe indicate a N$_2$ concentration of 3.5$\pm$0.8 \% at altitudes lower than 45 km \citep{vonZahn1983}. More recent measurements performed by MESSENGER's Neutron Spectrometer during the second flyby recorded a value of 5.0$\pm$0.4 \% for altitudes between 60 and 100 km, leading to the conclusion that the atmosphere of Venus might not be as well mixed as previously thought \citep{Peplowski2020}. 

Nitrogen isotope data indicate that the nitrogen of the terrestrial planets could originate from carbonaceous chondrites \citep[e.g.,][]{Marty2013}, likely from the main accretion phase \citep{Lammer2018}, perhaps as NH$_3$-ices or organic compounds like HCN, as speculated by \cite{Wordsworth2016}. The $^{14}$N/$^{15}$N ratio for the atmosphere of Venus is poorly constrained (273 $\pm$ 56, \cite{hoffman1979}) but the measured value remains close to that of Earth's atmosphere, indicating weak escape. This implies that on Venus, the N$_2$-rich atmosphere might have been shielded from escape. For example, it could either not have been outgassed early on, at a time when escape was still strong enough to remove it from the atmosphere. Alternatively, atmospheric N$_2$ may have been protected from escape by a thick CO$_2$ atmosphere efficiently cooling Venus' thermosphere. As a comparison with Earth, atmospheric escape simulations \citep[e.g.,][]{Lichtenegger2010} suggest that an early (older than 3.5 Ga) dense N$_2$-rich atmosphere could have escaped, in the absence of sufficient CO$_2$. It has been suggested \citep{Wordsworth2016} that, during the early stages of planetary evolution, a reducing atmosphere could have favoured reduced N species near the surface. High surface temperatures could have allowed atmospheric N to be dissolved in a reduced (below the IW buffer, since at more oxidized conditions, N$_2$ has very low solubility in melts) magma ocean and transported into the mantle \citep{Wordsworth2016}. Later during the planetary history, nitrogen could be released into the atmosphere as a consequence of magmatic degassing \citep{Lammer2018}. 

In volcanic gases, N$_2$ remains the dominant species compared to NH$_\mathrm{x}$ or NO$_\mathrm{x}$ \citep{Gaillard2014}. Calculations that use the C-O-H-S-N system and investigate the abundance of volcanic gas species in equilibrium with basalts indicate that the concentration of N$_2$ increases for large atmospheric pressures (larger than a few bars), while it becomes diluted in the volcanic gas for low pressures \citep{Gaillard2014}. Present-day Venus surface conditions suppress outgassing of N$_2$ less than H$_2$O and SO$_2$. If the majority of N has not already been outagssed from the interior of Venus (its atmosphere contains more N$_2$ than Earth's), then this would make nitrogen relatively more abundant in venusian volcanic gases. Another possibility is the exsolution of most of the present-day atmospheric nitrogen inventory at the end of the magma ocean phase \citep{Gaillard2022a}, for oxygen fugacities above IW-3, which would imply limited later volcanic outgassing over Venus' evolution. 

For reference, the current amount of N$_2$ in the Earth's mantle and crust is estimated between 0.32 bar and 5.6 bar \citep[][and references therein]{Catling_Zahnle2020,Wordsworth2016}, which still poorly constrains this reservoir. On present Earth \citep{Som2016}, nitrogen is thought to be approximately in balance between sources (half volcanic outgassing and half weathering by O$_2$) and the sink (burial, with a touch of subducted flux). Some data also suggest large changes in Earth's N$_2$ partial pressure over geological times \citep{Som2016}, possibly due to changes in volcanic activity or the presence/absence of oxygen \citep[see also][]{Westall2022}. However, the comparison between Venus and Earth is complicated by the fact that the nitrogen cycle is modulated by Earth's biosphere \citep{Jacob1999, Zerkle2017}.

Sulfur dioxide is the third most abundant gas in the atmosphere of Venus. In contrast to N$_2$, SO$_2$ is extremely chemically reactive and is involved in complex reaction chains. The solubility of SO$_2$ also varies strongly with pressure, which would affect its abundance in released volcanic gases on Venus (possibly reducing its concentration, compared to Earth, \cite{Gaillard2014}). Together with water and CO$_2$, SO$_2$ is one of the most important greenhouse gases in the atmosphere of Venus. On Earth and Mars, it is released into the atmosphere during volcanic outgassing. Venus' volcanism is thought to be a probable source for SO$_2$ that in turn is involved in atmosphere-surface reactions \citep{Zolotov2018} with calcium-bearing materials to form anhydrite (CaSO$_4$). Since these reactions would deplete the SO$_2$ in the atmosphere, reduce the production of sulfuric acid (H$_2$SO$_4$) and the formation of clouds, an SO$_2$ source likely in form of volcanic activity needs to be active at present-day. To be able to maintain the measured SO$_2$ concentration in the atmosphere of Venus, it has been suggested that an eruption rate of about 1 km$^3$/yr with lava compositions similar to those observed at the Venera 13, 14 and Vega 2 landing sites was necessary to account for the sink by reaction with surface calcium mineral \citep{fegley1989,Fegley2009}. However, it has since been proposed \citep[see][]{Zolotov2018} that equilibrium involving plagioclase could sustain present-day SO$_2$ atmospheric concentrations at present-day without recent volcanism (see section \ref{atm-surface-reactions}).

Models that calculate the gas composition of volcanic gases, indicate that sulfur becomes highly soluble in surface lavas with increasing atmospheric pressure and may be released only in small proportions into a dense atmosphere, such as that of present-day Venus \citep{Gaillard2014}. For this reason \cite{Head2021} estimate, based on observation of the total surface lava production, that it is unlikely that the present-day SO$_2$ atmospheric abundance could be solely maintained by extrusive volcanism.

\subsubsection{The CO$_2$ atmosphere: role and origins}\label{sec:CO2:origins}

Venus’ thick present-day atmosphere is dominated by CO$_2$, containing an amount of carbon comparable to Earth's combined atmospheric and crustal reservoirs \citep[e.g. ][]{DonahuePollack1983,wedepohl1995composition,Lecuyer2000,hartmann2012geochemical}. Given the uncertainties on the reservoirs, it is often suggested that most of Venus’ present-day total CO$_2$ inventory is now contained in its atmosphere \citep{Lammer2018} and that its mantle is therefore mostly degassed, but see Section \ref{models_melt_nobles} above. Due to the strong greenhouse effect from CO$_2$ the surface temperature is 737 K at the reference planetary radius of 6052 km, which is approximately 500 K higher than its equilibrium temperature \citep{DePater_Lissauer2013}. Understanding the evolution of CO$_2$ in the atmosphere is not only crucial for predicting the potential for the existence of liquid water on early Venus, but also provides insights into the divergent evolution of Earth and Venus. As discussed in \cite{salvador2022}, the solidification of a magma ocean may have outgassed large amounts of CO$_2$ into the early atmosphere, although it is difficult to constrain \citep[e.g.,][]{Bower2021,Gaillard2022a}. The present day atmosphere could be a combination of an early atmosphere resulting from magma ocean solidification, outside contribution from impactors \citep[both early and late]{Gillmann2020} and a later contribution from subsequent long-term magmatic mantle outgassing \citep{Lammer2018}.
 
While mantle composition first governs the availability of elements to be released, it also determines the chemical composition of the gases. The oxidation state of the mantle (linked to the oxygen fugacity parameter) controls how much C partitions into the melt during mantle partial melting \citep[for instance, ][ find that it is suppressed for oxygen fugacities below IW+2]{Ortenzi2020}. Then, as the magma rises to the surface, redox conditions also affect surface gas speciation during outgassing. Oxidized conditions (above IW+1) favor oxidized species such as CO$_2$, more reduced conditions (below IW+1) favor species such as CO \citep[e.g.,][]{Ortenzi2020}, or even CH$_4$ in more extreme situations \citep{wogan2020abundant}. Due to the lack of data, the oxidation state of Venus’ mantle is unconstrained. Only surface measurements of the mean FeO/MnO ratio by Venera 13, 14 and Vega 2 \citep{Surkov1984, Surkov1986}, have led to interpretations that its oxidation state lies between Earth and Mars \citep{Schaefer_Fegley2017}. As discussed before (section \ref{outgass}), early outgassing may occur under reducing conditions, possibly leading to the formation of a CO-rich atmosphere. If water is present in the atmosphere, or possibly if O accumulates due to hydrodynamic escape (see section \ref{hydrogen:oxygen:escape}), CO may convert efficiently into  CO$_2$, consistent with present-day observation. Primitive evolution scenarios are discussed further in \cite{salvador2022}.

Because of overall low CO$_2$ solubility, the decompression of ascending magma on Earth causes formation of gas bubbles deep below the surface \citep{Burton2013}. Although Venus' elevated atmospheric pressure slightly affects CO$_2$ degassing, its effect on degassing of H$_2$O and SO$_2$ is much stronger, and CO$_2$ is likely the most abundant volcanic gas on Venus \citep{Gaillard2014}. Surface pressure, in the thick modern-day CO$_2$ atmosphere, also affects the species that can be outgassed, as mentioned above. The existence and evolution of the thick CO$_2$ atmosphere significantly affects the ability of Venus to release water.

The accumulation of CO$_2$ in the atmosphere after solidification of the magma ocean depends on the tectonic mode and may have occurred either gradually over time or has been substantially enhanced by one or several catastrophic resurfacing events \citep{Lopez1998,BullockGrinspoon2001}. Based on Magellan data \citep{Schaber1992,Basilevsky1997}, it has been suggested that at least one global resurfacing event may have occurred 200-1000 Myr ago (the catastrophic evolution hypothesis). This was later self-consistently simulated with numerical models and \citep[see for instance]{armann2012,Rolf2018}, and is further discussed in \cite{Herrick2022}. If such an event occurred, it would have been accompanied by large-scale melting and therefore outgassing. The amount of CO$_2$ released by one such resurfacing event depends on the carbonate complexes concentration in the lava and thus the composition (and oxygen fugacity) of the mantle. Assuming Earth-like composition of mafic (basaltic) melts, one such global event could be responsible for a release of 5.6 $\times10^{19}$ kg of CO$_2$ (approximately 9 bar) \citep{Lopez1998} - under more reducing conditions in Venus' mantle the CO$_2$ release would be much lower. However, long-term continuous activity and outgassing could potentially lead to the same volatile build-up observed at present-day, without requiring any catastrophic event, and still satisfy $^{40}$Ar measurements \citep{Namiki_Solomon1998}. Some modelling efforts propose that it is unlikely that the bulk of the venusian atmosphere could come solely from mantle outgassing
\citep[][and compare with \cite{Weller2021,weller2022,Westall2022}]{Morschhauser2009,Gillmann2014,Gillmann2020}, either due to large initial exsolution from the magma ocean or because models suggest insufficient later CO$_2$ outgassing. Other sources may thus be required. However, the composition of the mafic melts (possibly water-poor tholeiitic) that could have formed most of Venus' plains could have allowed an efficient CO$_2$ degassing. Alkaline mafic melts ( in the Venera 13 case) could also release substantial amounts of CO$_2$ upon degassing. A relatively efficient CO$_2$ outgassing over Venus’s history is consistent with the abundances of CO$_2$ (gram per gram) in the Venus’ atmosphere versus Earth's crust \citep{Pollack1979,Lecuyer2000}.

In addition to the tectonic mode, the melt production and thus magmatic degassing rate of CO$_2$ depends on the temperature-depth gradient. First, the temperature-depth gradient controls the melting region below Venus’ lid. This gradient can depend significantly on the ratio between melt intrusion and melt extrusion \citep{Rozel2017,Lourenco2018}. On one hand, more extrusion directly translates into more outgassing, but on the other, more intrusion could locally affect the thermal profile in the lid and cooling rates, possibly easing melting. While it is likely that the more straightforward former effect dominates, possible feedback between the two processes has not yet been investigated in depth. Additionally, the temperature-depth profile is also affected by Venus’ surface temperature. It is unlikely that Venus' surface temperature has been constant throughout time: solar evolution \citep{Claire2012} could have contributed to a slow increase, or it could have varied widely with changes in the state and composition of the atmosphere. For example, water concentration and the presence or absence of clouds are major factors \citep{way2020venusian,Turbet2018b}. In particular a runaway greenhouse would have substantially increased the surface temperature, which would also have an effect on partial melting and degassing as discussed above.  

Altogether, in a reasonable but still unproven scenario, outgassing of CO$_2$ from the mantle into the atmosphere was likely stronger during the early evolution and then gradually decreased with time. This can be explained by the fact that the rate of mantle convection directly affects the degassing rate, as it controls the rate at which new mantle material enters the source region of the partial melt \citep{kite2009,o2014mantle}. After crystallization of the magma ocean, mantle temperatures are high and likely near the solidus \citep[see][]{salvador2022}, mantle viscosity is low, and thus mantle convection is particularly strong, presumably promoting rapid early degassing \citep{o2014mantle,Gillmann2014,tosi2017habitability}. Rapid outgassing is likely supported by the low $^{40}$Ar in the atmosphere, possibly indicating that a large part of $^{40}$Ar was not yet formed when outgassing occurred \citep{kaula_constraints_1999}, although precise estimations would require knowledge of both initial K content of the mantle and Venus' volcanic history. However, there were possibly one or several massive outgassing events related to catastrophic resurfacing, which eventually contributed to Venus’ thick, CO$_2$-rich atmosphere as observed today. Further, it could be that a substantial CO$_2$ atmosphere existed prior to the outgassing events and was released during the magma ocean solidification (a few bars to a few dozen bars).
A clear understanding of the complete outgassing sequence and timing and the respective contribution of the different processes at play to build-up the current atmosphere is thus still lacking.

\subsubsection{Constraining models of melt production with noble gases.}\label{models_melt_nobles}

In principle, noble gases in the atmosphere of Venus are clues to the outgassing history. Over the course of Venus' history several isotopes of noble gases have been produced by extinct ($^{129}$I, $^{244}$Pu) and extant (e.g., $^{238}$U, $^{40}$K) radioactive nuclides present in the interior of the planet. When a portion of Venus' crust or mantle melted these gaseous daughter nuclides migrated into the magmatic melt and were eventually degassed into the atmosphere. Outgassing produces radiogenic excesses on top of the primordial isotopic composition. While it remains difficult to put constraints on the efficiency of degassing of radiogenic gases from silicate reservoirs, models of outgassing of radiogenic noble gases suggest that, for Earth, degassing efficiency does not play the major role \citep{hamano_earth-atmosphere_1978,allegre_rare_1987}. Because the aforementioned radioactive nuclides have very different half-life times ($t_{1/2}(^{129}\rm I)=16\,Myr$, $t_{1/2}(^{244}\rm Pu)=82\,Myr$, $t_{1/2}(^{40}\rm K)=1.25\,Gyr$ and $t_{1/2}(^{238}\rm U)=4.47\,Gyr$), studying the abundances and relative proportions of their daughter products ($^{4}He, ^{40}Ar, ^{86}Kr, ^{129,131-136}Xe$) has the potential to give important constraints on the degassing history of Venus. A simplistic view is that relatively high rates of melt production over time should ultimately lead to higher abundances of radiogenic noble gases in the atmosphere today.

\begin{figure}
\centering
\includegraphics[width=12cm]{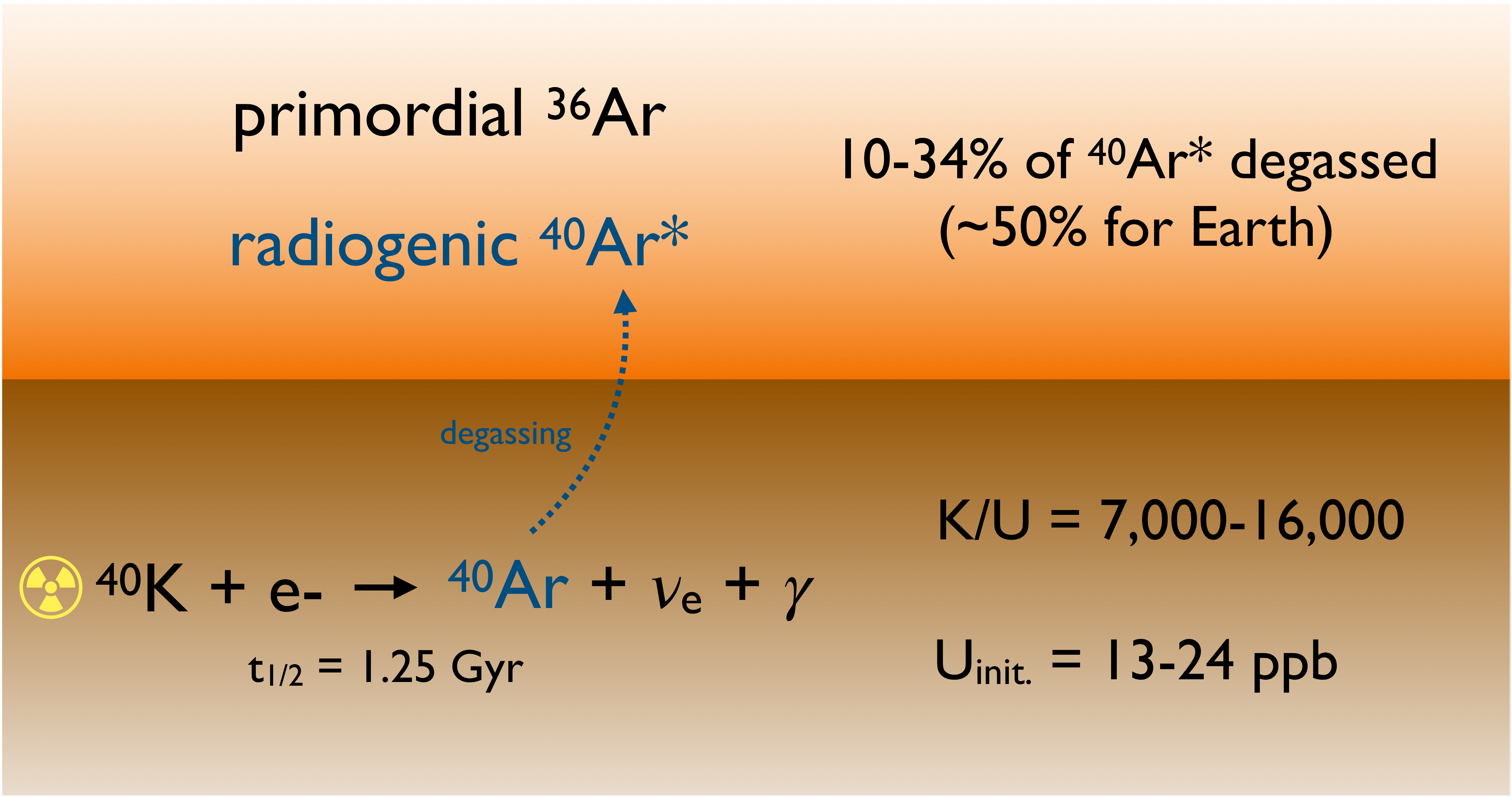}
\caption{Schematic explanation of how the degassing state of Venus (and Earth) has been estimated. Radiogenic $^{40}$Ar has been produced inside terrestrial planets by the radioactive decay of $^{40}$K. Simple mass balance considerations conclude that less radiogenic argon has been degassed from Venus interior to its atmosphere compared to Earth. Results are dependent on the assumed K/U ratio and the adundance of U in bulk silicate Venus, which are poorly constrained.
See \citet{kaula_constraints_1999,orourke_thermal_2015,allegre_rare_1987} and refs. therein for details on the method to evaluate the degassing state of terrestrial planets.}
\label{Fig:ArgonExplanations}
\end{figure}

About radiogenic contributions to the atmosphere of Venus, only the atmospheric $^{40}$Ar/$^{36}$Ar ratio for Venus is known with reasonable precision. This ratio, combined with estimates of the atmospheric abundance of $^{36}$Ar, gives an atmospheric abundance of $^{40}$Ar of 3.3 $\pm$ 1.1 $\times$ 10$^{-9}$ times the total mass of Venus while $^{40}$Ar is 12.7 $\pm$ 1.3 $\times$ 10$^{-9}$ times the total mass of Earth \citep{kaula_constraints_1999}. The $^{40}$Ar/$^{36}$Ar ratio is 1.11$\pm$0.02 \citep{istomin_venera_1983}, which is much lower than for Earth's atmospheric argon ($\approx$300, \citealt{ozima_noble_2002}). This difference has been interpreted as evidence that Venus experienced less outgassing than Earth through time (Fig. \ref{Fig:ArgonExplanations}).

Four missions (Venera 9 and 10, and Vega 1 and 2) used gamma rays to measure the abundances of K and U in 
surface material of their landing sites, which ranged from $\sim$0.3--0.5 wt\% and $\sim$0.5--0.7 ppm, respectively \citep{Surkov1987}.

The inferred K/U elemental ratio for bulk silicate Venus was $\sim$7000 by mass (see \citet{orourke_thermal_2015} for details). If this K/U ratio is correct and the bulk abundance of U is Earth-like, then Venus outgassed only about 10-34\% of the radiogenic $^{40}$Ar that would have been produced by the radioactive decay of $^{40}$K ($t_{1/2}$=1.25 Gyr) over the course of 4.56 Gyrs \citep{kaula_constraints_1999,orourke_thermal_2015,namiki_volcanic_1998,volkov_modeling_1993}. By comparison, Earth degassed $\approx$50\%  according to this metric \citep{allegre_rare_1987} although a recent study revisited the meaning of this result and suggested that half of the $^{40}$Ar budget has been subducted back into the Earth's mantle \citep{tucker_earths_2022}. If the K/U ratio for Venus is actually higher and closer to Earth-like values (i.e., up to ~16,000, \citealt{arevalo_ku_2009}), then Venus could be only ~10–12\% degassed, meaning that the vast majority of radiogenic $^{40}$Ar is still stored in the interior \citep{orourke_thermal_2015}. One idea is that less crustal production has occurred at Venus than at Earth. 

Speculatively, a basal magma ocean in the mantle could act as a ``hidden reservoir" of incompatible elements such as noble gases \citep{Jackson2021}. Such a basal magma ocean may exist today inside Venus \citep{ORourke2020}, or could have solidified recently so that its noble gas inventory has not been fully mixed throughout the mantle. On the other hand, though, one should note that the large inventory of N$_2$ in Venus' present-day atmosphere compared to Earth could be interpreted as evidence of comparatively strong outgassing during Venus' evolution (see section \ref{surfcond}).

Drawing any definite conclusions about the outgassing history of Venus from the present-day abundance of atmospheric $^{40}$Ar alone is very difficult. As stated above, all models rely on strong assumptions about the K/U ratio and the abundance of U in bulk silicate Venus \citep{lammer_constraining_2020}. For example, multiplying the bulk abundance of K by two in a model would mean that the best-fit rate of crustal production is halved. Furthermore, argon is assumed not to escape from the atmosphere, meaning that the timing of its outgassing is uncertain. \cite{kaula_constraints_1999} showed that a single, catastrophic resurfacing event could outgas the entirety of Venus’ atmospheric $^{40}$Ar---under a certain set of assumptions. However, several less dramatic outgassing episodes could release the same amount of $^{40}$Ar. \cite{Namiki_Solomon1998} demonstrated that steady (i.e., not catastrophic) outgassing, perhaps at different rates before and after a given transition time, is consistent with the present-day atmospheric abundances of both $^{40}$Ar and $^{4}$He. Building on these studies, \cite{o2014mantle} and \cite{orourke_thermal_2015} found that models with continuous stagnant lid convection in the mantle also yield acceptable trajectories of $^{40}$Ar outgassing for much or all of Venus’ history. 

Overall, new measurements of the chemical composition of rocks at the surface of Venus are needed in order to narrow down the range of estimates for the K/U ratio of bulk Venus. Precise knowledge of the bulk U abundance is required to translate the K/U ratio into the bulk K abundance that underpins outgassing models. Additionally, the I-Pu-U-Xe isotope systematics of Venus could be explored if $^{129,131-136}$Xe/$^{130}$Xe ratios of the Venus atmosphere are determined, which would allow us to put chronological constraints on degassing using only elemental ratios of refractory elements (U, Pu) and not absolute elemental abundances. These new investigations would help eliminate degeneracies between models of outgassing from the interior of Venus and would allow to build a coherent picture of Venus' geodynamics through time (see \cite{widemann2022} for a more details about future mission projects).

\subsection{\bf The sinks}\label{sec:sinks}
This section focuses on two types of sinks. The first, atmospheric escape has long been considered a dominant pathway for removing volatile elements from an atmosphere. We detail here the current state of escape-related observations (space \& surface based) of Venus including modelling efforts for both the present-day and past history (cumulative effect with time). A brief comparison with Earth and Mars is also included. This includes a review of issues related to extrapolating escape. The lack of solid global direct models of escape leads to uncertainties in escape rates under past conditions possibly different from the present-day observed atmospheric state of Venus (beyond solar energy input variation).

The second type of sink discussed is related to surface interactions as a possible alternative means of volatile loss on Venus. We consider the role of mineral reaction buffering for the atmosphere (for major gases like CO$_{2}$ or for minor ones, like SO$_{2}$). We next consider the possibility of mechanisms trapping volatile species, such as the oxidation of fresh basaltic material (to sulfates, iron oxides, and pyrite), as a possible way to remove an unknown amount of oxygen from the atmosphere. Further, consequences of crustal recycling are considered. We highlight different styles of recycling, e.g., plate tectonic-like, delamination, vertical advection, and crustal sequestration versus return to mantle.

\subsubsection{Hydrogen and oxygen escape during Venus' history}\label{hydrogen:oxygen:escape}

Observations by spacecraft and many theoretical studies have found that enhanced solar/stellar X-ray and extreme ultraviolet (XUV) radiation, solar wind plasma and coronal mass ejections (CMEs) result in forcing of the upper atmospheres of planets with no intrinsic magnetic field like Mars and Venus. The short wavelength radiation and the precipitating plasma flux can ionize, chemically modify, heat, expand, and erode upper atmospheres during a planetary lifetime \citep{Jeans1925, Chamberlain63,Chamberlain67,Opik63,Bauer04,Lammer2013}.

In addition to impact-related erosion (which we discuss later in this review), one can separate two main categories of atmospheric escape processes: (i) thermal escape and (ii) non-thermal escape. There are generally three thermal escape conditions possible: 

\begin{itemize}
    \item \textbf{Jeans escape:} loss of mainly atoms which populate the high energy tail of a Maxwell distribution \citep{ Jeans1925, Chamberlain63}.
    \item \textbf{Hydrodynamic escape:} very efficient escape of the bulk gas of the upper atmosphere where the atmosphere remains a collisional fluid as it passes through the transonic point \citep{Chamberlain63,Chassefiere1996b, Lammer2016, Owen16, Fossati17}. In such an extreme condition, nearly all atoms in the upper atmosphere exceed the escape velocity due to heating by XUV radiation and/or the surface temperature of a magma ocean. A special case of  hydrodynamic loss is the so-called boil-off where a whole atmosphere is not gravitationally bound to a planetary body and the thermal energy of the atmospheric gas overcomes the gravitational potential. This condition can occur when a hot magmatic low mass planetary embryo or protoplanet dissipates its primordial atmosphere during/after disk evaporation. There are also close-in exoplanets that could have experienced boil-off escape \citep{Owen16, Fossati17, LammerBlanc18}.
    \item \textbf{Slow hydrodynamic outflow:} a hybrid condition between Jeans and hydrodynamic escape, means that the lower thermosphere starts to expand hydrodynamically so that the exobase level is moved to high altitudes but the expanding gas cools adiabatically so that the escape can be described with a shifted Maxwellian \citep{Lammer2008,Tian08a,Tian08b}.
\end{itemize}

Depending on the atmospheric species and their mixing ratios, the exosphere can expand beyond an atmosphere protecting magnetopause for higher XUV fluxes than that of the present Sun \citep{Tian08a, Tian08b, Lichtenegger2010, Lammer2018}, so that non-thermal escape processes become relevant. For present solar activity conditions most non-thermal escape processes are relevant on non-magnetized planets. The most efficient known non-thermal atmospheric escape processes can be separated into two categories, which are ion loss processes and the loss of neutrals \citep[][, and references therein]{Lammer2013}. Ion loss processes are:

\begin{itemize}
    \item \textbf{Ion pick up:} planetary atoms are ionized and accelerated by electric fields within the solar/stellar wind plasma flow around a planetary obstacle.
    \item \textbf{Detached plasma clouds:} at non-magnetic planets plasma instabilities can cause wave structures at plasma boundaries such as the ionopause, where bubbles filled with cool ionospheric plasma can be detached.
    \item \textbf{Cool ion outflow:} At non- or weakly magnetized planets, planetary ions can also be accelerated by electrical fields to escape velocities throughout the tail.
    \item \textbf{Polar outflow/wind:} Ions that are accelerated via electric fields so that they reach escape energy and are lost over magnetospheric cusps of magnetized planets.
\end{itemize}

Besides these non-thermal ion escape processes, photochemical and particle interaction processes exist, such as:

\begin{itemize}
    \item \textbf{Dissociation:} Dissociative recombination and electron impact dissociation of molecular ions as well as photodissociation of neutral molecules than can yield neutral atoms with excess energies larger than the escape energy so that they can be lost or populate the exosphere with a suprathermal ``hot’’ atom population.
    \item \textbf{Charge exchange:} Charge exchange of solar wind protons with exospheric neutral particles can result in the loss of heavy neutral atoms that are ionized by the process which then produces an energetic neutral hydrogen atom from the former solar wind proton.
    \item \textbf{Atmospheric sputtering:} Precipitating solar wind ions or ionized exospheric atoms that are back-scattered into the upper atmosphere can act as sputter agents for atmospheric particles that are lost from the planet if their energy overcomes the escape energy.
\end{itemize}

One can expect that hydrodynamic thermal escape processes could have been involved in the loss of Venus’ primordial H$_2$-He-dominated atmosphere at the time when the disk dissipated \citep{Hayashi1979} and a few 10s to 100 million years after the planet's origin, if proto-Venus accreted primordial gas from the disk \citep{Gillmann2009,Lammer20a, Lammer20b}. Furthermore, hydrodynamic escape of hydrogen should also have played a role if Venus (i) produced water due to interaction of a primordial atmosphere with an underlying magma ocean \citep{salvador2022,Lammer21}; (ii) outgassed a huge amount of hydrogen and/or water vapour before and after its final magma ocean solidified or (iii) if the planet experienced a runaway greenhouse effect that evaporated a huge amount of H$_2$O \citep{Chassefiere1996a,Chassefiere1996b,Gillmann2009,Lammer2018}.

Hydrodynamic escape is most likely related to the escape of light species such as H, H$_2$ and He. It is also able to cause considerable O escape, directly during the first few hundred million years, and drag neutral O and heavier noble gases, even when they cannot escape directly \citep[for instance][]{hunten1973, zahnle_mass_1990, Gillmann2009, Odert2018}. However, this escape process depends on a complex interplay between the planet's gravitational potential and the atmospheric mixing ratio related to the availability of potential IR-cooling molecules in the planet’s thermosphere \citep[e.g.,][]{KastingPollack1983,Yoshida2020}. 

Here, we focus on the escape processes which have affected atmospheric evolution over most of the past 2.5 -- 3.0 Ga, the time range for which the increase in solar activity can likely be covered by the maximum activity during the present-day solar cycle. As mentioned before, thermal escape from Venus was important during its early history, but became negligible by present-day. Photochemical escape is only relevant for suprathermal H atoms \citep[e.g.,][]{Cravens1980,McElroy1982b,Rodriguez1984,Hodges1981,Hodges2000}. The most important escape process on Venus today is thus non-thermal escape, i.e. ion and energetic neutral escape. The main pathways for the non-thermal escape is either through the induced magnetotail of Venus (the nightside elongation of the induced magnetosphere), through pickup ions in the solar wind or through sputtering of neutral atoms via impinging on the atmosphere of the pickup ions \citep[][and references therein]{Lammer2006}. Thus far, there are no measurements on how much sputtering contributes to total escape from Venus, but simulations suggest that it is around 25\% of the total escape \citep{Lammer2006}. Measurements and simulations indicate that today the largest portion of the non-thermal escape is due to magnetotail escape \citep[e.g.,][]{Lammer2006, Masunaga2019}, which is what we focus on in this section.

Investigating the present non-thermal escape from Venus allows us to investigate the evolution of the D/H ratio and of oxygen back to the estimated time of the average age of surface, i.e., $\sim$ 0.2 -- 1.0 Ga \citep[e.g.,][]{Strom1994,Basilevsky1997,Kreslavsky2015,Bottke2016,McKinnon1997,Grinspoon1993}. If the observed present-day D/H ratio can be reproduced assuming the surface age as starting point for its initial reservoir - which should have likely been derived from chondrites, comets, the solar nebula or a mixture of these - all of them having significantly lower D/H ratios than present-day Venus' water \citep[e.g.,][]{Robert2000}, then Venus' present water was obtained during that time frame. In case the D/H ratio cannot be reproduced, it is an indication that the present water inventory is at least partially a remnant from before the average age of the present day surface. 
However, it neglects any additional unknown sinks for D and H that preferentially remove H from the atmosphere besides atmospheric escape.

\begin{figure}
\centering
\includegraphics[width=12cm]{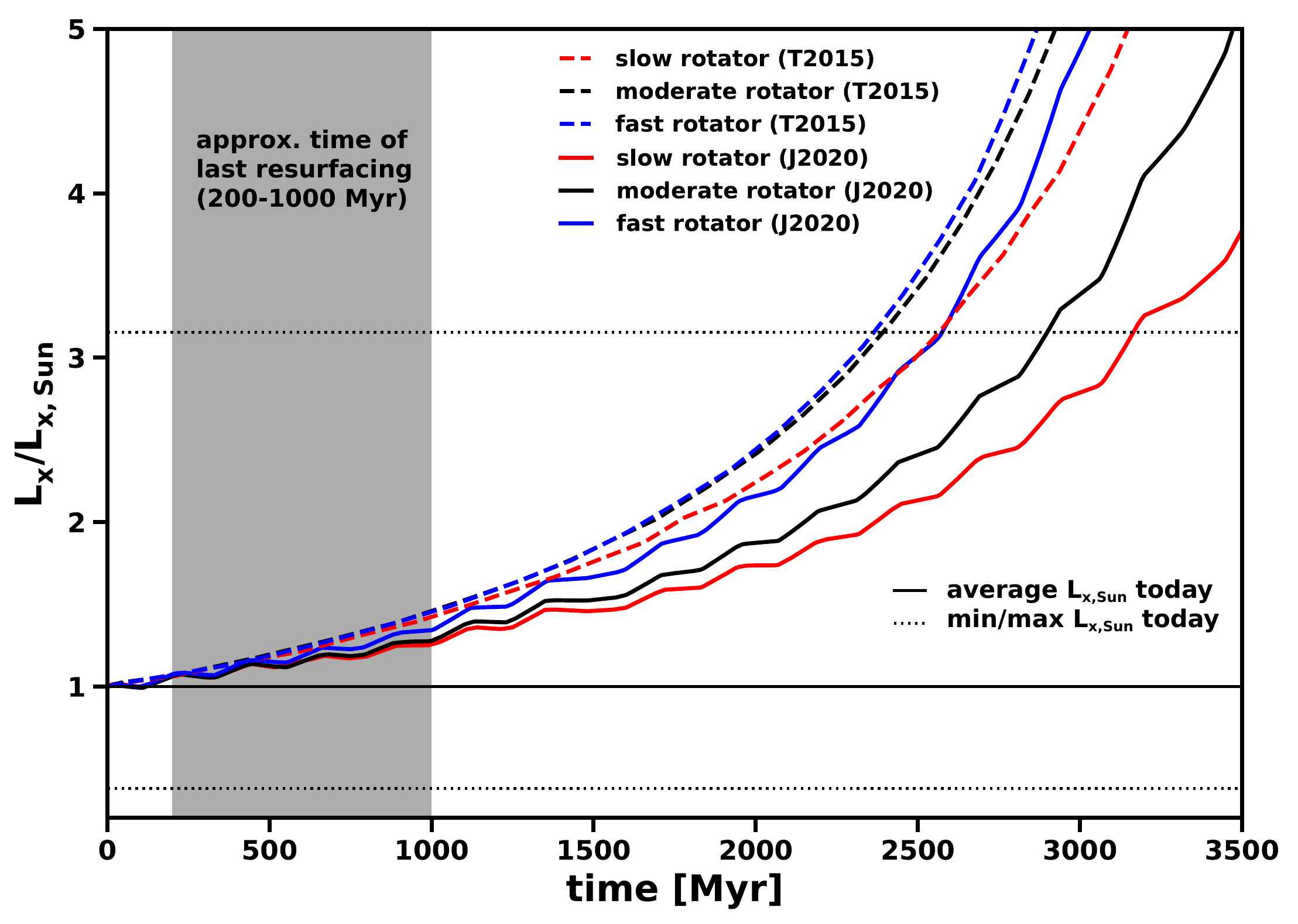}
\caption{The evolution of X-ray activity for Sun-like stars normalized to the Sun's present-day value of $10^{27.2}$\,erg\,s$^{-1}$\,cm$^{-2}$ according to \citet{Tu2015} (dashed lines) and \citet{Johnstone2020} (solid lines). Red, grey, and blue depict the tracks of initially slow, moderate, and fast rotating Sun-like stars. The evolutionary tracks by \citet{Tu2015} were employed to fit the Sun's average X-ray flux at present-day, while those by \citet{Johnstone2020} show the respective averages of the rotational tracks for all measured Sun-like stars. The grey shaded area shows the approximate average age of Venus' surface. As can be seen, this age falls clearly within a time period that can be covered by the variation of the X-ray flux over the solar cycle (solid line for average value, dashed lines for minimum and maximum).}
\label{Fig:Lx}
\end{figure}

The most recent measurements of non-thermal escape of H$^+$ and O$^+$ from Venus comes from the ASPERA-4 plasma package instrument \citep{Barabash2007b} on board Venus Express. The Venus Express measurements were performed over almost an entire 11 year long solar cycle (2006--2014), which allows us to reconstruct the escape of these ions during the past $\sim$ 2.5 -- 3.0 Ga assuming an identical atmospheric structure and lack of intrinsic magnetic field. For this we compare the observed solar XUV activity (with X-ray and EUV emissions from 1-10 nm, and 10-118 nm, respectively) with the reconstructed XUV history inferred from Sun-like stars with younger age by \citet{Tu2015} and \citet{Johnstone2021a}. This can be seen in Fig.~\ref{Fig:Lx} where the X-ray part of the spectrum can be directly obtained from observations of other Sun-like stars \citep[e.g.,][]{Airapetian2021}, whereas the EUV flux is normally assumed to directly scale with the respective X-ray emissions \citep{Tu2015,Johnstone2021a}. Such reconstructions are typically divided into different rotational evolution tracks (slow, moderate, and fast), since a star's XUV flux is correlated with the rotation rate of the respective star. For G-type stars these rotational tracks typically converge after $\sim$1 billion years. This makes it difficult to reconstruct the detailed activity history of the Sun. However, the XUV flux variation over the present-day Sun's solar cycle only reaches maximum values that can be compared to the XUV flux evolution of G stars for time periods at which the rotational tracks mostly converge. See Fig.~\ref{Fig:Lx}, which compares the maximum flux of the present solar cycle with the evolution of slow, moderate, and fast rotators.

Several studies have inferred escape rates of H$^+$ and O$^+$ using the ASPERA-4 measurements. Although the exact numbers of escape rates found for each study vary, depending on method and time period used for the analysis, in general the results and conclusions agree with each other \citep{Nordstrom2013}. On average today's H$^+$ and O$^+$ ions have escape rates at a ratio close to the stoichiometric ratio of H$_2$O of 2:1 \citep{Barabash2007a, Fedorov2011, Persson2018}, with an average O$^+$ escape rate of $(3-6) \times 10^{24}$ s$^{-1}$ \citep{Futaana2017}. However, the ratio varies with the solar cycle \citep{Persson2018}, and it is yet unclear whether O accumulates in the atmosphere since its actual abundance is not precisely measured \citep[with the mole fraction of O likely being $<$50 ppm, see][]{Johnson2019}. In case it does not accumulate in Venus' atmosphere, another oxygen sink has to be envisioned. It may additionally escape along with the hydrogen by a non-mass-dependent non-thermal escape process such as ion escape through moderately XUV-exposed upper atmospheres. On the other hand, some oxygen is consumed through oxidation of atmospheric gases (CO, COS, H$_2$S, S$_2$) and surface minerals to produce CO$_2$, SO$_2$, Fe$_3$O$_4$, Fe$_2$O$_3$ and sulfates \citep{Zolotov2018}. 

\begin{figure}
\centering
\includegraphics[width=12cm]{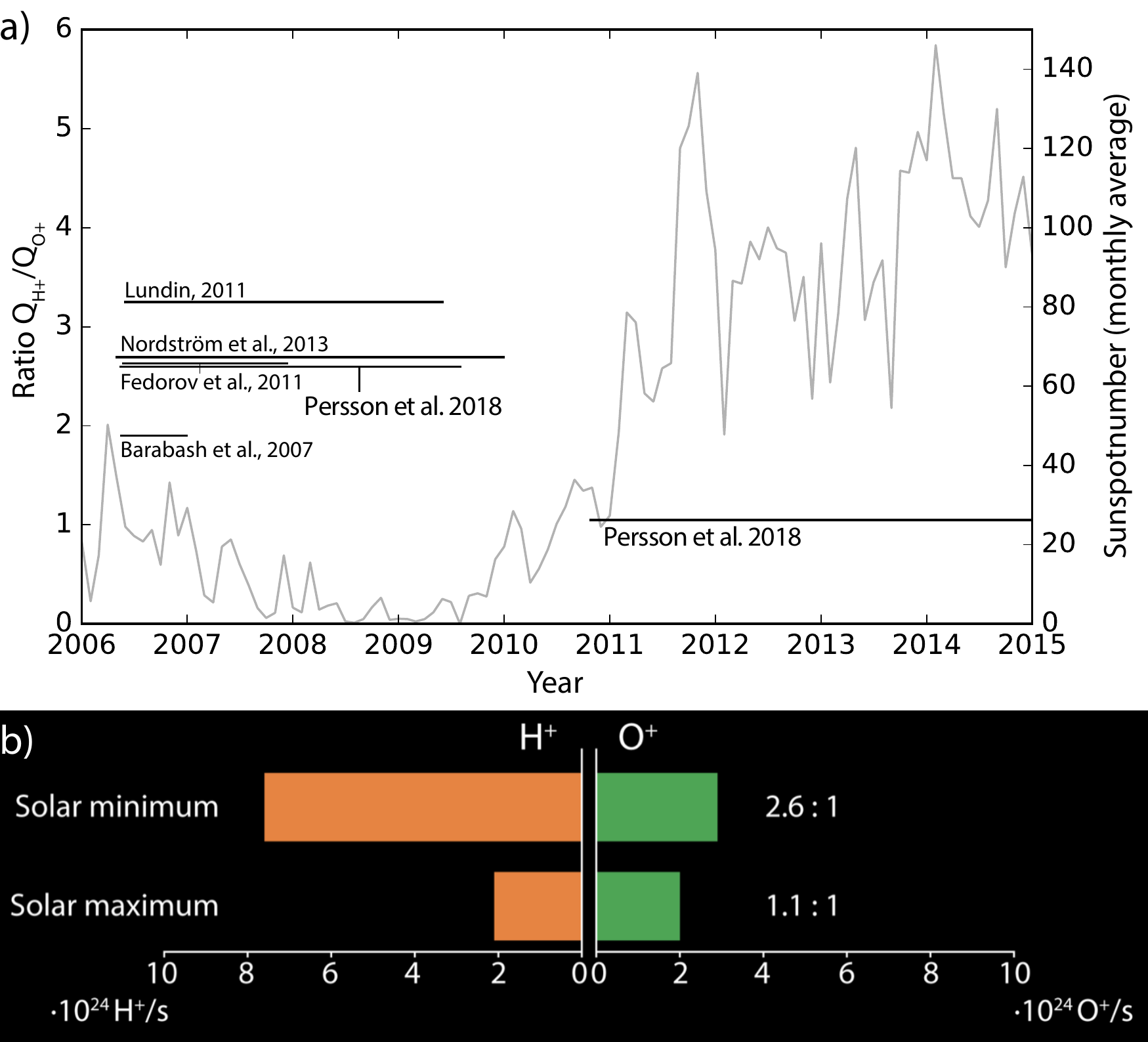}
\caption{a) Black lines indicate the ratio of H$^+$ and O$^+$ escape rates from different studies for the respective time periods in comparison to the results of \citet{Persson2018} based on the analysis of Venus Express ASPERA-4 data over the whole solar cycle. The gray curve depicts the monthly average sunspot number (\url{https://www.sidc.be/silso}). Figure adapted from \citet{Persson2018}. b) Illustration of the H$^+$ and O$^+$ ratios and escape ratios for solar minimum and maximum.}
\label{Fig:HO-ratio}
\end{figure}

Using all Venus Express measurements, \citet{Persson2018} found that the H$^+$ escape varied from $\approx7.6 \times 10^{24}$ H$^+$ s$^{-1}$ at solar minimum to $\approx 2.1\times 10^{24}$ H$^+$ s$^{-1}$ at solar maximum, while the O$^{+}$ escape rate exhibited smaller variation with $\approx2.9\times 10^{24}$ O$^{+}$ s$^{-1}$ at solar minimum to $\approx 2.0 \times 10^{24}$ O$^+$ s$^{-1}$ at solar maximum. The significant variations in the H$^+$ escape, as illustrated in Fig.~\ref{Fig:HO-ratio}b, over the solar cycle provide a change in the ratio between H$^+$ and O$^+$ escape rates from 2.6:1 at solar minimum to 1.1:1 at solar maximum. This means that more H$^+$ ions are lost, but at solar minimum only 1.1 times as much as O, and at maximum about 2.6 times. However, on average the value is in agreement with \citet{Barabash2007a} and is close to 2:1, indicating that no O gets enriched in Venus' atmosphere in relation to H, provided that both O and H originate from H$_2$O vapor. However, this assumption neglects any other sink for O, as we will discuss below.

The large decrease in H$^+$ escape from solar minimum to maximum is explained by an apparent change in the average direction of the proton flows in parts of the induced magnetotail. During solar maximum the protons have a larger contribution of ions flowing towards Venus instead of escaping through the magnetotail \citep{Kollmann2016,Persson2018}. On the other hand, \citet{Edberg2011} showed that in general the escape increases by a factor 1.9 from low solar wind dynamic pressure conditions to high solar wind dynamic pressure conditions. \citet{Persson2020} extended the study by further dividing the data set into ten bins using a combination of solar wind energy flux and solar XUV ranges. In agreement with \citet{Edberg2011}, they found a general increase in escape of O$^+$ as the solar wind energy flux increases, with variations between $\approx 9.0 \times 10^{23}$ O$^+$ s$^{-1}$ to $\approx 4.9 \times 10^{24}$ O$^+$ s$^{-1}$. A small decrease in the O$^{+}$ escape rates with an increase in the solar XUV flux was also found. In addition to the escape down the Venusian magnetotail, \citet{Masunaga2019} studied the escape of O$^{+}$ in the magnetosheath, and found that approximately 30\% of the total O$^{+}$ escape from Venus happens through the magnetosheath directly to the solar wind.

\begin{figure}
\centering
\includegraphics[width=12cm]{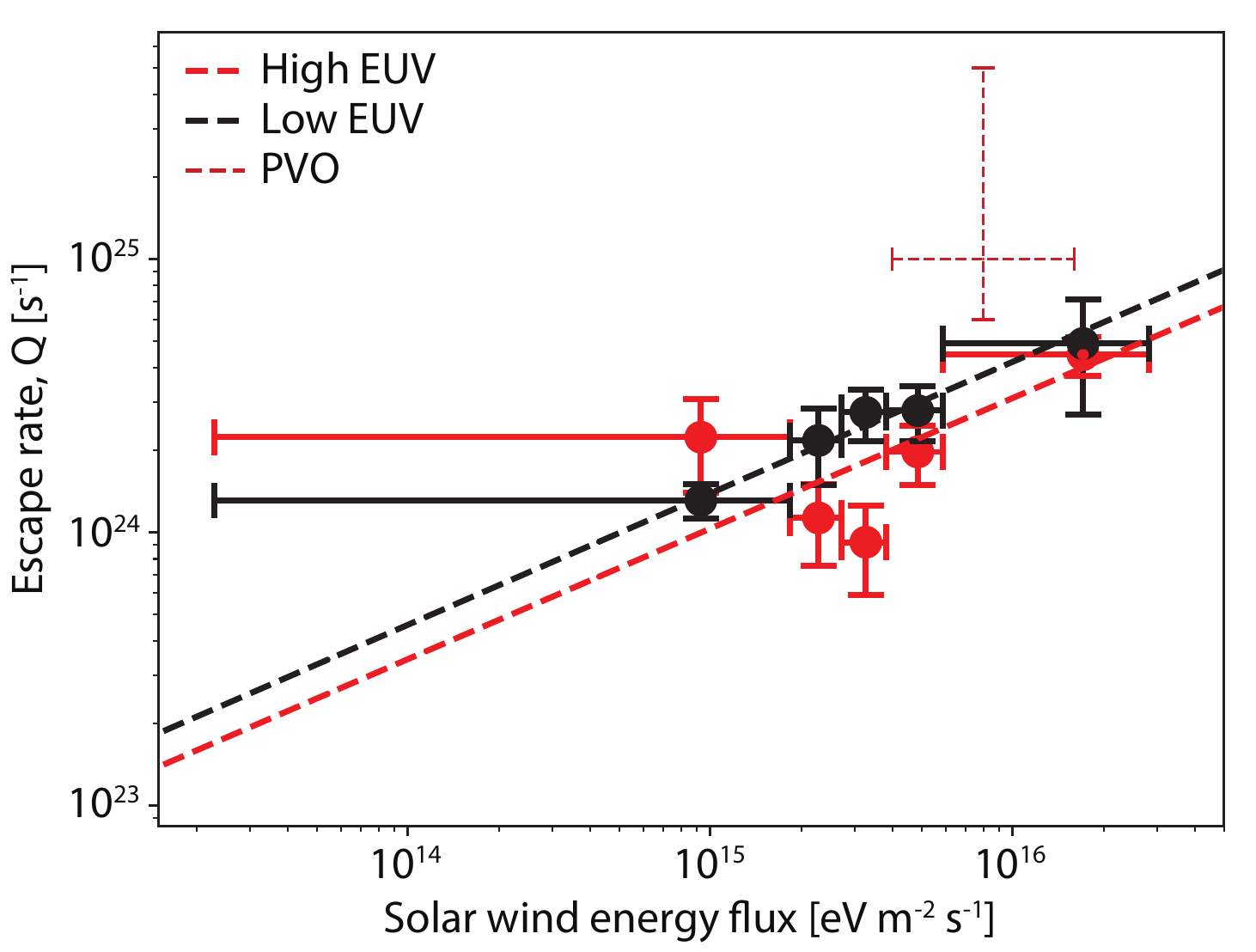}
\caption{O$^+$ escape rates from \citet{Persson2020} for five separated ranges of solar wind energy flux using high and low XUV flux including error bars. The dashed lines are best fits for high and low XUV, respectively. The red dashed cross shows the range of estimated escape rates based on PVO measurements \citep{Brace1987,Mccomas1986} and range of solar wind energy flux during the time PVO orbited Venus \citep{McEnulty2012}.}
\label{Fig:O-escape}
\end{figure}

Fig.~\ref{Fig:O-escape} shows the O$^+$ escape rate measured by Venus Express ASPERA-4 \citep{Persson2020} for five separate ranges of solar wind energy flux using high and low XUV flux which shows that O$^+$ escape diminishes with high solar activity due to the reasons mentioned above. An earlier mission that also studied the Venus plasma environment was the Pioneer Venus Orbiterv mission (PVO), which orbited Venus from 1978 until 1992 \citep{Colin1980}. Therefore, a comparison is included as the dotted-red cross, which shows O$^+$ escape estimates based on the PVO measurements of wave-like plasma irregularities above the Venusian ionosphere \citep[e.g.,][]{Brace1982,Russell1982}. Several studies assumed that such plasma clouds were detached from the ionosphere by plasma instabilities such as the Kelvin-Helmholtz and interchange instabilities and were therefore lost to space \citep{Wolff1980,Elphic1984,Terada2002,Arshukova2004}. Based on extrapolated occurrence rates of such clouds and assumed model parameters, \citet{Brace1982,Brace1987} and \citet{Mccomas1986} indirectly estimated the high O$^+$ loss rates depicted in Fig.~\ref{Fig:O-escape}, which likely overestimate the escape rates compared to Venus Express measurements. This, together with the higher solar wind energy flux and the stronger solar maximum during the PVO era, can potentially explain the differences found in escape rates between the two missions. 

Using the simple relationship between the O$^{+}$ escape rates and the solar wind energy flux in Fig. \ref{Fig:O-escape}, one can make a simple extrapolation backwards in time to estimate the total O$^{+}$ escape in time assuming the same atmospheric structure and lack of intrinsic magnetic field. This simple exercise gives a total loss of $\approx 3\times 10^{15}$ kg from now until 1 Ga, and $\approx 1.3\times 10^{16}$ kg in total from now to 3 Ga. Assuming that the oxygen originates from water, and converting this into a global water depth, the total water loss would be around a few millimeters at 1 Ga, reaching a few centimeters or a few decimeters at 3 Ga \citep{Persson2020}. These results indicate that oxygen originating from water was not lost through escape to space during the later evolution of Venus' atmosphere, but instead either was lost earlier during hydrodynamic escape or by oxidation of surface materials. However, it is important to note that these calculations assume that the atmospheric structure and the interaction between the Venusian ionosphere and the solar wind, and thus the non-thermal escape, remains unchanged over the considered time period. 

The Venus Express ASPERA-4 ion escape data obtained during the minimum and maximum solar activity conditions are in disagreement with the work by \citet{Hartle1993}. \citet{Hartle1993} modeled the H$^+$ loss rate from Venus' nightside due to acceleration by an outward electric polarization force related to ionospheric holes. These authors estimated this electric polarization force by assuming that electrons are more mobile than ions and therefore they may easily escape from the top of the ionosphere such that a charge separation occurs that leads to the polarization electric field force \citep{Hartle1993}. By using PVO data and ionospheric modeling they estimated a so-called hydrogen bulge sector which covers an area of $\approx 20\%$ of Venus' sphere that resulted in a planet wide averaged H$^+$ escape rate by this process of $\approx 7\times 10^{25}$ s$^{-1}$ \citep{Donahue1992a,Hartle1993}. This is $\approx 9 - 26$ times higher than the total H$^+$ ion escape rates inferred from the Venus Express ASPERA-4 data during minimum and high solar activity conditions.

This result is very relevant because it has consequences for the time span estimation that is required for escaping H and D to produce the present D/H ratio (HDO/H$_2$O) in Venus' atmosphere, as well as the characteristics of exogenous and endogenous hydrogen sources. According to \citet{Donahue1999}, unless the escape of D is very efficient on Venus, the present H escape rate averaged over a solar cycle should not exceed $5\times 10^{25}$ s$^{-1}$, if the present water vapor in Venus' atmosphere is a remnant of water deposited prior to the estimated average age of the surface \citep{Donahue1999}. If the H escape rate was larger during the past 100 million years then the planet's present atmospheric water would be a remnant of water outgassed only $\approx$500 Ma ago \citep{Grinspoon1993,Donahue1999}. Since the H$^+$ ion escape rates inferred from the Venus Express ASPERA-4 data during the solar cycle are much lower than $5\times 10^{25}$ s$^{-1}$, Venus' present atmospheric water is most likely a remnant that is older than the average surface age of Venus. Moreover, the results also indicate that the electric polarization force loss process via ionospheric holes as suggested by \citet{Hartle1993} overestimates the H$^+$ escape rates.

However, besides ion escape H and D also escape from Venus' upper atmosphere via the photochemical processes shown in the following equations,

\begin{eqnarray}\label{Eq:photo}
O^+ + H_2 \rightarrow OH^+ + H^*\hspace{1cm}\Delta E = 0.6 \hspace{1mm}eV,\\
OH^+ + e \rightarrow O + H^*\hspace{1cm}\Delta E = 8 \hspace{1mm}eV,\\
CO_2^+ + H_2 \rightarrow CO_2H^+ H^*\hspace{1cm}\Delta E = 1 \hspace{1mm}eV,\\
CO_2H^+ + e \rightarrow CO_2 H^*\hspace{1cm}\Delta E = 8 \hspace{1mm}eV,
\end{eqnarray}
and charge exchange processes,
\begin{eqnarray}\label{Eq:charge}
H^+ + O \rightarrow H^* + O^+,\\
H^+ + H \rightarrow H^* + H^+.
\end{eqnarray}

Present-day escape rates of suprathermal H$^*$ atoms that originate from the photochemical reactions shown above were studied by \citet{Lammer2006} with a Monte Carlo model. These authors obtained present-day suprathermal H$^*$ atom escape rates from the photochemical reactions given in Eq.~\ref{Eq:photo} on Venus' dayside of order $\approx 2.5\times 10^{24}$ s$^{-1}$. This is slightly lower than the earlier but simpler estimates of \citet{Mcelroy1982} of $\approx 4\times 10^{24}$ s$^{-1}$. Regarding the production of H$^*$ atoms via the charge exchange reaction shown in Eq.~\ref{Eq:photo}, the literature gives a wide range of escape rate values from $\approx 6\times 10^{25}$ s$^{-1}$ \citep{Hodges1981,Hodges1986} to $\approx 1\times 10^{25}$ -- $2.65 \times 10^{25}$ s$^{-1}$ \citep{Rodriguez1984}. A more detailed analysis of charge exchange related H$^*$ losses by \citet{Donahue1992a} and \citet{Hartle1996} revealed that the aforementioned studies used model inputs that did not fit the Venus International Reference Atmosphere (VIRA) empirical model that was developed based on the PVO ion mass spectrometer and electron temperature probe data. \citet{Donahue1992a} used ion profiles for their charge exchange model that were fitted to VIRA and obtained an H$^*$ escape rate of $\approx 2.65 \times 10^{25}$ s$^{-1}$. An analysis of charge exchange related H$^*$ for solar minimum to solar maximum conditions by \citet{Donahue1992b} resulted in escape rates from Venus' dayside of $\approx 1.3\times 10^{25}$ -- $2.65 \times 10^{25}$  s$^{-1}$, or on average $\approx 2.0 \times 10^{25}$ s$^{-1}$ which is higher than the H$^+$ escape rate of $\approx7.6 \times 10^{24}$ H$^+$ s$^{-1}$ as measured by ASPERA-4 (see above). Although there may also be some H$^*$ atoms produced on Venus' nightside, total H$^*$ escape rates that reach values $\ge 5\times 10^{25}$ s$^{-1}$ are very unlikely as long as the solar activity was not higher than at solar maximum. Taking the sum of ion and suprathermal escape for H, this yields a total loss of $\approx 2.8 \times 10^{25}$\,s$^{-1}$.

Suprathermal escape of oxygen atoms, on the other hand, is negligible, since the escape energy of O is almost twice as high as the energy of the photochemically produced hot atoms \citep{Lichtenegger2009,Groeller2010}. This imbalance between suprathermal H and O escape consequently alters the aforementioned loss ratio of H to O of $\approx$ 2:1 to a ratio that significantly deviates from 2:1. If one assumes that the majority of hydrogen originated from dissociated water, this would either suggest an accumulation of O in the atmosphere or an additional surface sink. However, H could also originate from other molecules than H$_2$O such as H$_2$SO$_4$, etc.

The findings of lower H$^+$ ion escape rates by \citet{Persson2018} compared to previous estimates by \citet{Hartle1993} and \citet{Donahue1999} indicate that Venus' present D/H ratio is most likely a remnant of an ancient water inventory that was mixed with more recently outgassed water vapor. D/H ratio has been modified by the non-thermal atmospheric escape processes discussed above on timescales of $\approx$200 - 1000 Myr (the average age of the surface) or more.

Even though no study has thus far estimated thermal and non-thermal loss of H and D at Venus even further back in time, it can be expected that at least thermal escape rates of its atmospheric species were significantly higher due to the elevated XUV flux from the young Sun, as different studies on the early Earth and exoplanets might indicate. Thermal losses of water, but also of CO$_2$ and N$_2$, were more significant earlier on \citep[e.g.,][]{Tian2008,Tian2009,Tian2018,Johnstone2020,Johnstone2021b}. This can be exemplified by a simple comparison between Earth and Venus. As has been shown by \citet{Johnstone2021b}, Earth would not have been able to maintain an N$_2$--CO$_2$ atmosphere with even 99\% CO$_2$ for an x-ray surface flux of 
$\geq
10$\,erg\,s$^{-1}$\,cm$^{-2}$, i.e., prior to $\approx$3.5 Ga if the Sun would have been a fast rotator. A nitrogen-dominated atmosphere with only 10\% CO$_2$ would have been unstable, i.e., a 1 bar atmosphere would escape to space within less than 10 million years for a surface flux of 
$\geq 5$\,erg\,s$^{-1}$\,cm$^{-2}$, i.e., until $\approx$3.0 Ga for a fast, and $\approx$3.5 Ga for a slow rotator, respectively \citep{Johnstone2021b}. Such high fluxes start to dissociate CO$_2$ \citep{Tian2009,Johnstone2021b}, the main infrared cooler in the upper atmosphere \citep[e.g.,][]{Roble1995}, resulting in hotter \citep[e.g.,][]{Cnossen2020} and more extended thermospheres and exobase levels. In extreme cases, the upper thermosphere can even hydrodynamically expand so that the gas flow cools adiabatically with the exobase level reaching several planetary radii, again resulting in massive atmospheric outflows due to the low gravitational potential at the exobase level \citep{Tian2008,Tian2009,Lammer2008,Johnstone2021b}. Oxygen as an additional constituent and heating agent in the upper atmosphere is not needed for an expansion under such high XUV fluxes \citep{Kulikov2006,Johnstone2019,Johnstone2021b} 

that are susceptible to strong thermal escape \citep[e.g.,][]{Tian2008,Johnstone2021b}. It is also important to note that besides the incident flux the mixing ratio of CO$_2$ with other species such as N$_2$ is the main factor that determines the upper atmosphere structure, while the density of the underlying atmosphere has negligible effects. Since Venus receives approximately twice as much XUV than the Earth and has only about 80\% of its mass, such atmospheres would have potentially been susceptible to escape to space at Venus until later times.

Finally, one should note that for such high escape rates, i.e., when the loss becomes hydrodynamic, H will not (or only marginally) be fractionated from D, since both species will escape together with the bulk atmosphere. Only when hydrodynamic escape transitions to Jeans escape D/H fractionation will start to become important \citep[see e.g.,][]{Lammer2020}. 
Therefore water loss from Venus cannot be simply estimated from the fractionation of D/H. This exercise is even more difficult considering that the starting D/H ratio for Venus' hydrogen is unknown. The extent of water loss would be a $\geq$60 m deep global ocean through non-thermal escape if one assumes carbonaceous chondrites as initial source of H$_2$O \cite[e.g.,][]{Donahue1999}. On the other hand it would be $\geq 420$ m in case the Venus' initial D/H ratio was defined by water that was produced due to surface interactions between a magma ocean and a primordial hydrogen-rich atmosphere \citep{salvador2022} that had a $\approx$7 times higher solar-like D/H ratio of $21\pm5\times10^{-6}$ \citep{Geiss1998,Robert2000}. 

If one assumes comets as a main reservoir, the water layer would have been even thinner, potentially ranging from $\approx$30\,m to only $\approx$4\,m, depending on the cometary D/H ratio \citep[see, e.g.,][for different D/H ratios within comets]{Robert2000}.
Several planned in-situ measurements of the HDO/H$_2$O ratio by the DAVINCI entry probe will constrain the D/H ratio in the bulk atmosphere along a vertical profile.

\subsection{Surface-gas reactions, mineral buffering and atmospheric evolution}
In this section, we discuss gas-mineral type reactions that deplete atmospheric volatiles and could control atmospheric composition before and during the last global resurfacing event, at present, and in the future. The mass of chemically active elements in the atmosphere (S, Cl, and F, but not C) is small compared to masses of those elements in permeable near-surface crustal materials, based on surface material compositions data from Venera/Vega landers \citep{Surkov1984,Surkov1986} and calculated by \cite{zolotov1992}. It follows that atmosphere-crust type reactions should have affected atmospheric composition throughout history. Such reactions trapped volcanic, impact-generated, chemically and photochemically produced O-, C-, H-, S-, Cl-, and F-bearing gases, as well as excessive O-bearing compounds (e.g., CO$_2$, SO$_2$, O$_2$) formed via net O accumulation after H escape. In turn, gaseous products of weathering reactions affected the atmosphere. In some cases, the gas-trapping reaction reached chemical equilibria that maintained (buffered) atmospheric abundances of the involved gases. 

\subsubsection{Atmospheric-surface reactions on ancient Venus}\label{atm-surface-reactions}
If Venus had an early aqueous history, atmospheric concentrations of chemically active gases (CO$_2$, H$_2$O, SO$_2$, etc.) were likely controlled by aqueous processes such as gas-water partitioning, dissolution and precipitation of solids, as well as erosion, transport and sedimentation. As on Earth, aqueous reactions would have allowed trapping of gases in secondary minerals such as salts (carbonates, chlorides, fluorides, etc.), phyllosilicates, oxides, and hydroxides \citep{Zolotov2009}. These likely occurred in a permeable surface layer on continents and were affected by local temperature, humidity and by water erosion.
A geological time-scale carbonate-silicate cycle \citep{walker1981negative} could have regulated the concentration of atmospheric CO$_2$ and thus surface temperature through the greenhouse effect of CO$_2$ (see Section \ref{sec:CO2:origins}). 

In the case of ‘runaway’ greenhouse warming without any condensation of surface water \citep{Ingersoll1969}, lower erosion rates and higher temperatures would have favored the formation of secondary mineral assemblages that could have controlled the concentrations of gases. In the absence of any major supply of gases and rocks, concentrations of certain atmospheric gases could be controlled by gas-solid equilibria rather than ongoing reactions. For example, the equilibrium assemblage of ferric and ferrous oxides in surface materials could control the CO/CO$_2$ ratio in the lower atmosphere (g denotes a gaseous species),

\begin{equation}\label{eqn:8}
\text{2Fe}_{3}\text{O}_{4} \ \text{(magnetite)} + \text{CO}_{2}\text{,g = 3Fe}_{2}\text{O}_{3} \ \text{(hematite) + CO,g}
\end{equation}

\noindent because the reaction equilibrium constant at a fixed temperature is expressed as 
K$_1$ = ($a$\ce{Fe2O3})$^3$ $f$\ce{CO}/(($a$\ce{Fe3O4})$^2$ $f$\ce{CO2}) $\sim$ $f$\ce{CO}/$f$\ce{CO2} $\sim$ $x$\ce{CO}/$x$\ce{CO2} 
where $a$, $f$, and $x$ stand for activity, fugacity, and mixing ratio, correspondingly. 
A change in gas concentration, e.g., through volcanic degassing, will cause the gas to interact with the corresponding mineral in order to restore its equilibrium concentration. Changes in temperature mean changes in the reaction constant, and therefore would facilitate forward or backward reaction to the new equilibrium concentrations of gases. For reaction \ref{eqn:8}, CO$_2$ degassing will cause the partial oxidation of the magnetite in the rocks to restore the CO$_2$/CO ratio. In contrast, heating will decrease the CO/CO$_2$ ratio in the atmosphere. As in laboratory gas-mineral buffers, natural buffers in permeable near-surface materials could operate until exhaustion of a reactant. In the case of Eq.~\ref{eqn:8} on Venus, the CO/CO$_2$ ratio would be controlled until complete oxidation of the rock’s magnetite to hematite.
Without liquid water, the massive atmospheric CO$_2$ may not be controlled by silicate-carbonate equilibria because of the inefficient carbonation of dry near-surface rocks \citep{Tanner1985} and other factors that are notable for the current epoch (Section~\ref{gas-solid:interactions}). However, hot and dense CO$_2$ is an efficient oxidizer of ferrous minerals and glasses \citep{berger2019experimental} and sulfide sulfur S(II). Corresponding gas-mineral reactions (e.g., Eq.~\ref{eqn:8}) could have affected $x$CO$_2$ throughout the anhydrous history. 

Volcanic S-, Cl-, and F-bearing gases (SO$_2$, S$_2$, H$_2$S, HCl, etc.) are highly reactive with rocks and atmospheric/surface dust. In other words, they are efficiently trapped in secondary minerals (salts, sulfides) and may not accumulate in the atmosphere without re-supply. If volcanic Cl and F are provided by the degassing of HCl and HF (this is not the case in H-depleted gases; \citealt{Zolotov2002}), rapidly formed secondary minerals would maintain $x$HCl and $x$HF at trace levels, that depend on temperature, mineralogy of buffering equilibria, and H$_2$O,g content \citep[e.g.,][]{Lewis1970,Fegley_Treiman1992}. 

Whether atmospheric SO$_2$-silicate reactions led to buffering sulfate-bearing mineral assemblages (e.g., CaSO$_4$-plagioclase; \citealt{Zolotov2018}) depended on the relative rates of SO$_2$ volcanic degassing, fresh rock supply, and gas-solid reactions. Although SO$_2$ trapping in minerals was unavoidable, slow SO$_2$-silicate reactions limited by the rates of metal (Ca, Na) diffusion \citep{King2018,berger2019experimental} could have maintained atmospheric $x$SO$_2$ above values of corresponding gas-solid equilibria, as at present (Section~\ref{gas-solid:interactions}).

Atmospheric $x$H$_2$O on early Venus could have been controlled by hydration-dehydration reactions involving salts, hydroxides, and phyllosilicates (serpentine, clay minerals, micas, and amphiboles). The hydrated mineralogy could have affected $x$H$_2$O either after the cessation of an aqueous epoch or during a greenhouse evolution without water condensation. At a constant temperature, net consumption of H$_2$O gas through H escape and oxidation of surface materials could have caused a step-wise decrease in $x$H$_2$O buffered by sequentially changing mineral assemblages in surface and subsurface materials. Each dehydration reaction, for example, could have maintained $x$H$_2$O until exhaustion of a dehydrating mineral (Eqs. \ref{eqn:9}, \ref{eqn:10}).

\begin{equation}\label{eqn:9}
\begin{aligned}
\text{(Mg,Fe)}_3\text{Si}_{2}\text{O}_{5}\text{(OH)}_{4} & \text{(serpentine)}  \rightarrow &\\
\text{(Mg,Fe)}_2\text{SiO}_{4} & \text{(olivine)} + \text{(Mg,Fe)SiO}_{3} \text{(pyroxene) + 2H}_{2}\text{O,g} \\
\end{aligned}
\end{equation}


\begin{equation}\label{eqn:10}
\begin{aligned}
\text{(Mg,Fe)3Si}_{4}\text{O}_{10}\text{(OH)}_{2} & \text{(saponite)} \rightarrow & \\
3 \text{(Mg,Fe)SiO}_{3} & \text{(pyroxene)} + \text{SiO}_{2} \text{(silica)} + \text{H}_{2}\text{O,g}
\end{aligned}
\end{equation}

 Micas and amphiboles could be the last H-bearing phases before the complete decomposition of H-bearing minerals that is expected at present \citep{Zolotov1997}. If surface temperature increased, the dehydration of minerals (e.g., Eqs. \ref{eqn:9},\ref{eqn:10}) would have caused a positive feedback because the released H$_2$O is a strong greenhouse gas. The negative feedback, through slowing mineral dehydration in an increasingly H$_2$O-rich atmosphere, could not have prevented a rapid and thorough dehydration, nor the corresponding temperature spike. It is unclear if subsequent consumption of H$_2$O,g through H escape and oxidation of surface/atmospheric species could have allowed temporary formation of micas or amphiboles on lower-temperature highlands.

The high D/H ratio in the current atmosphere \citep{Donahue1982,DeBergh1991} may imply H escape and net oxidation of rocks by remaining net water’s oxygen \citep[e.g.,][]{KastingPollack1983,Lewis_Prinn1984}, but see Section \ref{hydrogen:oxygen:escape} for a more detailed picture. The oxidation of exposed surface materials as an alternative to atmospheric O escape has been considered in recent publications
\citep[][]{Gillmann2009,Gillmann2020,Wordsworth2018,krissansen2021Venus,way2020venusian,Warren_Kite2021LPSC} though the pathway of post magma ocean O consumption remains to be better quantified. 
In one pathway, photochemically produced oxygen was consumed through oxidation of atmospheric gases 
(CO $\rightarrow$ CO$_2$; COS, S$_2$, H$_2$S $\rightarrow$ SO$_2$; H$_2$ $\rightarrow$ H$_2$O) followed by oxidation of rocks (e.g., Eq. \ref{eqn:8}) by formed O-bearing gases. In a parallel pathway(e.g., Eqs. \ref{eqn:11},\ref{eqn:12}), the oxygen was consumed through rock-H$_2$O,g reactions leading to oxidized solids (Fe(III) oxides, pyrite, sulfates) and subsequent H$_2$ loss \citep[e.g.,][]{Khodakovsky1982}. However, the accumulation of atmospheric O$_2$ above trace amounts is unlikely because of its reactivity with reduced atmospheric gases and crustal solids, if O$_2$ ever reached the surface. The amount of oxygen consumed by rocks would have depended on the unknown amount of water that supplies oxygen through photo-dissociation in the upper atmosphere and water-rock reactions followed by H escape.
\begin{equation}\label{eqn:11}
\begin{aligned}
\text{3FeO} \ \text{(in silicates) + H}_{2}\text{O,g}  \rightarrow \text{Fe}_{3}\text{O}_{4} \ \text{(magnetite) + H}_{2}\text{,g}
\end{aligned}
\end{equation}

\begin{equation}\label{eqn:12}
\begin{aligned}
\text{2Fe}_{3}\text{O}_{4} \ \text{(magnetite) + H}_{2}\text{O,g}   \rightarrow \text{3Fe}_{2}\text{O}_{3} \ \text{(hematite) + H}_{2}\text{,g}	
\end{aligned}
\end{equation}

 Using Earth-like FeO concentration \cite{Lecuyer2000} proposed that 10$^{21}$ kg of water (roughly one Earth ocean) could be removed by the oxidation of a global layer of $\approx$50 km of basaltic crust into hematite \citep[also see,][]{KastingPollack1983,Lewis_Prinn1984}. For evolution scenarios, this must be nuanced by actual volcanic production rates required to bring fresh basalt to the surface and bury older material. Earth production ($\approx$20 km$^{3}$yr$^{-1}$; \citealt{Morgan1998}) is assumed to be a maximum for recent Venus, but no data exist for $\geq$1 Ga Venus. The efficiency and degree of crustal oxidation is also uncertain, but would be limited once the surface is solid. Experimental studies using Venus-like conditions and atmosphere composition \citep[e.g.,][]{berger2019experimental} showed that the alteration was limited to surface oxidation or a coating at the surface of olivine, but that it occurred rather fast: the Fe content in a 10s to 100s of nanometers thick layer could be oxidized within about 15 days \citep{teffeteller2022}. Other important factors would include the mineralogy of formed oxidized phases, kinetics of oxidation reactions, grain size and crystallinity of altering materials, and the depth of permeable surface layers. By contrast, oxidation of the liquid material before it solidifies could be much more efficient. This is especially true during the magma ocean phase \citep{Gillmann2009,Schaefer_Wordsworth2016,Warren_Kite2021LPSC,salvador2022}. However, some simulations have shown that significant O$_2$ atmospheres can still remain due to H$_2$O photolysis once the magma ocean freezes over \citep{Wordsworth2018,krissansen2021oxygen}.
 
 In addition to oxidation of ferrous silicates (Eqs. \ref{eqn:8}, \ref{eqn:11}, \ref{eqn:12}), formation of pyrite at the expense of ferrous silicates and/or sulfides and sulfatization of Ca-bearing silicates, glasses, and putative ancient water-deposited carbonates by SO$_2$ (\ref{eqn:13}) could have provided an additional sink of atmospheric O throughout the anhydrous history, including the current epoch \citep{Zolotov2018}.

\begin{equation}\label{eqn:13}
\begin{aligned}
\text{CaO \ (in silicates and/or carbonates) + 1.5SO}_{2}\text{,g} \rightarrow & \\
\text{CaSO}_{4} \ \text{(anhydrite)} & + \text{0.25S}_{2}\text{,g},
\end{aligned}
\end{equation}


In the hypothetical case of several global volcanic resurfacing events, the intensity of surface oxidation could have been sluggish between the events. During periods of limited supply of fresh volcanic materials to the atmosphere and surface, the oxidation rate might roughly be comparable to the rate of removing excess oxygen through non-thermal escape \citep[][see above]{Persson2020}. Whether gas-solid reactions buffered atmospheric gases between resurfacing events depended on volcanic activity. In the case of minute volcanic supply of rocks and gases, as suggested for the current epoch \citep{Schaber1992,Basilevsky1997} and limited physical weathering and erosion, buffering is likely. Secondary minerals (Fe oxides, pyrite, sulfates) in permeable surface materials formed through oxidation by CO$_2$, H$_2$O, and SO$_2$ (Eqs. \ref{eqn:8}, \ref{eqn:11}--\ref{eqn:13}), could have controlled an array of redox-dependent atmospheric gases. For example, the magnetite-hematite assemblage is likely a major player in controlling ratios of reduced (COS, CO, S$_2$, H$_2$S, H$_2$) and oxidized (CO$_2$, H$_2$O, SO$_2$) gases at Venus’ present and future (Sections~\ref{gas-solid:interactions},~\ref{atmosphere-lithosphere}). Because of gas-solid type redox equilibration in the near-surface atmosphere, further oxidation is only driven by a supply of excess O, which becomes limited for recent history and without endogenic and/or exogenic (comets, carbonaceous chondrites) supplies of H$_2$O gas to the atmosphere.

Global volcanic resurfacing events, if they occurred, likely eliminated existing buffering reactions due to changes in surface composition, volcanic outgassing, and related greenhouse warming \citep{Solomon1999}. Elevated concentrations of volcanic SO$_2$, COS, S$_2$, CO, CO$_2$, H$_2$O, HCl, and HF in the atmosphere, increased greenhouse temperatures, and exposure of fresh volcanic materials facilitated trapping of the gases in minerals. HCl and HF gases would have needed to be efficiently trapped and would likely have reached minute gas-solid equilibrium concentrations at the corresponding temperatures. Elevated concentrations of degassed H$_2$O,g \citep{berger2019experimental} could have favored oxidation of solids (e.g., ferrous species to Fe oxides; \citealt{Warren_Kite2021LPSC}) and H$_2$ formation (Eqs.~\ref{eqn:11},\ref{eqn:12}). High temperatures could have facilitated oxidation of ferrous and sulfide compounds by hot and dense CO$_2$ (Eq.~\ref{eqn:8}). 

Likewise, a rapid interaction of hot SO$_2$ with silicate minerals and glasses \citep{Renggli2018} could cause sulfatization of the exposed surfaces of rocks, ash, and dust via sulfur disproportionation reactions exemplified by the sulfatization of Ca in silicates and glasses (e.g., Eq.~\ref{eqn:13}). As on Earth \citep{Delmelle2018} alterations of lava surfaces and pyroclastic products could have occurred while the materials were still hot, though they cool faster on Venus than on Earth due to its dense atmosphere \citep{Frenkel_Zabalueva1983}.

Oxidation of surface materials and trapping of atmospheric HCl and HF could have led to the secondary mineralogy on the current surface (Section~\ref{gas-solid:interactions}). However, the oxidation potential of the atmosphere could be limited because of overall reduced nature of planetary volcanic gases \citep{Gaillard2014} and because of the accumulation of reduced products (CO, COS, H$_2$, S$_2$, H$_2$S) of gas-solid reactions (Eqs.~\ref{eqn:8}, \ref{eqn:11}-\ref{eqn:13}). A low mass (up to a $\%$) of degassed compounds (H$_2$O, SO$_2$, HCl, HF), compared to that of erupted rocks, implies that alteration of exposed rocks would be limited. The burying of earlier altered lava flows by fresh ones would remove trapped volcanic volatiles from the atmosphere-surface system, and favor further trapping but not gas-solid equilibration. Consumption of degassed H$_2$O through rapid oxidation (Eqs.~\ref{eqn:11},\ref{eqn:12}) and H escape would cool the atmosphere. Limited resurfacing during hundreds of Myr could favor co-evolution of trace atmospheric gases with buffering mineral assemblages that partially formed during and shortly after more ancient and intense volcanic activity.

\subsubsection{Current gas-solid interactions and atmospheric buffering}\label{gas-solid:interactions}
Current Venus’ atmosphere-surface system is a checkpoint to constrain effects of gas-solid reactions and corresponding equilibria on atmospheric composition. Since the early 1960s, atmosphere-surface interactions and buffering reactions have been considered in multiple journal publications \citep[e.g.,][]{Mueller1964,Lewis1970,Barsukov1982,Klose1992,Semprich2020} and book chapters \citep[e.g.,][]{Fegley_Treiman1992,Fegley1997,Wood1997,Zolotov2018}. This Section provides a summary of the current views.

\begin{figure}
\centering
\includegraphics[width=9cm]{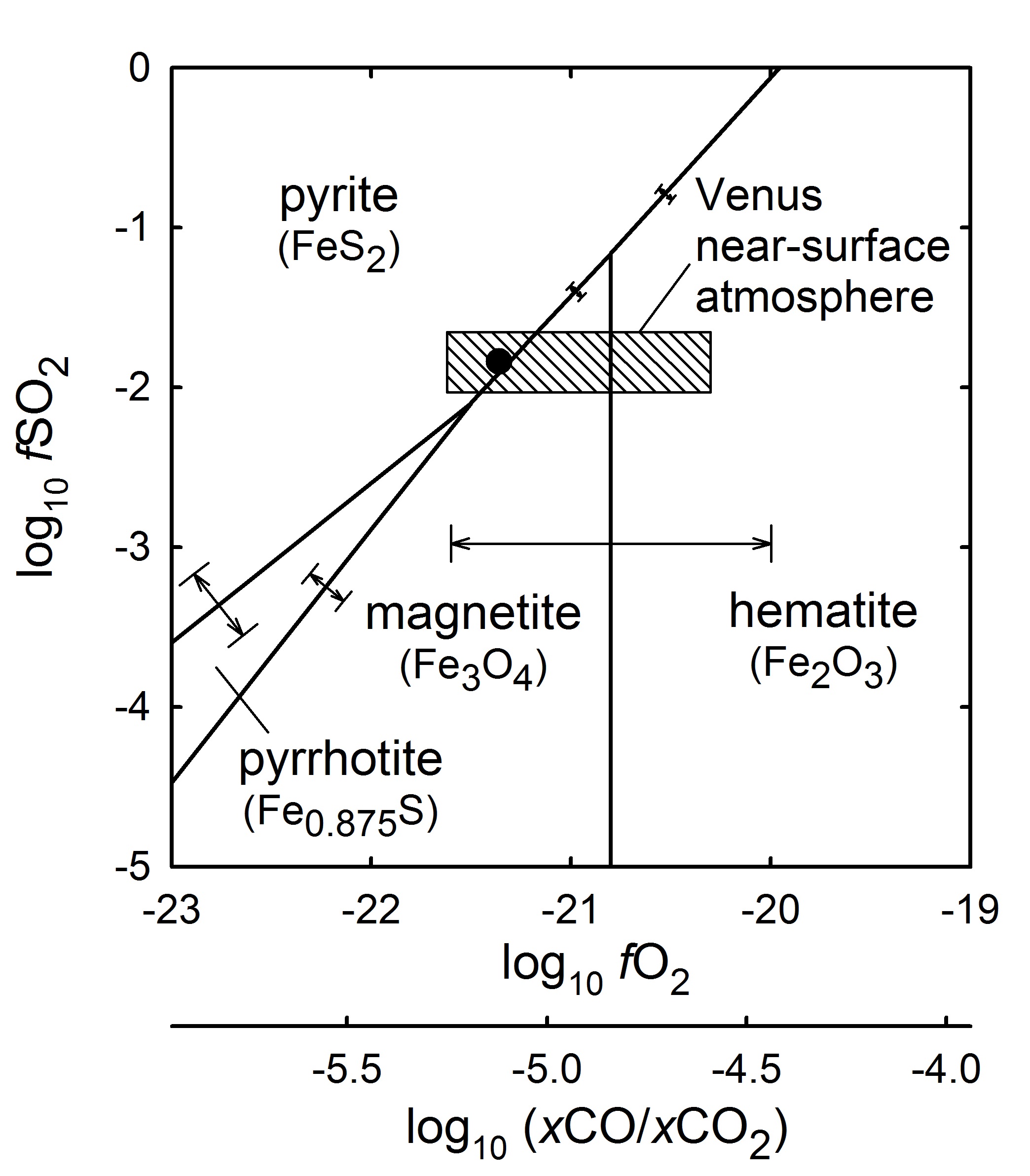}
\caption{Stability fields of iron oxides and sulfides at the conditions of modal Venus radius (740 K, 95.6 bars). The Venus box corresponds to mixing ratios of SO$_2$ (95-228 ppm) and CO (5-23 ppm) based on Pioneer Venus and Venera entry probe data with uncertainties. The lowest CO value reflects an extrapolation of the atmospheric gradient toward surface \citep{Fegley1997b}. The circle symbol corresponds to $x$SO$_2$ = 150 ppm and $x$CO = 17 ppm. The $f$O$_2$ data are for the CO-CO$_2$ gas equilibrium. The error bars reflect uncertainties of thermodynamic data of minerals. The figure shows that Venus gas composition is close to conditions at iron oxide and oxide-sulfide equilibria. Modified from \cite{Zolotov2019}.
}
\label{Fig:MZ1}
\end{figure}

Although the composition of the sub-cloud atmosphere remains uncertain \citep{Marcq2018}, existing data do not exclude gas-gas type equilibration at Venus’ lowlands at $\approx$740 K \citep[e.g.,][]{Fegley1997,Zolotov1996,Krasnopolsky2007}. Partial pressures of equilibrated gases match those at the hematite-magnetite (Hem-Mt) (Eq. 1), hematite-magnetite-pyrite (Hem-Mt-Py) and magnetite-pyrite (Mt-Py) mineral equilibria \citep[e.g.,][Fig. \ref{Fig:MZ1}]{Fegley1997,Zolotov2018}.

The match does not exclude gas-solid equilibration and these minerals are expected weathering products formed during \citep{Warren_Kite2021LPSC} and after a global volcanic resurfacing event. The presence of hematite is consistent with the near-infrared reflectance and color of surface materials at the landing sites of Venera 9, 10 and 13
\citep{Pieters1986,Shkuratov1987,Yamanoi2009}. A hematite-rich layer could be at least a few micrometers thick in order to be optically thick \citep{Gilmore2017}. The current working hypothesis is that Hem-Mt-Py surface mineralogy buffer key ratios of reduced and oxidized gases, CO/CO$_2$, SO$_2$/(COS, H$_2$S, S$_2$), and H$_2$/H$_2$O. If the buffering works, current atmospheric gases are not involved in corresponding weathering reactions, changes in atmospheric composition will be compensated by gas-solid reactions (e.g., Eq.~\ref{eqn:8}), in one direction or another, and major changes in surface mineralogy (e.g., via volcanism) will affect atmospheric composition and its redox state. In contrast to thermodynamically favorable oxidation of Fe(II) in surface materials to magnetite by atmospheric CO$_2$, H$_2$O and/or SO$_2$, formation of hematite will only be driven by net oxidation of the atmosphere that currently occurs through a slow H escape (see Section \ref{hydrogen:oxygen:escape})

It is unclear whether minerals buffer abundances of S-bearing gases. On the one hand, atmospheric $x$SO$_2$ is high enough to allow alteration of Ca-bearing pyroxenes to CaSO$_4$ (Eq.~\ref{eqn:13}), as inferred from calculations of chemical equilibria \citep[e.g.,][]{Barsukov1982,Fegley_Treiman1992,Zolotov2018} and supported by modeling experiments \citep[e.g.,][]{berger2019experimental}. The high but variable S/Ca ratio in three Venera and Vega probes of surface basaltic material \citep{Surkov1984,Surkov1986} indicates the ongoing trapping of atmospheric S. On the other hand, plagioclase, a major mineral in basalt, is more resistant with respect to Venus’ SO$_2$ and a plagioclase-anhydrite assemblage could have a buffering effect on SO$_2$ \citep{Zolotov2018}. The low current rate of H escape and a comparable rate of O escape (Section \ref{hydrogen:oxygen:escape}) imply minute (if any) net supply of oxidants to the atmosphere. This does not contradict the buffering of atmospheric redox conditions by secondary minerals. 

The minute abundances of atmospheric HCl and HF are insufficient to alter surface minerals and suggest a major anterior trapping of the gases \citep{Zolotov2018}. Although the mineralogy of secondary Cl- and F-bearing phases remains unclear \citep{Lewis1970,Fegley_Treiman1992}, buffering of atmospheric HCl and HF by gas-solid equilibria is highly plausible. In contrast, buffering of H$_2$O gas by H-bearing minerals is unlikely because of their instability at the current surface \cite{Zolotov1997} and a thorough thermal dehydration related to extreme greenhouse heating during volcanic resurfacing events. Likewise, mineral buffering of CO$_2$ is unlikely because of the instability of carbonates and carbonate-silicate equilibria with respect to greenhouse warming \citep{Hashimoto_Abe2005}, because of inhibited carbonation without aqueous fluid \citep{Tanner1985}, because of the instability of Ca carbonates and silicates (except plagioclase) with respect to Venus’ SO$_2$, and other reasons \citep{Zolotov2018}. A match of Venus’ surface $f$CO$_2$ with $f$CO$_2$ at the carbonate-silicate equilibrium 

\begin{equation}\label{eqn:14}
\begin{aligned}
\text{CaCO}_{3} \ \text{(calcite) + SiO}_{2} \ \text{(quartz) = CaSiO}_{3} \ \text{(wollastonite) + CO}_{2}\text{,g}	
\end{aligned}
\end{equation}


\noindent \citep[e.g.,][]{Mueller1964,Nozette_Lewis1982,Fegley_Treiman1992} is probably accidental.

\begin{figure}
\centering
\includegraphics[width=12cm]{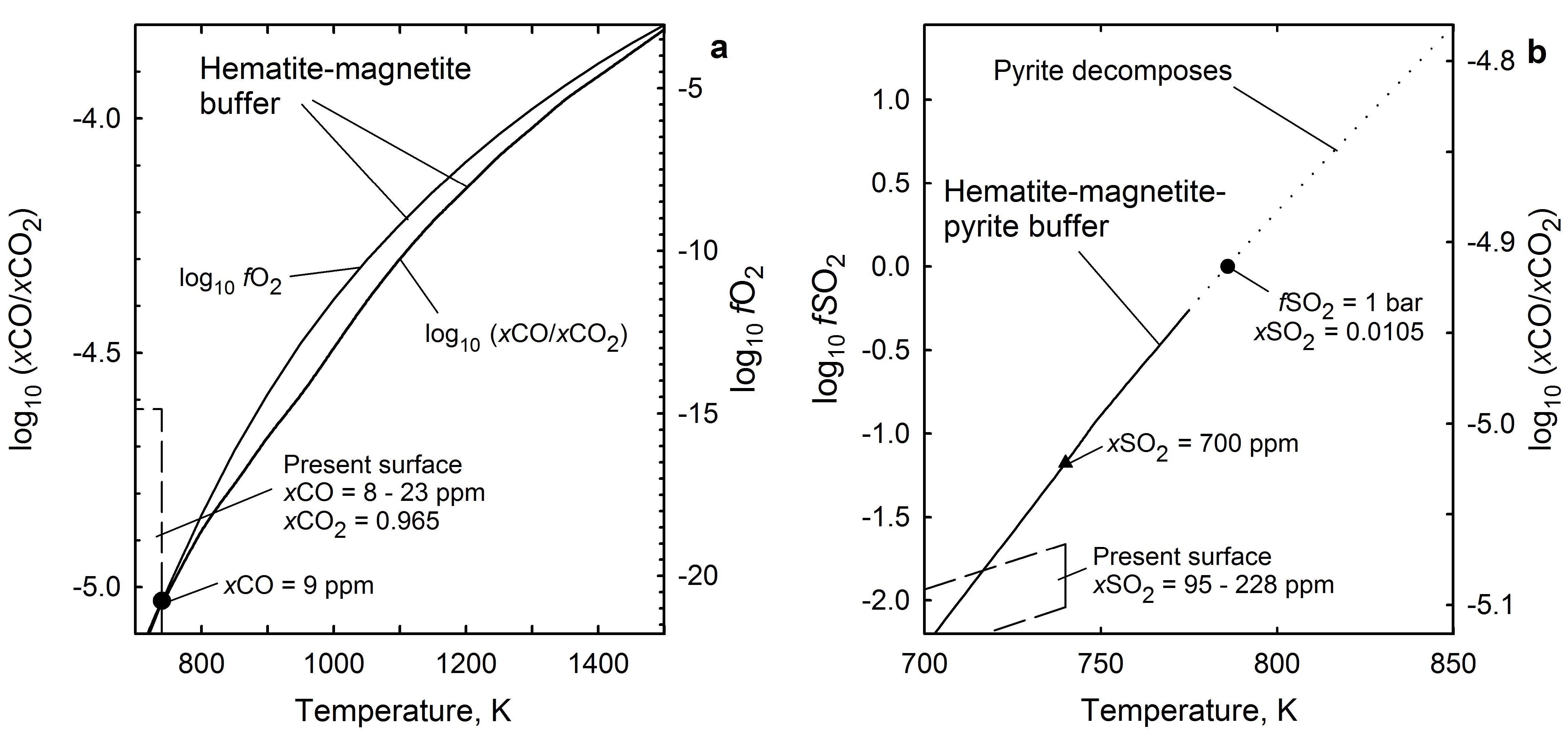}
\caption{Possible future changes in concentrations of trace atmospheric gases controlled by mineral-gas equilibria at 95.6 bar. In (a), CO/CO$_2$ mixing ratio and $f$O$_2$ are controlled by the hematite-magnetite equilibrium (Eq.~\ref{eqn:8}) during warming. The area shown by dash lines corresponds to the present near-surface CO/CO$_2$ ratios (see Fig.~\ref{Fig:MZ1} for $f$O$_2$). The circle symbol shows conditions at which CO extrapolated toward the surface \citep{Fegley1997b,Zolotov2019} closely match the equilibrium (Eq.~\ref{eqn:8}). In (b), fugacity of SO$_2$ is controlled by the hematite-magnetite-pyrite equilibrium at the CO/CO$_2$ ratio at Eq.(~\ref{eqn:8}). The fugacity of SO$_2$ in the current near-surface atmosphere reflects a range of measured SO$_2$ concentrations with uncertainties. The triangle symbol shows nominal conditions of the mineral equilibrium at 740 K (see Fig.~\ref{Fig:MZ1}). The circle symbol corresponds to partial pressure of SO$_2$ of 1 bar at 786 K. The dotted curve shows possible conditions of through decomposition of pyrite in a permeable surface layer. (This figure was developed by M. Zolotov)}
\label{Fig:MZ2}
\end{figure}

\subsubsection{Future co-evolution of the atmosphere-lithosphere system}\label{atmosphere-lithosphere}
\begin{figure}
\centering
\includegraphics[width=9cm]{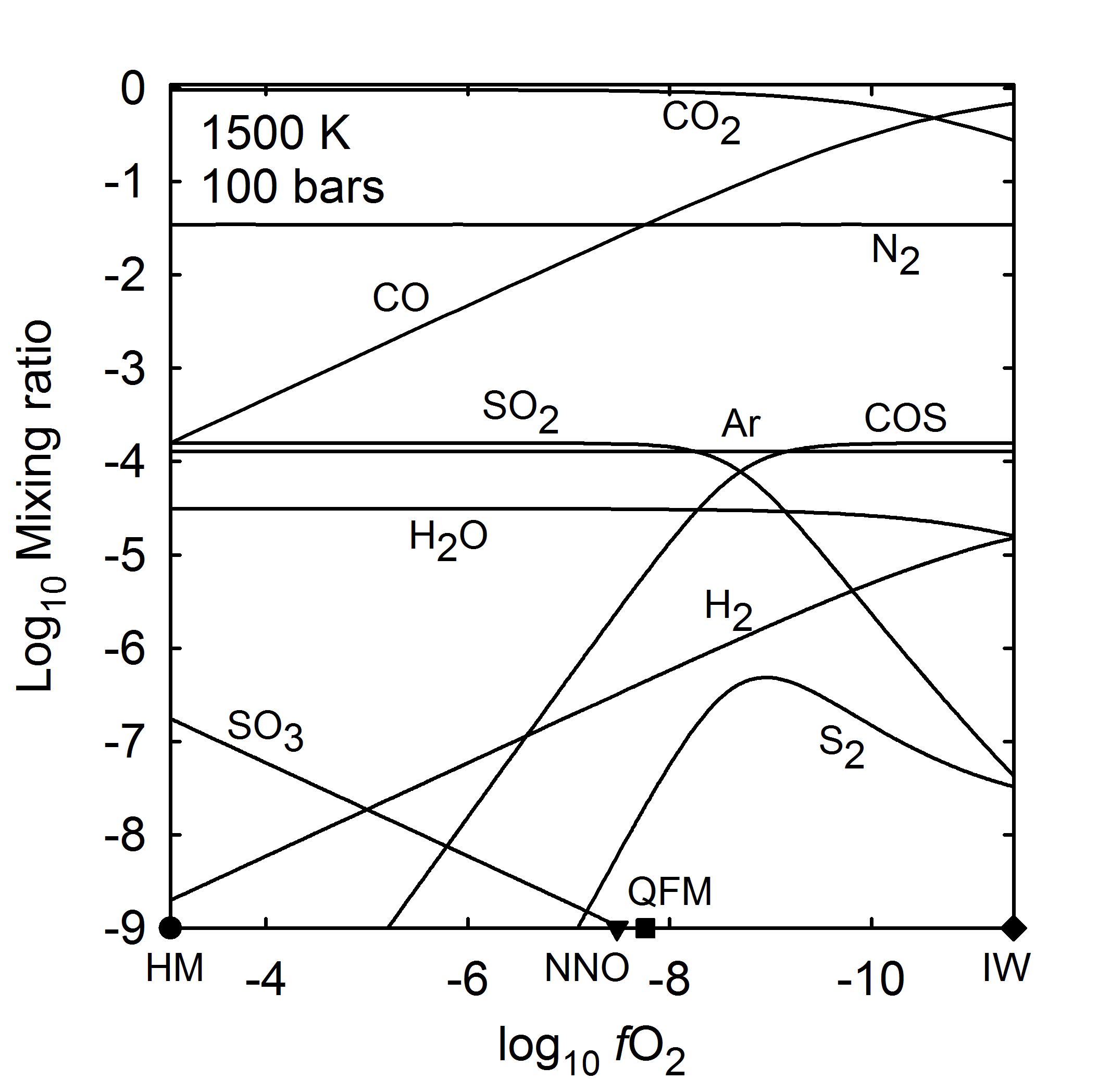}
\caption{Mixing ratios of C-O-H-S-N gases with bulk composition of the current atmosphere as functions of $f$O$_2$ at 1500 K. The Ar value corresponds to the current atmospheric $^{36}$Ar and $^{38}$Ar contents and tripled current $^{40}$Ar. The figure illustrates possible sequential reduction ($f$O$_2$ decrease) of the atmospheric composition after formation of a second magma ocean on Venus. HM, hematite-magnetite buffer; NNO, Ni-NiO buffer; QFM-quartz-fayalite-magnetite buffer; IW, iron-w\"ustite buffer. (This figure was developed by M. Zolotov)}
\label{Fig:MZ3}
\end{figure}

Future Venus will likely be characterized by the physical-chemical co-evolution of the atmosphere, surface materials, and shallow interior. It is unclear if the decreasing radiogenic heating in the interior will cause further major volcanic resurfacing and related greenhouse heating. If a global volcanic event occurs, physical-chemical processes would be similar to those during the last global resurfacing at 0.2--1 Ga (Section~\ref{atm-surface-reactions}). Intensive gas-solid type reactions would again proceed toward establishing gas-solid equilibria that will control minor chemically active gases. As in the present epoch \citep{Zolotov2018}, Fe oxides and sulfides, Ca sulfate, chlorides and fluorides will be major participants of buffering reactions, though actual gas concentrations will depend on evolving temperature-pressure conditions. Ultimately, exhaustion of radionuclides of U and K will terminate volcanic supply of fresh rocks and gases. That termination will further favor establishing and maintaining buffering gas-solid equilibria in permeable surface materials. Further evolution will be moderately affected by increasing solar luminosity and corresponding warming, gas-solid reactions at the surface, and atmosphere escape. The thermodynamically favorable oxidation of Fe(II) and S(II) in exposed solids by atmospheric CO$_2$, and traces of SO$_2$ and H$_2$O from volcanic gases suggest establishing ferric-ferrous equilibria and equilibria with participation of relatively oxidized S-bearing minerals, such as pyrite, and Ca and Na sulfates. In the likely case of hematite-magnetite and/or hematite-magnetite-pyrite equilibration, one would expect changes in CO$_2$/CO, SO$_2$/COS, SO$_2$/H$_2$S, and H$_2$O/H$_2$ ratios that reflect increasing $f$O$_2$ with temperature (Fig.~\ref{Fig:MZ2}). The suppressed gas-solid reaction rates at near-equilibrium conditions would not cause a thorough oxidation of exposed rocky materials to secondary phases. In other words, net reactions at the atmosphere-surface interface

\begin{equation}\label{eqn:15}
\begin{aligned}
\text{2FeO} \ \text{(in silicates, magnetite) + O } \text{(in\ CO}_{2}\text{, H}_{2}\text{O, SO}_{2}) \rightarrow & \\
\text{Fe}_{2}\text{O}_{3} \ \text{(hematite) +} \ \text{reduced gas}\ & \text{(CO, H}_{2}\text{, S}_{2})
\end{aligned}
\end{equation}


and sulfatization reactions (Eq. \ref{eqn:13}) will not proceed toward completion, especially given the limited expected O enrichment of the atmosphere by escape of H and O from the atmosphere (Section~\ref{hydrogen:oxygen:escape}). As in the present epoch, the elevated level of atmospheric SO$_2$ controlled by equilibria with pyrite and/or sulfates would not allow formation of carbonates of Ca, Na and K that affect atmospheric CO$_2$. Atmospheric HCl and HF will be controlled by equilibria with corresponding surface Cl- and Fe-bearing phases (chlorides, fluorides, amphiboles, phosphates; \citealt{Lewis1970,Fegley_Treiman1992}) and H$_2$O,g at corresponding temperatures. CO$_2$ and N$_2$ will remain the major atmospheric gases and the increasing greenhouse warming could slightly contribute to their abundances through the degassing of rocks. In several Gyr, the Sun will become a red giant and a significant greenhouse heating will cause shallow and then surface salt (e.g. NaCl), sulfide and then silicate melting. These events and corresponding melt-atmosphere interactions will reduce the atmosphere by magma that is rich in Fe(II) and S(II) melt complexes. One would expect a decrease in CO$_2$/CO and SO$_2$/(H$_2$S, S$_2$, COS) ratios in the atmosphere towards $f$O$_2$ expected for Venus’ mantle melts, that is ~2 log fO$_2$ units below the QFM buffer \citep{Wadhwa2008}. In other words, the gas-melt equilibria at the surface of a second magma ocean in Venus history will control the composition of its atmosphere, in which CO$_2$, CO and N$_2$ could be the major gases (Fig.~\ref{Fig:MZ3}).

\subsubsection{Silicate weathering and carbon cycling}\label{SilicateWeathering}

As hydrogen is fused into helium, the density of the Sun increases, and thereby both its fusion rate and luminosity. As a consequence, the incident insolation of Earth has increased by $\approx$30$\%$ throughout its history \citep{Gough1981}. However, liquid water is known to have existed on Earth’s surface at least since the early Archean, which requires a substantially higher amount of greenhouse gases in the early atmosphere \citep[e.g.,][]{kasting2003evolution}. An efficient feedback mechanism that regulates the amount of CO$_2$ in Earth’s atmosphere against long-term changes in the surface temperature is the long-term carbonate-silicate cycle \citep[e.g.,][]{Berner_Caldiera1997}: in this cycle, CO$_2$ reacts with rainwater to form carbonic acid, which dissolves silicate rocks. Weathering products are washed via rivers into the ocean and lakes, where calcium carbonate is precipitated. Burial of calcium carbonates acts as a sink for carbon. In subduction zones, part of the carbon is then thermally decomposed and returned to the atmosphere, while the remainder is subducted into the mantle. Outgassing of CO$_2$ at mid-ocean ridges completes the cycle. Importantly, the rate of silicate weathering depends on the concentration of CO$_2$ in the atmosphere and on the surface temperature, which efficiently regulates the climate against fluctuations of atmospheric CO$_2$ and against increasing incident insolation \citep{walker1981negative}. \cite{Way2022Exo} also offers an in-depth look at weathering and volatile cycles, in the context of exoplanetary studies.

If liquid water existed on early Venus \citep[e.g.,][]{Kasting1984,Kasting1988,Abe2011,way2016venus}, silicate weathering may have served as a sink for atmospheric CO$_2$ \citep[e.g.,][]{Kasting1984}. This could have involved either dissolution of silicates on land and the subsequent precipitation of carbonates in soils, lakes and ocean basins (so-called continental weathering), or the sub-oceanic circulation of carbon-bearing fluids and precipitation of carbonates from in-situ dissolution products (seafloor weathering) \cite[e.g.,][]{driscoll2013divergent}. In either case, cations produced from rock weathering combine with carbonate ions to remove carbon dioxide from the atmosphere-ocean system \citep{walker1981negative,brady1997seafloor}. 

On Earth, the precise dependence of seafloor weathering on dissolved inorganic carbon (and thereby indirectly atmospheric CO$_2$), bottom ocean temperature, and ocean pH is uncertain \citep{brady1997seafloor,sleep2001carbon,krissansen2017constraining}. In scenarios for early Venus, the functional dependence of continental silicate weathering on atmospheric CO$_2$ or surface temperature may be important for the climate, since it determines the efficiency at which CO$_2$ could be removed from the atmosphere and thereby the potential for maintaining habitable conditions.

Recent models considering the thermodynamic, energetic, and lithological limitations to continental weathering suggest the weathering response to changes in climate and atmospheric CO$_2$ may be strongly dependent on geologic factors. For example, the chemical weathering rate of granite is significantly lower than that of basaltic crust, which becomes particularly relevant for high surface temperature \citep{hakim2021lithologic}. In addition, if weathering is controlled by a thermodynamic limit, climate becomes increasingly sensitive to land fraction and outgassing \citep{graham2020thermodynamic}. With limited exceptions \citep{zolotov2020water}, the implications of such weathering models for a potentially habitable early Venus have not yet been explored, but point at possible consequences for the CO$_2$ evolution in the atmosphere of Venus if liquid water was present.

The fate of carbonated crust has important implications for Venus’ climate evolution. On Earth, carbonate sediments and carbonatized basalt are subducted with the oceanic lithospheric plates. Depending on the geotherm of subduction zones, subducted carbonates either release CO$_2$ into the atmosphere via arc volcanism, or remain stable and are recycled into the mantle. It has been suggested that plate tectonics on early Venus, if it ever existed, could have similarly stored carbon in the mantle and thus preserved habitable surface conditions for billions of years \citep{way2020venusian}. 
As discussed in \citep{Rolfetal2022}, the existence of plate tectonics on early Venus is debated. However, plate tectonics may not be a requirement for early Venus habitability \citep{foley2018carbon,honing2019,tosi2017habitability}. In the absence of plate tectonics, silicate weathering may still occur, and carbonate sediments would be buried by new lava flows. Ultimately, these carbonates would heat up as they move downward in the crust and eventually become unstable.

To what extent CO$_2$ released by metamorphic decarbonation reactions in the buried crust makes its way back into the atmosphere is unclear. At shallow depth, cracks likely form a path for CO$_2$ to the surface. However, deeper in the crust increasing pressure and temperature may lead to sealing voids, which would cause CO$_2$ to sink towards the mantle as the crust gets buried by new lava flows. In submerged metamorphosed rocks, CO$_2$ fugacity could have been buffered by newly established silicate-carbonate equilibria. If rising melt comes into contact with the trapped CO$_2$, the latter would likely rise with the melt to the surface, similar as at arc volcanoes on Earth. Melting in CO$_2$-rich source regions could have caused the formation of alkaine silicate melts \citep[e.g.,][]{kargel1993} that are common on oceanic islands on Earth. Parent melts of K-rich Venera 8 and Venera 13 surface probes could have formed this way, and carbonate-rich lavas (carbonatites) might be responsible for the formation of Venus canali \citep{kargel1994}. Altogether, part of the released CO$_2$ from decarbonation reactions of buried carbonates could return to the atmosphere. However, the precise fraction is difficult to assess and depends on mechanical processes in the crust and the path of rising magma. It also depends on the depth and temperature where carbonates become unstable as well as on the crustal burial rate and porosity, which are both uncertain. However, without plate tectonics the rate at which carbonates are recycled into the mantle is certainly limited compared to Earth.

The resulting carbon cycle for a stagnant lid Venus would differ from that of present-day Earth since the mantle would not serve as an efficient inorganic carbon sink \citep{foley2018carbon,honing2019}. For a stagnant lid Venus, the concentration of carbon in the combined crust-atmosphere reservoir would increase with time, since mantle degassing supplies CO$_2$ to this reservoir on the long-term. This would ultimately enhance the rate of metamorphic crustal decarbonation and raise the atmospheric carbon budget. In addition, without compensating cloud albedo feedbacks \citep{way2020venusian}, increasing solar luminosity makes it increasingly difficult to preserve liquid water on the planetary surface. Above a critical absorbed solar radiation threshold, all surface water evaporates and aqueous weathering ceases. Crustal decarbonation would result in a substantial rise of atmospheric CO$_2$ and surface temperature. Mantle outgassing would then release CO$_2$ over time, although the exact amount (in the range of 10$^{19}$-10$^{20}$ kg of CO$_2$) is highly dependent on mantle composition and oxygen fugacity.

\begin{figure}
\centering
\includegraphics[width=7cm]{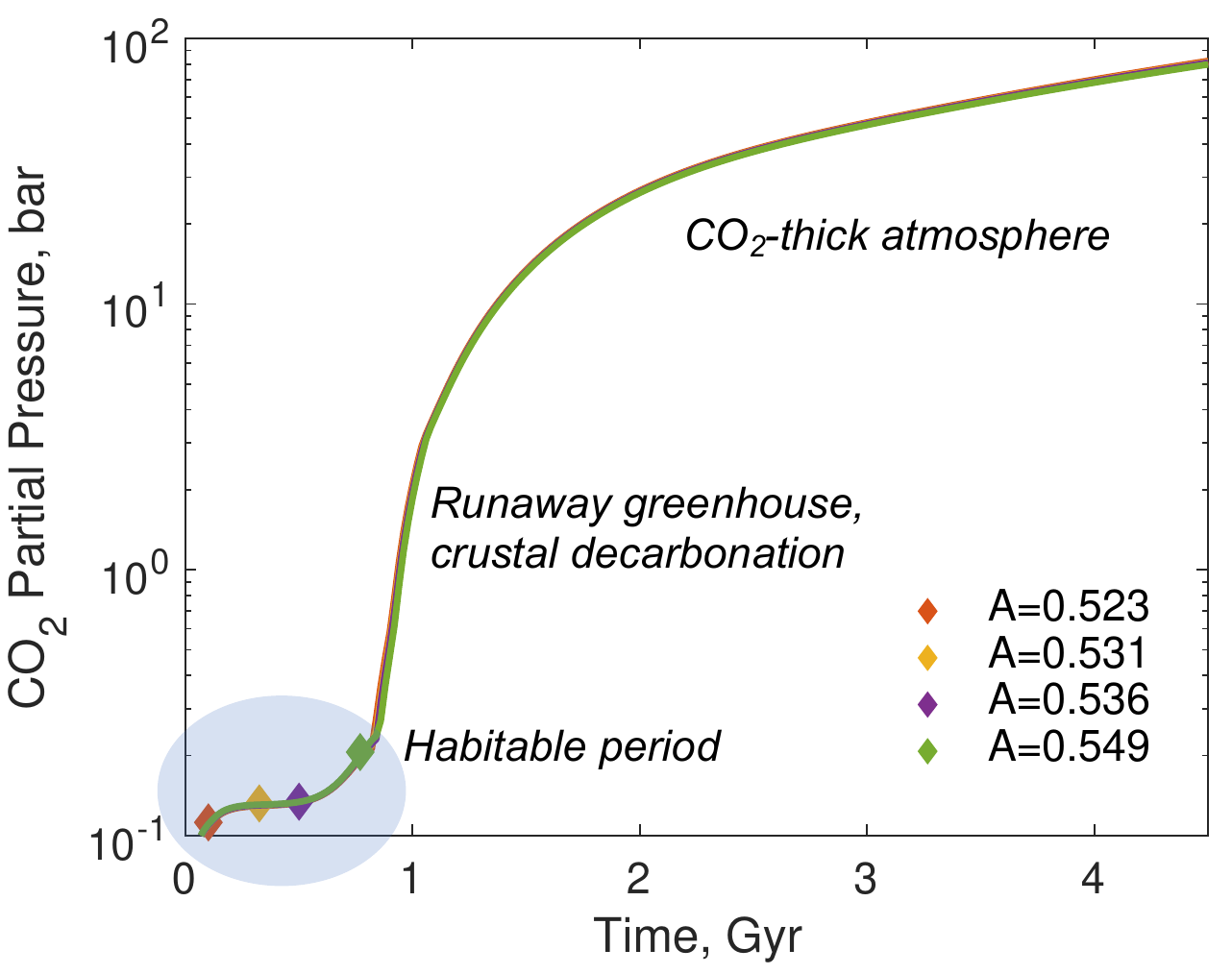}
\caption{Modelling results from a coupled interior-atmosphere model of a stagnant lid Venus assuming different planetary albedos, after the solidification of the magma ocean \citep{honing2021}. The model assumes an initially thin atmosphere containing only the present-day N$_2$ inventory. Diamonds depict control runs from a 3-D general circulation model \citep{way2016venus,way2020venusian} at the given time (corresponding to incident insolation) and atmospheric CO$_2$ pressure, yielding the indicated albedo.}
\label{Fig:earlyw}
\end{figure}

The time up to which aqueous silicate weathering and carbonate formation could keep a stagnant lid Venus habitable depends on the planetary albedo, which in turn depends on the atmospheric composition and the rotation rate of the planet. 3-D general circulation models by \cite{way2020venusian} indicate that cloud feedbacks of a slowly rotating early Venus would result in a planetary albedo between 0.5 and 0.6, sufficiently high to potentially allow for liquid surface water on early Venus. Coupled atmosphere-interior models that include silicate weathering, carbonate burial and metamorphic decarbonation \citep{honing2021} indicate that for planetary albedos in this range an early stagnant lid Venus could have been habitable for up to 1 Gyr, followed by evaporation of water and dramatic rise of atmospheric CO$_2$ (Fig. \ref{Fig:earlyw}).

Given the right conditions, the silicate weathering thermostat could have conceivably maintained habitable surface conditions on Venus for several hundreds of millions or even billions of years, even under a stagnant lid regime. Further work, however, is required to address uncertainties in order to better understand a potential habitable period. For example, the fraction of CO$_2$ released by metamorphic crustal decarbonation that makes it to the surface is a major uncertainty in coupled atmosphere-interior models. In addition, volatile partitioning during magma ocean solidification controls the post-crystallization atmosphere \citep{Elkins-Tanton2008,Massol2016,hier2017origin,salvador2017relative,Nikolaou2019,Bower2021,Gaillard2022a,Gaillard2022b}.
The composition of this initial atmosphere may in turn dictate whether the precipitation of water oceans, and therefore CO$_2$ drawdown and climate regulation via silicate weathering is permissible.

\section{It is not only about volatile exchanges: Other important mechanisms}\label{sec:other}

This section highlights that while important, volatile exchanges are not the only way the atmosphere and the solid planet can affect each other, nor the only complex process that can affect long-term evolution and surface conditions on several scales. \newline

\subsection{Magnetic fields}
\label{magneticfields}
The links between magnetic field and other processes or layers of terrestrial planets are perhaps the canonical example of complex feedback cycles on a planetary scale. A terrestrial planet's self-generated magnetic field has its origins within a large volume of electrically conductive fluid---typically the liquid metallic core, where a magnetic dynamo effect can be powered by the convection of this electrically conducting fluid. Especially early in the history of Venus-sized planets, liquid silicates in a basal magma ocean may also convect vigorously enough to produce a dynamo \citep{Ziegler2013,Stixrude2020,Blanc2020,ORourke2020}. 

Core convection can be thermally or compositionally driven, e.g., by the exclusion of light elements from an inner core \citep[e.g.,][]{Nimmo2015} or the exsolution of buoyant element components from the liquid core into the mantle \citep{ORourke2016,Badro2016,Hirose2017}. However, even chemical convection is ultimately caused by the heat loss from the core. Therefore, the existence of a dynamo is governed by the capacity of the mantle to extract and transport heat \citep[see][for more details]{Rolfetal2022}. In turn, the core offers an evolving boundary condition that affects mantle evolution, and conversely, evolving lower-mantle conditions may influence core dynamics. Heat sources in the core include radiogenic heating, gravitational energy and latent heat of crystallization of iron. At present, it is still uncertain whether the generation of a strong magnetic field (here we consider the intrinsic magnetic fields of Mercury and Ganymede to be weak) requires plate tectonics to operate in the mantle. In the Solar System at present-day, Venus, Earth and Mars might suggest so, but the number of samples is too small to draw conclusions: only present-day Earth exhibits plate tectonics and incidentally possesses a strong magnetic field. Plate tectonics is indeed more efficient at transporting and removing heat from the planetary interior than stagnant lid or single plate convection. While Venus shows no sign of a strong self-generated magnetic field at present-day, numerical simulations of its evolution suggest that, for similar parameters to Earth's (state of the core, composition, etc.), it is possible that Venus could have sustained a magnetic field until recently, even without long-term plate tectonics \citep{DriscollBercovici2014,ORourke2019}. However, according to thermal evolution models, a solid inner core may or may not exist at present day, because it depends on the current temperature and composition of the core which are unknown.

It is still unknown whether Venus had a magnetic field at any point during its history, but remnants of past magnetization could potentially remain to be discovered. Evolution of the planetary mantle and changes in convection regimes that could have occurred on Venus would directly affect the planet’s ability to generate a magnetic field.
On the opposite extent of the planetary domain, the magnetic field is known to affect non-thermal atmospheric escape (hydrodynamic escape remains unaffected), see Section~\ref{hydrogen:oxygen:escape}. It has long been considered as a shield that protects a planet’s volatile species from escape mechanisms \citep[i.e.,][]{Lundin2007}, which would have explained the differences between Earth, Mars and Venus. Earth would have been protected and remained habitable with plenty of water due to its magnetic field. Venus, lacking a magnetic field would have therefore lost its water to space. Mars, smaller and unprotected after an early dynamo phase of about 700 Myr, would have lost most of its primordial volatile inventory. This hypothesis has been challenged in the past decade, on the basis of the comparison of present-day estimates for escape rates from these three planets \citep{Strangeway2010}: numbers suggest that all three loss rates are comparable and do not directly reflect the alleged larger losses of Venus and Mars. Such an observation has been explained by considering the cross section of the area that interacts with incoming radiation for those three planets \citep{Egan2019}. 

In the case of Venus and Mars, it consists of little more than the planetary diameter plus a relatively limited atmospheric extent. In the case of Earth, that would include the magnetosphere, leading to a much larger amount of energy intercepted by the planet, thus fuelling escape, despite the protection provided by the magnetic field to the lower atmosphere. Numerical studies \citep[][in the case of Mars]{Kallio_Barabash2012}, and semi-analytical work \citep[][for various planets]{Gunell2018} have found that some specific magnetic field strengths could be associated with peak escape rates that vary depending on the planet and solar wind pressure \citep{gronoff2020atmospheric}. In some cases the magnetic field could enhance atmospheric losses. For example, in the case of Venus with a magnetic field, losses increase by a factor of 2-5 \citep{Gunell2018}. 

Other investigators \citep[i.e.,][]{Tarduno2014} have pointed out possible reasons that said observations only partially describe the actual situation. They suggested that a possibly large \citep{Seki2001} return flux of volatile species (downward) was not taken into account and led to overestimates of the escape rate for magnetized planets. They also raised the question of the response of the atmosphere to high solar wind pressure events that could affect overall escape differently on ``shielded" or unmagnetized planets \citep{Wei2012}. Finally, they pointed at the difference between atmospheric loss and the loss of a key species: while Venus’ atmosphere remains dense the loss of water by non-thermal mechanisms alone could have had tremendous implications for its climate and geodynamical history. However, water can also be lost neutrally through the hydrodynamic escape of its H atoms: for strong thermal escape, a magnetic field cannot prevent the desiccation of a planet.

The issue has certainly not been resolved yet: it remains unknown whether the presence of a magnetosphere would result in decreased or increased atmospheric losses. It is not a clear-cut choice between shielded planets versus unprotected ones. One certainty is that magnetic fields have a strong and complex influence on loss processes. For example, at Earth the ion escape processes are different from those at Venus and Mars due to Earth's intrinsic magnetic field. The geometry and strength of the field affect both how much of the planetary atmosphere is lost and where it is lost \citep{Tarduno2014}: for internally generated magnetic fields, losses seem to occur at the cusps of the field, near poles, while in the case of non-self-generated magnetic fields, losses could occur in the magnetotail. This seems to be supported by simulations for high-obliquity exoplanets around M-type stars \citep{Dong2019} exhibiting higher loss when the cusps face the stellar wind. Moreover, atmospheric structure also plays a role in atmosphere escape as discussed in Section \ref{hydrogen:oxygen:escape} and \cite{salvador2022}.

Present-day observation of the remanent magnetization from a past intrinsic magnetic field on Venus would affect our vision of both the interior of the planet and its atmospheric evolution \citep{ORourke2019}. No mission has yet conducted a magnetometer survey below the ionosphere, so a large amount of crustal remanent magnetism could await detection. A future mission that, for example, included an aerial platform dwelling in the clouds could detect regions of magnetized crust as thin as $\sim$200~m if the magnetization intensity is comparable to typical values observed on Earth, Mars, and the Moon.

\subsection{The role of impacts}
Meteoritic and cometary impacts are a common feature to all terrestrial bodies in the Solar System and probably beyond: all known planets feature impact craters to some degree \citep[e.g.,][]{zahnle1999,neukum2001}. Impacts are also catastrophic and sudden events that affect planetary evolution at many levels: interior, surface, atmosphere. In Earth's history they have been linked to major extinction events \citep[e.g.,][]{Keller2014}.
They have been suggested as possible causes for changes in surface conditions and even interior dynamics. Micrometeorites ($<$2 mm) dominate the extra-terrestrial flux to present-day Earth \citep[40,000 tons/year $\pm  50\%$,][]{love1993direct,peucker2000} and probably on other terrestrial planets \citep[][for Venus, specifically]{frankland2017co}. However, impacts of km-sized objects have much stronger consequences for planetary evolution \citep{Schlichting2015}. 

Due to the young age of most of its surface, few studies have targeted the role and consequences of impacts on Venus. A young surface implies that, unlike the Moon, Mars or even Earth, most traces of impacts that occurred during Venus' history have been erased from the planet's surface. Only impacts occurring in the last $\sim$ 200 Ma--1 Ga are recorded \citep{Bottke2016,McKinnon1997}, which excludes the largest bodies (10s to 100s km radius) that are most likely to collide with a planet during its early history. Additionally, due to both its young surface and its thick atmosphere, Venus shows no sign of small craters ($<$1 km radius), as the small impacting bodies would vaporize or disintegrate during their atmospheric entry phase \citep{Herrick_Phillips1994}. This shows that at least during the last $\sim$ 200 Ma--1 Ga Venus must have had a thick atmosphere comparable to today's. The current cratering record is mostly used for estimating surface ages via crater counting and to estimate how surface features are affected by surface alteration (erosion, tectonics) or volcanic flows \citep[see][for a detailed discussion]{Herrick2022}.

Collisions with other solid bodies can have a multitude of consequences depending on the temporal and spatial scale. Short-term effects include ejecta blankets, thermal radiation, blast-wave propagation in the atmosphere, crater excavation and earthquakes \citep{Pierazzo_Artemieva2012}. Here, we will instead focus on long-term, global-scale consequences that affect the evolution of terrestrial planets. Impacts can (i) deliver material to a planet, such as gases (oxidizing and reducing gases, noble gases, etc.); they can (ii) erode its atmosphere, and finally they can (iii) directly affect the solid planet, sometimes down to its core. 

All those consequences can vary depending on the specific conditions of the collision (angle, velocity, impactor size, mass and composition). Impactors have also been proposed to explain both Venus' obliquity and rotation rate since the 1960s \citep{McCord1968}. \cite{WardReid1973} pointed out that an impactor less than 1\% the mass of the Earth's Moon' can drive Venus retrograde if the planet was initially spinning slowly prograde due to solid body tidal dissipation. See \citet[][Section 5]{way2020venusian} for a detailed discussion on Venusian tidal dissipation and the possible role of impactors on its rotational history, and \cite{Jacobson2022}. 

\subsubsection{Material Deposition}

The evolution of a terrestrial planet will first be affected by the mass of material delivered/retained during collision. The impact geometry and physics are essential to determining the repartition (atmosphere, surface, interior, or loss) of material delivery \citep{Golabek2018,Landeau2016}. A portion of the impactor can be lost back to space as ejecta, depending on the size/mass of the impactor, impact velocity and incident angle, gravity, and atmospheric properties, as evidenced by hydrocode simulations \citep{Shuvalov2009}. The remainder of the repartition of material delivered is complicated by the specific mechanism of the impact (e.g., impact angle). Part of the impactor can remain solid \citep{Pierazzo_Melosh2000}, while part of it can undergo partial melting or be vaporized \citep{Svetsov_Shuvalov2015}. 

Mass delivery by impacts plays a major role during the accretion phase, when the planet is forming, as detailed in \cite{Jacobson2022}. During later evolution, this effect becomes less important as the incoming mass flux decreases exponentially while the mass of the planet is increasing, leading to only minor consequences for both the solid planet and its atmosphere \citep{Sakuraba2019,Gillmann2020}. It is possible that large impacts can modify the climate of planets, as has been studied in the case of Mars \citep{Segura2002,Segura2008,Segura2012,Turbet2018b,Haberle2019}. In the case of Venus' post-accretion history, a stray $\approx$10 km sized impactor could still deliver more water to the atmosphere of Venus than its present-day measured H$_2$O content, with possible consequences on planetary surface conditions/temperatures \citep{Gillmann2016}. Negligible amounts of CO$_2$ and N$_2$ would be delivered compared to Venus' present day atmospheric inventory. Additionally, large impacts can deliver enough reducing material to convert H$_2$O-CO$_2$ dominated atmospheres to transient H$_2$-CH$_4$-CO dominated atmospheres, resulting in enhanced H escape \citep{genda2017, zahnle2020creation,Haberle2017}. The effects of such large impacts on Venus’ atmospheric evolution remain to be explored.

At a given mass delivered to the planet, the nature of impactors governs the deposition of material on rocky planets during collision processes. On present-day Earth, 80-86\% of finds from the influx of meteorites is made of ordinary chondrites \citep[from various collections,][]{harvey1989,krot2014,Dehant2019}. Carbonaceous chondrites are a rare ($\approx$4\% of the finds) type of chondrite \citep{Dehant2019}. Despite their name, only the some  subsets of carbonaceous chondrites (namely, CM, CR, and CI) are significantly enriched in carbon and H (up to a few percent mass; \citealt{Grady_Wright2003,Marty2012,Pearson2006}) relative to ordinary chondrites \citep{Gounelle2011}. Even rarer ($\approx$2\%) are enstatite chondrites (EH, EL), that are reduced specimens. They are comparatively dry, having formed in the inner Solar System. However, studies revealed that enstatite chondrites can contain significant amounts of H ($\approx$0.1-0.5 wt.\% water equivalent; \citealt{Pepin1991,Muenow1992,Piani2020}).

Beyond simple material delivery, impactor chemistry could prove critical in order to estimate the composition and state of post-impact atmospheres and mantles. It plays a key role in determining what species are released during impact outgassing (depending on how reduced the atmosphere is), and how material (such as iron) delivered by the impactor interacts with the planetary atmosphere. As thermochemical equilibrium modelling has shown \citep{Hashimoto2007,Schaefer_Fegley2010}, impacts could lead to the outgassing of reduced species (H, CH$_4$, CO) for all types of chondrites \citep{Hashimoto2007}, except the most wet carbonaceous chondrites \citep{Schaefer_Fegley2010}. However, \cite{Lupu2014} show that post giant impact atmospheres would most likely be dominated by water and CO$_2$.

Nevertheless, it has been suggested that reaction with iron metal could consume up to several water oceans and produce tens of bars of hydrogen \citep{genda2017}. However this is debated, as another recent study \citep{Citron2022} suggests that most of the iron delivered by late accretion impactors is deposited in the crust and mantle which implies that only a fraction of the projectile iron can react with water and CO$_2$ on the planet. Consequently, it is unlikely that this mechanism could trap entire pre-existing Earth-like oceans of water. Therefore, both the chemistry and the physics of the impact should ideally be considered together to assess how they affect the distribution of material between surface and interior.

\subsubsection{Atmospheric erosion}
\label{imperosion}
A second effect of collisions is the loss of volatile species from the planetary atmosphere. Three main mechanisms are thought to cause atmosphere erosion during a meteorite impact \citep{Pham2011}: 
\begin{enumerate}
    \item Direct ejection of the atmosphere via atmospheric compression by the impactor upon entry, possibly including an aerial burst if the body disintegrates in an explosion \citep{Shuvalov2014}
    \item Hot vapor plume shock-wave above the impact location involving the vaporized projectile and target body
    \item Interaction of high speed ejecta with the atmosphere; particles are ejected into the atmosphere, accelerating and heating atmospheric molecules. 
    \item For giant collisions a further mechanism has been suggested: pressure waves cause vertical ground motion at the antipode of the impact location, accelerating atmospheric particles above the escape velocity \citep{Genda_Abe2003,Schlichting2018}.
\end{enumerate}

At first, simple parameterizations were developed to estimate atmosphere losses, such as energy calculations or simple geometrical models \citep{Cameron1983,Melosh_Vickery1989}. Early energy-based considerations suggested that large, possibly complete atmospheric losses were possible, if the impact was sufficiently large and/or fast enough. Considering the geometry of the collision, for example by limiting maximum escape to the portion of the atmosphere above a plane tangent to the planet at the impact location (the tangent plane model) reduced estimations to a much lower portion of the planetary atmosphere \citep{Melosh_Vickery1989,Pham2011,Schlichting2015}. Other approaches rely on full numerical hydrocode simulations of the entire impact process, from entry to vapor plume, such as those simulations performed with the SOVA hydrocode \citep{Shuvalov2009}. Hydrocode simulations contain limitations for impactor bodies with sizes larger than the thickness of the planetary atmosphere. For that reason, estimates of the loss through giant impacts are mostly done using the tangent plane model and ground motion estimates. This type of collision involves complex material redistribution of the solid bodies involved and can have important consequences for the atmosphere. 

Again, most studies are not targeted directly at Venus, but usually cover multiple bodies or are general enough to produce laws including planetary and atmosphere masses as parameters. In general, lower mass planets such as Mars tend to undergo more efficient atmospheric erosion than more massive ones, like Earth or Venus \citep{Pham2011}. Hydrocode results indicate that large impacts (\emph{R}$>$10km, up to a few 100s km) are relatively inefficient at removing planetary atmospheres, especially when volatile delivery is taken into account (see above), and yield results implying lower losses (but of the same order) than simple tangent plane models for Earth and Venus \citep{Shuvalov2009,Shuvalov2014,Gillmann2016}. Smaller collisions are even less efficient when only single events are considered, but erosion can then be considerably increased by airburst mechanisms \citep{Shuvalov2014}. Calculations by \cite{Schlichting2015} suggest that the bulk of atmospheric loss by impacts could be attributed to swarms of small impactors (1 km$<$\emph{R}$<$10 km), as they are sufficiently numerous to have a significant cumulative effect. This is because most of their energy is delivered to the atmosphere and used to power escape. Giant impacts that can theoretically cause very efficient erosion are rare. They require the right set of conditions: large mass relative to the target body (above 40$\%$) and high impact velocity (a few times mutual escape velocity). Such giant impacts could remove 50-90$\%$ of a preexisting atmosphere. However, slower, less massive collisions are more likely and would cause the loss of up to 20$\%$ of early atmospheres \citep{Schlichting2015,Genda_Abe2003}. Studies modelling the relative atmospheric loss and delivery by impacts can sometimes differ by orders of magnitude, due to the use of different equations of state and dynamical models \citep{Hamano_Abe2010,Melosh_Vickery1989,Newman1999,Shuvalov2009,Svetsov2007,Vickery_Melosh1990}.

\subsubsection{Energy transfer}
The third main way for impacts to affect planetary evolution revolves around the large amount of energy they can transfer to the solid part of the target. 
During the collision event, the kinetic energy of the impactor is transferred to the interior of the planet. A shock wave is generated that propagates hemispherically away from the point of impact, decreasing in amplitude with increasing distance \citep{OKeefe-Ahrens1977,Melosh1989}. Shock pressure followed by decompression cause heating inside the planet. The temperature increase is proportional to the maximum shock pressure \citep{Pierazzo1997}, possibly leading to melting and vaporization of lithospheric and even mantle material. 

Under the impact location, the shock pressure is uniform in a spherical isobaric core, leading to an isothermal central thermal anomaly, where planetary material is heated. The size of the isobaric core is governed by the that of the impactor, its radius being about 1-2 times the impactor radius \citep{croft1982,Monteux2007}. Outside the isothermal zone, the thermal anomaly’s amplitude decreases rapidly with distance from the isobaric core \citep{Pierazzo1997,Monteux2007}.
The temperature increase depends on the impact velocity, as well as the size of the target \citep[but not the impactor's;][]{Monteux2007}. As a consequence, larger target bodies experience higher impact-generated temperatures. One can expect an amplitude of a few hundred K for a Mars-sized planet, but upward of 1000 K for Venus or Earth \citep{Gillmann2016,Ruedas2017}. The thermal anomaly geometry is also affected by the impact angle \citep{Bierhaus2012}, and more energy is transferred in a head-on collision than a grazing impact. 
 
As a consequence, fast, head-on and larger impacts on large planets (like Venus) are more likely to have long term consequences \citep{Gillmann2016}; low energy events have mostly local or short-term consequences on the global-scale picture (although they could still affect surface conditions and habitability). 
Antipodal heating has been suggested to occur due to a convergence of pressure waves on the other side of the planet, but effects have been shown to be limited to displacement and fracturing, with no outright melting, outside of a giant impact scenario \citep{Melosh2000,Meschede2011}. 
The models briefly discussed here are inaccurate for larger events (radius above a few 10s to 100s of kilometers; \citealt{Manske2018}), and should not be used for giant impacts \citep{Nakajima-Stevenson2014,Nakajima-Stevenson2015,Cameron-Ward1976,Canup2004,Melosh1990,Pahlevan-Stevenson2007}. For these larger impacts, new scaling laws have been developed recently \citep{Nakajima2021}. Their effects, in particular during planetary accretion, are explored in \cite{Jacobson2022}.

An additional effect of impacts is the acceleration of target material, by the energy transfer and shock wave, up to velocities sufficient to displace it over long distances or even eject a small portion into space. This causes the excavation of a crater and projection of ejecta (vapor, for the larger planets, solid and molten rocks from both the planet and the impactor). Impactors larger than $\approx$100 km lead to the redistribution of mass several times the impactor mass over the surface of the whole planet \citep{Shuvalov2012}.

These two effects (ejecta and thermal anomaly) can have consequences on the evolution of the planet and its mantle. First, given the low thermal conductivity of ejecta, it has been proposed that layers created by their accumulation from several impacts could insulate the interior of planets and affect their thermal evolution \citep[][in the case of the Moon]{Rolf2017}. Melting and outgassing of the lithosphere (for smaller impacts, \emph{R}$<$100 km) or the mantle (for larger ones, \emph{R}$>$100 km) are other common effects of impacts. In that latter case, depletion of the mantle can be expected: large impacts could contribute to removing volatiles from planetary mantles \citep{Davies2008,Gillmann2016}. In particular, a succession of large impacts (\emph{R}$\approx$100s km) could be responsible for the efficient depletion of the upper mantle of Venus, especially in the absence of rehydration mechanisms \citep{Gillmann2016,Gillmann2017}. All sizes of impactors, however, contribute to outgassing and the release of volatiles from the mantle or lithosphere in the atmosphere, with potential effects on the climate. The impact-generated melting of crustal carbonate deposits (if they were formed in large quantities at some point of Venus' history) could result in the release of considerable amounts of CO$_2$, for example (see section \ref{sec:Retroactions}).

Depending on the size of the impacts, even mantle convection can be affected. The volume of mantle heated by the collision is the governing factor for long term consequences on mantle evolution, therefore, larger thermal anomalies will affect the mantle more strongly. Small impacts that do not penetrate through the lithosphere into the convecting mantle have negligible consequences on the mantle. On the other hand, large impactors can generate a thermal anomaly that reaches the deeper convecting parts of the mantle, and thus affect convection processes. 
Because it is hotter than the surrounding material, the material forming impact-induced thermal anomalies is positively buoyant and exhibits lower relative viscosity. The thermal anomaly tends to rise and flatten under the lithosphere, causing melting and resurfacing \citep{Watters2009,Roberts2014,Gillmann2016,ONeill2017,Padovan2017,Ruedas-Breuer2017,Rolf2017,borgeat2021}. As the thermal anomaly then spreads laterally over hundreds of thousands to millions of years, the upper mantle is pushed away from the impact location. This effect may be enhanced on larger planets like Venus, due to the larger thermal (and therefore viscosity) contrast between the upper mantle and the anomaly.

Given a sufficiently large impactor (a large thermal anomaly), hot material may accumulate at an antipodal position, where it forms downwellings or subduction events \citep{Gillmann2016,ONeill2017,borgeat2021}. Thus, large impacts, on an otherwise stagnant lid or single plate mantle, can break the lid and shift tectonics toward a mobile lid or plate-like regime. However, the convection regime change is temporary and reverts to the original stagnant lid when the impact flux diminishes or stops \citep[][based on numerical models]{Gillmann2016,borgeat2021}. Impact-triggered subduction events also affect the mixing of the mantle (in particular the upper mantle) by recycling crust into the interior. On the long term, especially strong impacts can affect the mantle down to the core-mantle boundary.

The consequences of impacts on the mantle can eventually affect the core of the planet: \cite{ONeill2017} suggest that vigorous impact-driven convection could strengthen the magnetic dynamo. On the other hand, the thermal anomaly of a sufficiently large collision (impactor radius upward of 800 km) can even reach down to the core-mantle boundary. This would place a high temperature layer on top of part of the core, possibly reducing the heat flux out of the core. In turn, core convection would shut down, and magnetic field generation would cease on long timescales ($\approx$1 Gyr; \citealt{Arkani-Hamed2009,Roberts2014}).

\section{\bf The evolution of Venus: shaped by a multitude of mechanisms}\label{sec:Retroactions} 

The conditions on a planet are not static, but change with time as evidenced throughout Earth's history. In turn, changes in mantle and atmospheric states affect the mechanisms at work, possibly creating positive or negative feedbacks. Due to its thick atmosphere and seemingly active mantle, Venus is a place of great interest to study the mechanisms and their interactions that have shaped the evolution of the planet and its surface conditions.
Changes in the convection regime are one of the prominent examples of feedback loops, due to their scale and possibility for far-ranging consequences. By itself, mantle dynamics change with time. But the convection regime is also affected by external conditions, such as the distribution of water between the atmosphere, the surface and the interior, or surface conditions (pressure, temperature, liquid water availability). Different convection regimes imply variations in the interactions with the atmosphere. For example, it has been suggested that a mobile lid regime favors liquid surface water, while surface temperature variations can lead to a change from stagnant lid to mobile lid convection or vice-versa \citep{Lenardic2008,Gillmann2016}. Conversely outgassing is affected by the convection regime and vigor. Degassing affects surface temperature via the greenhouse effect and this, in turn, could affect mantle processes such as convection style and velocity, possibly creating a cycle of interactions. Further examples of this include the recycling of surface material into the mantle and the question of formation, stability and recycling of carbonates.
\par

\subsection{Possible evolutionary pathways}
\begin{figure}
\centering
\includegraphics[width=12cm]{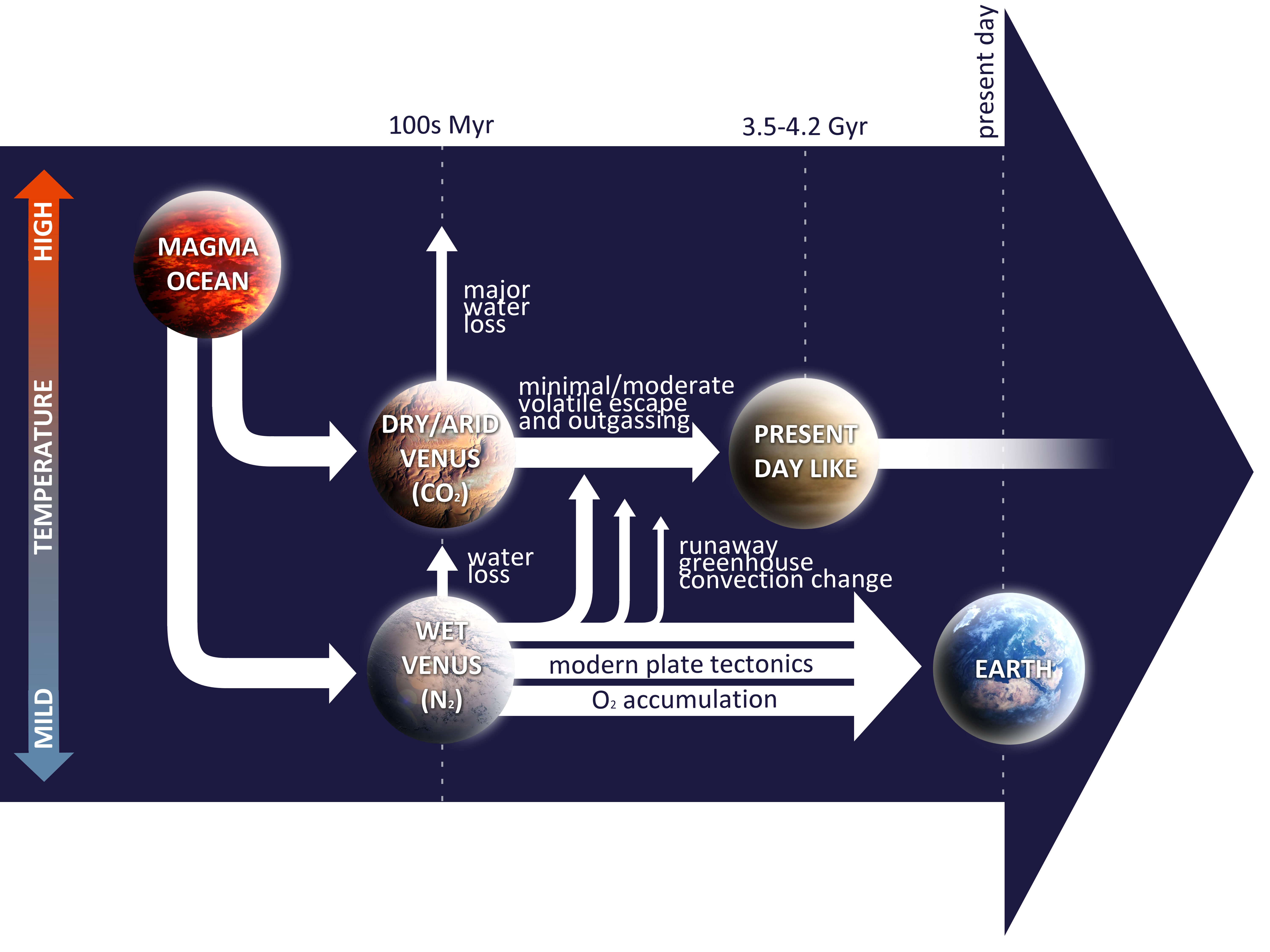}
\caption{Current understanding of the extreme tentative scenarios for the evolution of Venus' surface conditions, from its origins to present-day, compared to Earth. On top, Venus lost its surface water early on (desiccated Venus, or stifled outgassing scenarios), while on the bottom evolution, it evolved closer to Earth, retaining a larger portion of its water inventory, until its climate was destabilized. For now, both evolutionary pathways remain consistent with our global knowledge of the planet. \textit{Only general evolution trends are represented, Earth-related processes (modern plate tectonics and O$_2$ accumulation) are not attributed a specific time and only included for comparison with Venus}.}
\label{Fig:dualityofvenus.jpg}
\end{figure}

Improved understanding of atmosphere-interior couplings could shed light on the history of water on Venus, and as a result, on the planet's long-term history. Even 35-50 years after pioneering works on Venus’ volatile history coupled with solar and/or geological processes \citep[e.g.,][]{Ingersoll1969,Rasool1970,Walker1975,KastingPollack1983,Kasting1984,Kasting1988} there is still no consensus on the evolutionary pathways followed by Venus. We still do not know whether Venus was ever habitable in the past or if its surface has always been far too hot. Likewise, little is known about the past history of its mantle convection regime and if it ever supported a mobile lid or even Earth-like plate tectonics before its apparent present-day stagnant lid/single plate state. In our current approach to understand Venus's past, the fate of water particularly unites most of the unknowns. This is because it is intimately linked to fundamental questions such as surface conditions, past atmospheric escape, and interior convection, structure and composition. Based on that, three main end-member scenarios describe what could have led Venus to the hot and dry observable current state, assuming similar building blocks to the Earth's (see Fig. \ref{Fig:dualityofvenus.jpg}).

In a first scenario (the ``dry Venus" scenario, top path on Fig. \ref{Fig:dualityofvenus.jpg}), Venus was desiccated at the onset of its evolution, as early as the magma ocean phase. This scenario is described in \cite{salvador2022}, and identifies Venus as a Type-II planet, following the \cite{Hamano2013} classification. In contrast to Type-I Earth's short-lived ($\sim$1 Myr) magma ocean, Venus' vicinity to the Sun would result in a slow ($\sim$100 Myr) magma ocean solidification. The associated long-lived steam atmosphere overlying the molten surface would slowly lose its water en route through early intense hydrodynamic escape \citep{Gillmann2009}, and its desiccation would thus preclude water condensation at the surface \citep{Walker1975,MatsuiAbe1986,Lebrun2013,Turbet2021}. Remaining atmospheric oxygen could then be consumed by oxidation of the solid surface or magma ocean. Upper mantle rocks and basalts redox state could help us test the likelihood of this scenario. Overall, it induces a mostly dry planet (inside and outside). After only a few hundred million years, Venus would look rather similar to its present-day state, with a CO$_2$ and N$_2$ atmosphere that would slowly grow to its present-day level, through moderate exchanges (moderate outgassing and loss), in a rather straightforward evolution. In such a scenario, Venus' habitability would be ruled out early on and a liquid water ocean would have never existed.

In a second possible scenario, Venus could have followed a Type-I planet, fast cooling magma ocean evolution \citep{Hamano2013}, leading to a temperate Venus (bottom path on Fig. \ref{Fig:dualityofvenus.jpg}). In such a case, Venus would have cooled down fast enough to evolve in a similar way to Earth and allowed condensation of water on its surface early on, resulting in an Earth-like climate for an unspecified period of time \citep{way2020venusian}. Yet, some cooling mechanisms are required to buffer Venus high incident solar flux and allow for such a rapid cooling. The development of highly reflective clouds during the magma ocean cooling could be an example of such mechanisms \citep{salvador2017relative, way2020venusian}. While it has been shown that a concentration of clouds in the sub-stellar region could maintain temperate climates at insulation even higher than experienced by early Venus \citep{Yang2014}, thus favoring this temperate Venus scenario, details of 3D atmospheric circulation and resulting clouds distribution might prevent early ocean formation on Venus. To date, the lack of a self-consistent magma ocean-atmosphere coupled model accounting for realistic clouds does not allow one to make a definitive statement. It is discussed in length in \cite{salvador2022} and \cite{Westall2022}. As on Earth, temperate conditions could have favoured the extraction of CO$_2$ from the atmosphere by the formation of carbonates and possibly a limited carbon cycle. This would have allowed for a N$_2$ dominated atmosphere to sustain low CO$_2$ partial-pressures for long periods of time and temperate surface conditions akin to Archean Earth. At a later, but unknown time (between 1-4 Ga), geological events, such as the formation of large igneous provinces \citep{way2020venusian,Way2022}, could have released magmatic CO$_2$ into the atmosphere and, depending on the planetary albedo response, possibly started a late runaway greenhouse \citep[already suggested by][]{Pollack1971}, ultimately causing the decomposition of crustal carbonates and leading to the present-day conditions, as water is lost. External factors such as the increase of the solar luminosity with time or the occurrence of a large impact on a carbonate-rich crust could also be responsible for triggering a late runaway greenhouse.

In a third scenario (stifled outgassing) the interior of Venus would have retained its water initially due to inefficient outgassing of the magma ocean \citep{Ikoma2018, Solomatova2021, Salvador2021_sub} and magma ocean redox conditions (an oxidized magma ocean, \citealt{Gaillard2022a}). These conditions would have been sustained in its later evolution possibly due to high surface pressures \citep{Gaillard2014} preventing the outgassing of large quantities of H$_2$O. An initially wet magma ocean could cause part of its H content to partition into the core, leading to oxidized conditions \citep{Ringwood1977,Okuchi1997,Tagawa2021} and ultimately a FeO-rich (and water-poor) mantle. In this scenario, Venus would not have lost most of its water to space but could not outgas it, either due to its chemical composition/redox state or due to surface pressure. Additionally, both carbon and nitrogen components of the atmosphere would be a direct result of the early magma ocean outgassing \citep{Gaillard2022a}, with possibly a minor component outgassed later.

\subsection{The role of water during evolution}
As illustrated by the scenarios mentioned above, the fate of water is decisive for the evolution of Venus \citep[and has been recognized as such since early works, e.g.,][]{Ingersoll1969,Walker1975}. It should be noted that the history of water on Venus is intimately tied to Venus’ atmosphere-interior redox evolution. If water has not been lost during the magma ocean solidification and if cloud feedback maintained a high albedo due to Venus' slow rotation \citep{Yang2014,way2016venus}, the longevity of temperate surface conditions is constrained by the time required to lose leftover steam and remaining O after the runaway greenhouse initiation. Very recent habitability may not provide sufficient time for water/hydrogen and oxygen loss. But even for dry Venus scenarios, H loss to space is expected to leave behind substantial oxygen \citep{Chassefiere1996a,Gillmann2009,Schaefer_Wordsworth2016}. Reaching the oxygen-free modern atmosphere may require efficient dry crustal oxidation \citep[any oxidation of the crust that is not attributable to volatile degassing or water-rock reactions,][]{krissansen2021Venus}. As noted in Section \ref{atm-surface-reactions}, one interesting possible mechanism for this surface exchange is oxidation of dust particles from explosive volcanism, which could provide a significant oxygen sink \citep{Warren_Kite2021LPSC}. Furthermore, this mechanism could be enhanced if the total atmospheric mass was small during a temperate phase on early Venus, because in such a case, the lowered pressure overburden would potentially increase rates of explosive volcanism.

Another consequence of suppressed water outgassing on Venus would be the formation of a relatively water-rich lithosphere. Such a lithosphere would be rheologically weaker due to the presence of H$_2$O \citep{wang2020effects}, which might be inferred from future observations. In addition, a weak lithosphere might drip or delaminate more easily and result in a thinner crust than would be expected otherwise \citep{Anderson_Smrekar2006,James2013}. As in terrestrial seafloor basalts \citep{Holloway2004,Holloway2000}, interaction of remaining H$_2$O with ferrous Fe in melts could produce magnetite (Fe$_3$O$_4$) and H$_2$ (Eq. \ref{eqn:11}), that could be released into the atmosphere through physical or chemical weathering of lavas.

\begin{figure}
\centering
\includegraphics[width=13cm]{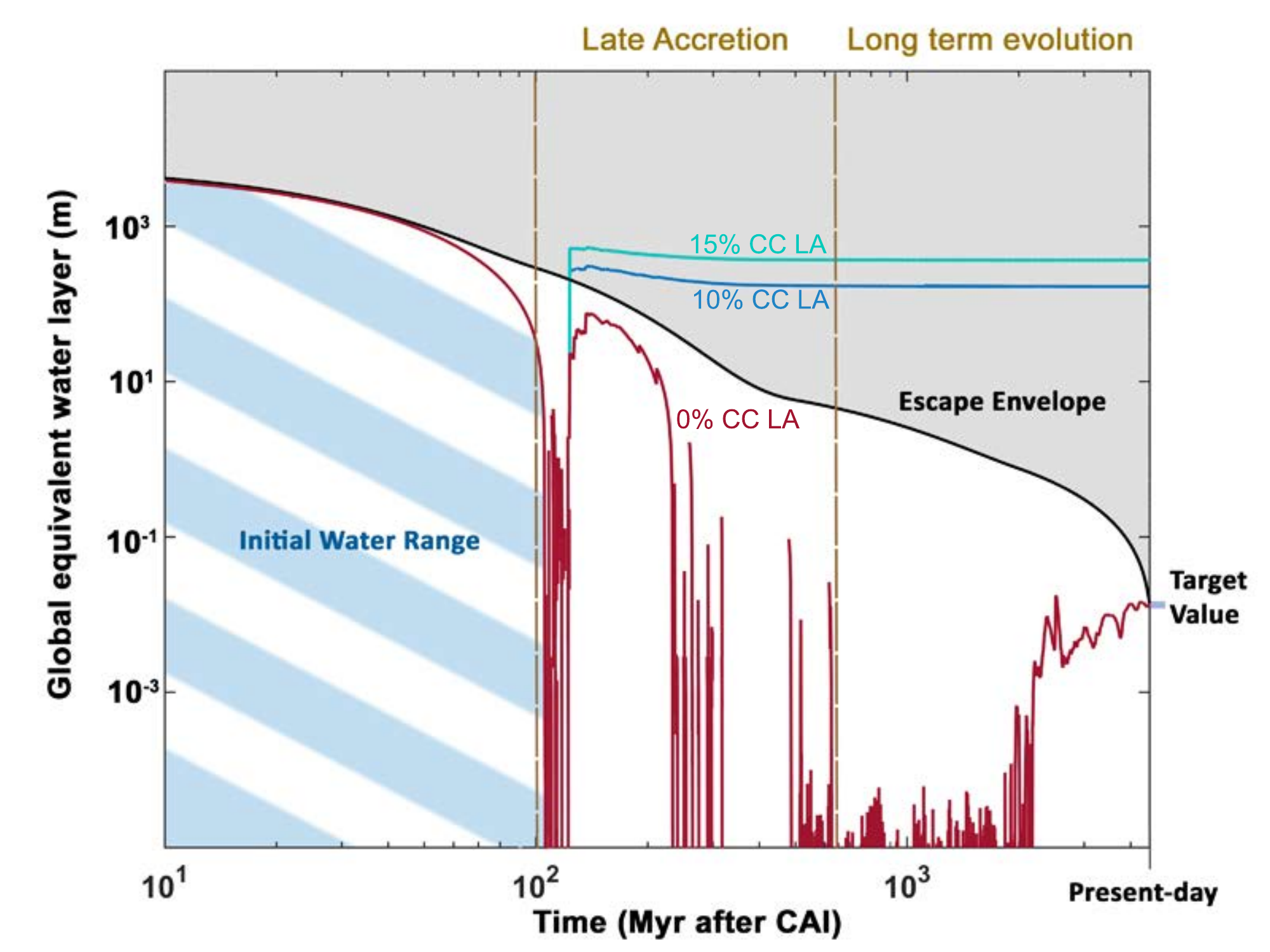}
\caption{Evolution of the amount of water at the surface of Venus since its accretion up to present-day, in equivalent global ocean layer depth \citep[adapted from][]{Gillmann2020}. The simulation includes the effects of impacts during late accretion (erosion, delivery, energy transfer), atmospheric escape and volcanism. Three compositions are proposed for the late accretion: fully composed of relatively dry enstatite chondrite-like material \citep[red line with 0.1$\%$ H$_2$O, 0.4$\%$ CO$_2$ and 0.02$\%$ N$_2$]{Piani2020}; with 10$\%$ wetter material akin to carbonaceous chondrites (blue curve noted CC; by mass, 8$\%$ H$_2$O, 4$\%$ CO$_2$ and 0.2$\%$ N$_2$); and with 15$\%$ CC (light blue curve). Only the drier late accretion is consistent with the present-day water inventory in the Venus atmosphere (target value). The escape envelope is the maximum amount of water that could have been present in the atmosphere at a given time and still be consistent with present-day measurements in the absence of all mechanisms other than atmosphere escape. Atmospheric escape is calculated based on \cite{Kulikov2006}, and is higher than estimates from \cite{Persson2020}. The initial water range illustrates the uncertainty on the water inventory during accretion that is not resolved by the model. Time is measured from the formation of calcium-aluminum rich inclusions (CAIs).
}
\label{Fig:VenusEvolution_CG20.pdf}
\end{figure}

In fact, the fate of water is central to most aspects of the evolution of Venus. Any evidence of the history and state of water through the ages would not only improve considerably our understanding of the evolution of Venus' climate, but of other linked processes as well, both at the surface and in the interior of the planet. Tracking water has thus been proposed as a tool to investigate Venus past and its possible states \citep{Gillmann2020,Warren_Kite2021LPSC}. It has been used to constrain possible accretion scenarios (see Fig. \ref{Fig:VenusEvolution_CG20.pdf}), the past climate or more recent changes. Central to this approach is the estimation of volatile species fluxes over time (for non-thermal escape for instance, \citealt{Persson2020}). For this reason, it depends on our understanding of a wide array of mechanisms. In some aspects we have made progress in the last decades, but the unconstrained parameter space is still large. It is also important to keep in mind that processes evolve with time. Different past conditions could have influenced the way they affect volatile species fluxes. Therefore each step in understanding mechanisms at work in the evolution also informs further attempts to illuminate Venus' history.

\subsection{The link with origins}

The question of very early Venus has long been identified as a major aspect determining the planet's evolution \citep[e.g.,][]{KastingPollack1983,Kasting1984}. Building upon this, recent works have shown that the evolutionary pathway that a planet like Venus can follow can be decided by its ``initial" era, or rather the conditions of its transition from the magma ocean (MO) phase to the solid planet phase. The time required for MO solidification \citep{Hamano2013}, and the early loss of water \citep{Gillmann2009} could be responsible for the distinct evolutionary paths of Earth and Venus. Climate simulations of early Venus have also demonstrated that the initial state of water (liquid vs. vapor) has a strong influence on the post-MO conditions \citep{way2020venusian,Turbet2021}. \cite{Jacobson2022}, \cite{salvador2022} and \cite{Avice2022} offer a more comprehensive picture of primitive Venus.

Initial distribution of water between the mantle and atmosphere is controlled in part by the crystallization of the post-accretion magma ocean. Given its insolation, early Venus could be located at the boundary between dichotomous magma ocean evolution and lifetimes \citep{Hamano2013,salvador2017relative,Lebrun2013}. If the insolation absorbed by early Venus was less than the runaway greenhouse limit, then the magma ocean likely persisted for a few million years until Venus had cooled sufficiently for water to condense at the surface as occurred for Earth. However, if the insolation absorbed exceeded the runaway greenhouse threshold, then a magma ocean could have persisted for around 100 Myr or more, until almost all water was photodissociated during the slow solidification process, with most hydrogen lost to space while the oxygen was partially lost to space and partially consumed by the MO \citep[e.g.,][]{Schaefer_Wordsworth2016} and by reducing atmospheric gases. Which of these scenarios occurred depends on the radiative and cloud feedbacks in the atmosphere. 

Some GCM modeling suggests that cloud feedback on a slowly rotating early Venus could have maintained a temperate climate due to high subsolar cloudiness leading to high albedo values \citep{Yang2014,way2016venus,way2020venusian}. However, it has been argued that a hot magma ocean on Venus could not plausibly transition to a temperate state because a thick steam atmosphere would inhibit day-side cloud formation whereas night-side clouds would limit radiative cooling, thereby preventing the condensation of surface water \citep{Turbet2021}. This would ultimately maintain a hot climate and lead to a ``dry Venus" scenario (Type-II planet in \citealt{Hamano2013}), when early loss of water is taken into account. The major difference between these two modeling efforts is the initial state of the water: liquid surface layer \citep{way2016venus,way2020venusian}, or in the atmosphere as steam \citep{Turbet2021}.

In addition to its dynamics and resulting radiative transfer and clouds distribution, the initial composition of the atmosphere is of fundamental importance to decipher whether early Venus was wet or dry. It is influenced by both the magma ocean chemical composition and thermal evolution during the crystallization which control the partitioning of water and other volatile species between the interior and the atmosphere. The cloud feedback invoked to keep early Venus hot and ultimately dry \citep{Turbet2021} assumes an early \ce{H2O}-, \ce{H2O}--CO$_2$-, or \ce{H2O}--N$_2$-dominated atmosphere (or all three species). Such oxidizing, \ce{H2O}--\ce{CO2}-dominated early atmospheres are generally considered \citep[e.g.,][]{Elkins-Tanton2008, Hamano2013, Lebrun2013, Massol2016, salvador2017relative, Nikolaou2019, Bower2019} in analogy to modern volcanic gas composition \citep[e.g.,][]{Holland1984}, resulting from the equilibrium with their mantle sources and with hot surfaces of different compositions \citep{Lupu2014}. Yet, despite geological evidence, the origin of life on Earth seems to require a highly reduced (rich in \ce{CH4}, \ce{H2}, and \ce{CO}) early atmosphere \citep[e.g.,][]{Urey1952, Miller1959, Johnson2008}. Alternative scenarios involving impact degassing, volatiles trapping in solidifying mantle or varying the magma ocean oxygen fugacity have thus emerged and been proposed to match such requirements and enlarged the possible range of early atmosphere compositions \citep[e.g.,][]{Hashimoto2007, hier2017origin, zahnle2020creation, Solomatova2021, Gaillard2022a,Bower2021}.
Extensive studies remain to be done in order to constrain how the dynamics of such atmospheres would affect the heat and clouds redistribution and thus influence the magma ocean cooling and resulting surface conditions.

Incidentally, some early atmosphere composition possibilities also give rise to the ``stifled outgassing" scenario for Venus, where the early outgassing of CO$_2$ is not followed by water release at the end of the magma ocean phase. Instead, in the resulting C-poor, deep magma ocean, H incorporation (due to H being increasingly siderophile at higher pressures, while C is less so) into the core could lead to desiccation of the magma ocean \citep{Gaillard2022b} and formation of a FeO-rich mantle, through oxidation of remaining Fe-metal in the mantle. The possibly oxidized conditions on Venus (relative to Earth) due to H-loss \citep{zahnle_rise_2013} may have played a role in the growth of a dry magma ocean atmosphere on this planet compared to an H-bearing one on Earth \citep{Gaillard2022a}. 

Furthermore, some dynamical aspects might also favor an incomplete to minimal degassing scenario \citep{Ikoma2018}. Indeed, while the vigorous convection has been thought to promote efficient mantle mixing and thus rapid early degassing, a recent study showed that the convective magma ocean dynamics and detailed patterns may significantly reduce the amount of magma reaching the surface, thereby significantly reducing the outgassing rates \citep{Salvador2021_sub}. In addition, extreme magma ocean conditions are still out of the range of the experimental data constraining volatile behavior in magma and large uncertainties remain.
It should be noted that in the above scenarios it is usually assumed that Venus had at least one global and deep magma ocean. On Earth, this likely resulted from the moon-forming impact. For Venus, however, it is not clear whether there was a global and deep magma ocean, resulting from the combination of additional heat sources, or rather local magma ponds that contributed to the early atmosphere \citep{salvador2022, Jacobson2022}. In the latter case, H incorporation into the core is less likely.

On the other hand, impacts could bring a wide array of volatiles (both possibly reduced and oxidized) to early atmospheres, making the primitive atmospheric conditions even more complex. Additionally, one has to reconcile these results with the case of Earth, where evidence suggests that surface liquid water was present as early as 4.3 Ga ago \citep{mojzsis2001}. The composition of the early atmosphere will also be determined by the redox state of the terminal magma ocean, which is poorly constrained for both Earth and Venus. Recent experimental work on iron speciation in silicate melts has suggested that the terminal magma ocean atmospheres of Earth and Venus could have been CO-dominated with a comparatively minor H$_2$-H$_2$O component \citep{sossi2020redox,Bower2021}. In such a case, one could hypothesise that oxygen remnants from hydrodynamic escape could oxidise CO to produce the more recent CO$_2$ atmosphere. It is also not clear exactly what types of clouds would be produced in such atmospheres \citep[e.g.,][]{Herbort2021} nor their feedbacks on the climate.

The cloud dynamics of such atmospheres have yet to be explored, but the rapid escape of H to space would presumably result in comparatively cool CO-dominated atmospheres (CO is a poor greenhouse gas). The possible radiative and dynamical effects of photochemical hazes \citep{he2018photochemical} or atmospheric chemical reduction caused by giant impacts \citep{zahnle2020creation} have also yet to be fully investigated. 

\begin{figure}
\centering
\includegraphics[width=13cm]{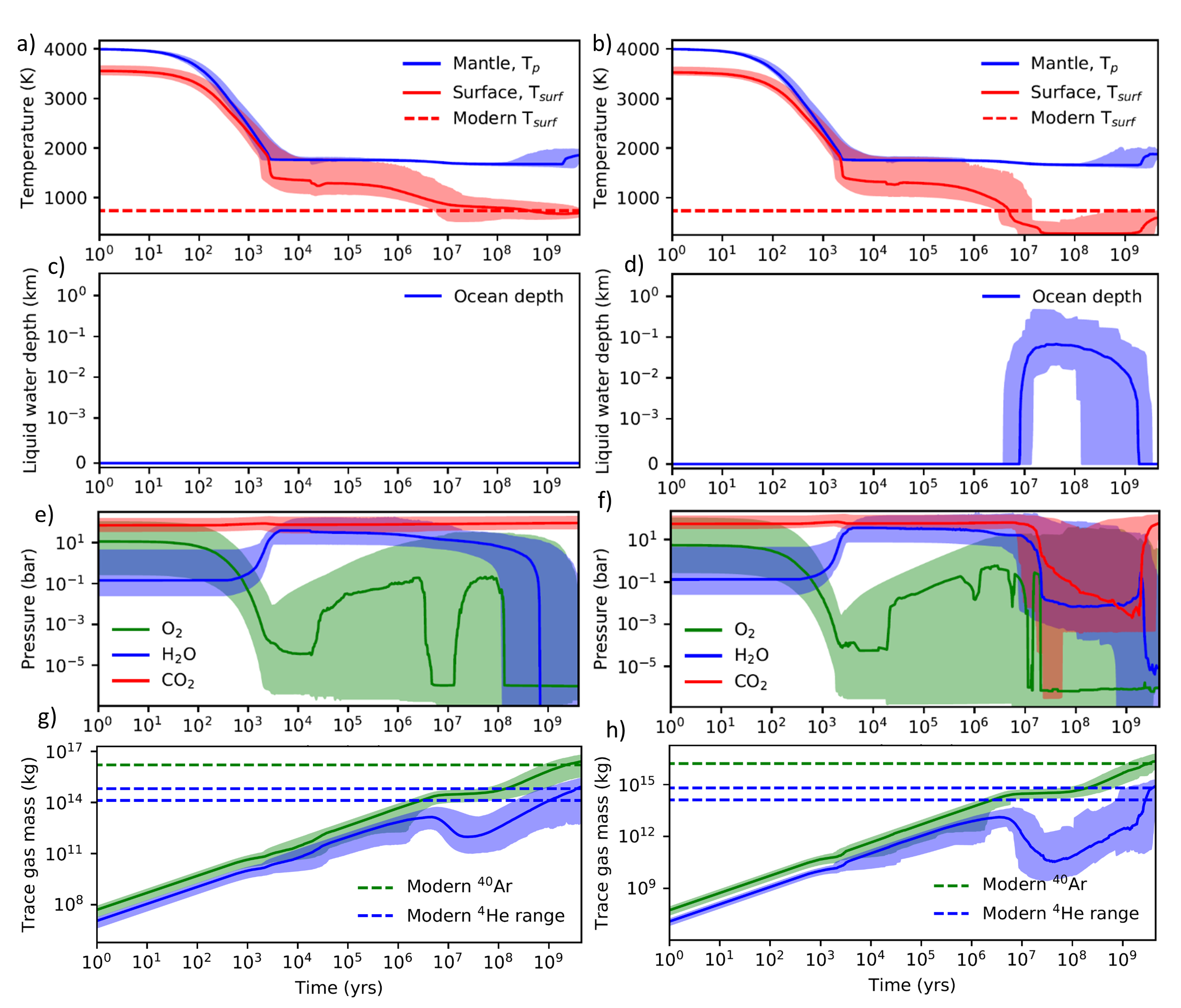}
\caption{Two scenarios for the coupled atmosphere-interior evolution of Venus. Solid lines are median values, shaded regions denote 95\% confidence intervals, and the time evolution spans from post-accretion magma ocean to the present-day (4.5$\times$10$^9$ yrs). Left-column plots show low albedo model runs that recover the modern Venus atmosphere and were always dry (no surface water condensation), whereas right-column plots show high albedo model runs that recover modern Venus and were temporarily wet. Subplots (a) and (b) show the evolution of mantle potential temperature (blue) and surface temperature (red), with the modern mean surface temperature denoted by a dashed red line for reference. Subplots (c) and (d) show the global average liquid water depth (blue). Subplots (e) and (f) show the evolution of atmospheric composition, including partial pressure of carbon dioxide (red), oxygen (blue), and steam (green). Subplots (g) and (h) show the accumulation of atmospheric $^{40}$Ar (green) and $^{4}$He (blue) from degassing. Dry scenarios show a transition from magma ocean to runaway greenhouse (a), and the gradual loss of water to space (e). Transiently habitable model runs experience temperate conditions with ${\approx}$100 m global oceans (d) where atmospheric CO$_2$ is drawn down by silicate weathering (f), before Venus reenters a runaway greenhouse and loses remaining water to space. The model successfully recovers modern atmospheric $^{40}$Ar abundance, plausible atmospheric $^{4}$He abundance (dashed lines), and modern heat flow (not shown). Adapted from \cite{krissansen2021Venus}.}
\label{Fig:VenusEvolution_KT21_v2.png}
\end{figure}

Fig. \ref{Fig:VenusEvolution_KT21_v2.png} illustrates the possible dichotomous evolution of water on Venus within a coupled atmosphere-interior model starting with a fully molten mantle \citep{krissansen2021Venus}. The differing evolutionary scenarios are attributable to assumed initial volatile inventories and fixed albedo. The subplots on the left show the long-lasting hot climate and resulting dry surface conditions scenario described above \citep{Turbet2021}. Here, cloud feedback does not allow for buffering the excess (above the runaway greenhouse threshold) of received solar insolation which maintains Venus in a runaway greenhouse state, thus preventing it from cooling down to temperate climates where ocean condensation could occur i.e. a low albedo is assumed. Atmospheric water vapor is gradually lost to space during the slow magma ocean solidification and the atmosphere is mostly desiccated. Once the surface has solidified, it then evolves in a similar way to the dryer models described in \cite{Gillmann2020} (see Fig. \ref{Fig:VenusEvolution_CG20.pdf}). The panels on the right illustrate a post-magma ocean wet scenario where liquid water condenses at the surface immediately after magma ocean crystallization, and CO$_2$ cycle/silicate weathering feedback maintains a temperate surface i.e. a high albedo is assumed. Eventually, increasing solar luminosity triggers a transition to runaway greenhouse, carbon degassing from the mantle returns CO$_2$ to the atmosphere (crustal decarbonation is neglected), and all remaining water is lost to space. Here, the longevity of surface liquid water is up to 3.5 Gyrs because both plate tectonics and a temperature-dependent silicate weathering feedback are assumed. This differs from the shorter-duration wet scenario illustrated in Fig. \ref{Fig:earlyw} which assumes a stagnant-lid early Venus and CO$_2$-dependent silicate weathering. Both evolutionary scenarios (wet or dry post-magma ocean surface conditions) can explain modern Venus conditions, which is to say they recover the observed bulk atmospheric composition (Fig. \ref{Fig:VenusEvolution_KT21_v2.png}, bottom panels), but also produce reasonable values for atmospheric $^{40}$Ar \& $^4$He (Section \ref{models_melt_nobles}), and plausible surface heat flow values \citep{krissansen2021Venus}. This highlights the need for new discriminants (see below).

\subsection{Feedback processes}
The role of models based on physics and experimental results becomes especially important when considering the interactions between various mechanisms governing the evolution of terrestrial planets. This has been already highlighted at various points in this chapter, for example with uncertainties related to the interpretation of the atmospheric D/H ratio. Understanding the evolution of Venus (and other terrestrial planets) becomes even more complex if feedback processes are taken into account. Those are interactions between several mechanisms or domains that affect one another and can result in positive or negative control of the given processes. As an example of positive retroaction, we have already described (Section \ref{atmosphere-lithosphere}) that in a scenario where carbonates were formed, destabilization of the carbonate reservoir by increasing temperatures in the lithosphere could cause carbon outgassing, followed by further increase of temperature due to enhanced greenhouse effect, until all the reservoir has been depleted. In an Earth-like situation, one could even imagine a more complete carbon cycle \citep{krissansen2021Venus}. The same mechanism would apply to thermal dehydration of crustal materials and water.

One important feedback that has been observed in models directly affects both the climate and the mantle evolution of Venus: the effects of surface temperature variations on mantle convection regime \citep[see][]{Rolfetal2022}. This is especially relevant in the case of Venus due to its high observed surface temperature, and the possibility that, at some point in its past, surface conditions could have been much cooler, with the potential for large variations ($\approx$400 K) during its history.
A simple attempt to model such a feedback was made by \cite{Phillips2001}, using a parameterized model. They obtained a positive feedback where higher surface temperatures led to higher partial melting, larger outgassing and in turn even higher surface temperatures because of the release of greenhouse gases. It is now suspected that the situation is more complex. Increases in temperature weaken the lithosphere, leading to a decrease in both lithosphere and mantle viscosity. Convective stresses in the lithosphere would therefore be lower at higher surface temperatures \citep{Lenardic2008}. It is also expected that Venus's present-day lithosphere is more ductile than Earth's and that pores and fractures are healed more efficiently due to its higher temperature and lower viscosity \citep[e.g.,][]{Bercovici2014}. As a consequence, higher surface temperatures would favor a stagnant-lid like regime, while lower surface temperatures would allow for a mobile lid regime more akin to plate tectonics, where the lithosphere can be fractured by convective stresses \citep{Weller2015}. Therefore the surface temperature could result in a different mantle convection regime on Earth and Venus. While it has not been fully investigated yet, it is likely that the ``plutonic squishy lid" convection regime highlighted in recent work \citep{Rozel2017,Lourenco2020} would be affected by changes in the surface temperature. One could imagine that this regime would be favoured by temperatures on present-day Venus, as fracture healing and ductile behaviour may prevent a larger part of the magma from reaching the surface. A simple model for changes in intrusive/extrusive volcanism ratio could well suggest a complex interaction between three evolving convection regimes rather than the two end-members studied in past works.

The role of surface temperatures has been observed in long-term modeling of the coupled evolution of Venus \citep{Gillmann2014} as part of a negative feedback cycle. If the greenhouse effect decreases due to the loss of water, a mobile lid regime emerges accompanied by high volcanic production and outgassing. The water reservoir in the atmosphere is replenished, leading to a corresponding increase in surface temperatures that suppresses the mobile lid and favors a return to stagnant lid convection. 

On the other hand, \cite{Noack2012} observed a similar stabilizing feedback but with a different mechanism: higher surface temperatures allowed for the mobilization of the surface, by reducing the viscosity contrast between the mantle and lithosphere, which removes heat from the mantle efficiently, and ultimately reduces volatile concentrations in the atmosphere and surface temperature. 

As illustrated by these modelling efforts, the consequences of surface temperature variations go beyond changes solely in mantle convection. In turn, different tectonic regimes lead to variations in outgassing efficiency and regulate surface conditions, thus completing a feedback cycle. Beyond this, changes in surface conditions can affect (aqueous) weathering and the state of water, with possibly far reaching consequences. If surface conditions are modified enough to allow for new processes to dominate, large changes in atmosphere composition cannot be ruled out, as suggested previously \citep{krissansen2021Venus}. 

In the same way, a change in tectonic regime can also affect how heat is removed from the interior of the planet, with possible consequences on heat fluxes at both the surface and core mantle boundary. Magnetic field generation might therefore be affected by surface changes \citep{o2018prospects}. While the exact contribution of the magnetic field to the protection or erosion of the atmosphere is still debated \citep[e.g.,][as discussed in section \ref{magneticfields}]{gunell_why_2018,gronoff2020atmospheric}, it is widely accepted that magnetic field modifications would affect atmospheric escape mechanisms and thus, the long-term atmospheric composition.
By considering the planet as a coupled system, estimating long-term evolution becomes a complex game that involves identifying the dominant feedback mechanism at specific conditions.

\subsection{Reaching present-day Venus}
All possible scenarios for the past evolution of Venus must reach the planet's present-day state. First, it means that the current bulk atmospheric composition must be met: 11$\times$10$^{18}$ kg of N$_2$ \citep[3.5 $\pm$ 0.8 $\%$,][]{johnson_goldblatt2015}, 10$^{16}$ kg of H$_2$O \citep[30 $\pm$ 15 ppm,][]{KastingPollack1983,Lecuyer2000}, 4.69$\times$10$^{20}$ kg of CO$_2$ \citep[96.5 $\pm$ 0.8 $\%$,][]{fegley2014}, virtually no O$_2$ \citep{oyama_et_al1980,fegley2014}, and 150 $\pm$ 30 ppm SO$_2$ \citep{KastingPollack1983,Lecuyer2000}. Note that actual abundances vary with altitude. Other common constraints are given by $^{40}$Ar or $^{4}$He abundances and isotope ratios of other noble gases, topography, geoid or the age of the surface and volcanic eruption rates. Conversely, understanding late evolution places constraints and gives clues on early scenarios by highlighting what initial conditions can be compatible with present-day. The most drastic changes in surface conditions would take place when branching out of an early temperate state into hot and dry present-day Venus \citep[see also][]{Avice2022}. 

If cloud feedback did enable a wet early Venus, then the subsequent evolution and ultimate longevity of such clement surface conditions were also controlled by atmosphere-interior interactions. Indeed, this habitable period could persist for billions of years barring destabilizing outside or internal events \citep{way2016venus,way2020venusian}. Even under a stagnant lid tectonic regime, the weathering feedback could have maintained temperate surface conditions over long periods of time \citep{honing2021}. The era of a temperate climate could have ended as buried carbonates became unstable at depth and released CO$_2$ into the atmosphere. The rising surface temperature would have moved the decarbonation depth even closer to the surface, leading to a catastrophic outgassing of CO$_2$, thus establishing a strong greenhouse effect \citep{honing2021}.

The stagnant lid also limits surface heat loss, which may increase interior temperatures and further destabilize crustal carbonates. Against this destabilizing feedback, the drawdown of carbon dioxide by silicate weathering (Sections \ref{atm-surface-reactions} and \ref{SilicateWeathering}) could maintain habitable surface climate for a duration on the order of 1 Gyr according to model calculations \citep{honing2021} before the transition to runaway greenhouse would occur (Fig. \ref{Fig:earlyw}). However, the exact timing of this mechanism is extremely model-dependent and is affected by initial mantle temperature, the rate at which inorganic carbon in the lid is transferred to the atmosphere, and the sensitivity of silicate weathering to climate. As discussed in Section \ref{SilicateWeathering} the dependence of silicate weathering on atmospheric composition and climate is highly uncertain, even for Earth. A pure CO$_2$-dependence, as assumed in Fig. \ref{Fig:earlyw}, is a worst-case scenario for extending habitability since it provides no stabilizing thermostat to offset gradual CO$_2$ outgassing through time. Early plate tectonics or episodic subduction could also extend the duration of a wet Venus surface via efficient transport of carbonates to the deep mantle, thereby limiting the decarbonation feedback as Venus’s surface warms.

It is unlikely that CO$_2$ accumulation on Venus could solely originate from regular volcanic outgassing, especially in a short period of time. Calculations indicate that between 10 \citep{Lopez1998} and 100 \citep{Head2021} times the current total observable crust inventory would be needed for a complete Venus atmosphere build-up. Models \citep[e.g.,][]{Gillmann2020} suggest that around 10 bars of CO$_2$ could be outgassed during the late evolution (although this depends heavily on composition). 

Such late massive outgassing would need to be checked against constraints brought by the various radiogenic noble gas isotopes (e.g., $^{4}$He, $^{40}$Ar, $^{129,131-136}$Xe). Current estimates for argon suggest less degassing of Venus compared to Earth \citep{kaula_constraints_1999}, and are inconsistent with large scale recent volcanism.

The main issue remains the fate of water in this scenario. Atmospheric oxygen loss cannot account for the escape of more than a few tens of centimeters of global equivalent layer of water over the last $\sim$4 Gyr \citep{Persson2020}, assuming present-day atmospheric composition. Solid surface oxidation could account for roughly the same amount of water loss as atmospheric escape, as discussed earlier in this work \citep{Gillmann2020}. Therefore the mechanism could only be significant for a long-lived liquid magma at the surface, such as in a late intense runaway greenhouse event \citep{zahnle2007emergence}. Additionally, present-day observations suggest that hydrogen and oxygen escape is close to the stoichiometric ratio (2:1) of water molecules \citep{Barabash2007a}. If oxygen and hydrogen originate from water dissociation in Venus’ atmosphere, this stoichiometric escape could suggest that no other substantial sink alters the oxygen repartition at present-day.
An interesting alternative has recently been proposed by \cite{Warren_Kite2021LPSC} where fine ash resulting from explosive volcanism could be fully oxidized due to their larger surface to mass ratio. First results from their parameterized model indicate a large amount of oxygen (stored in H$_2$O and CO$_2$) could thus be removed from the atmosphere. However, if this takes place after the modern day thick atmosphere, their models needs to be reconciled with the fast cooling of volcanic ash at Venus' surface conditions due to heat transfer to the dense atmosphere \citep{Frenkel_Zabalueva1983}. Explosive volcanism is thought to require volatile contents $>$3-5 wt$\%$, several wt$\%$ higher than typical Earth magmas ($<$1 wt$\%$) \citep{Head1986, Head2021}. This aspect also requires further experimental investigation in order to assess the efficiency of this mechanism, which may vary by orders of magnitude, especially compared to solid basalt oxidation. Further surface observation of Venus may yield clues to understand whether a significant amount of ash can be formed under Venus past and present-day conditions. Until now, only very limited pyroclastic activity has been identified on Venus, with most volcanic activity being effusive in nature \citep{Campbell_Clark2006,Ghail_Wilson2015,Grosfils2000,Grosfils2011,Keddie1995,McGill2000}.

In contrast, both the desiccated Venus scenarios (dry Venus and stifled outgassing) have a straightforward evolution from early Venus to present-day, since in those cases, the bulk of the atmosphere was generated early on by oxidized magma ocean outgassing. Competition between moderate/marginal outgassing and loss processes would have kept water and oxygen abundances low and only allowed for a slight increase in CO$_2$ and N$_2$ pressure over time. Long-term stagnant lid like (or single plate) convection would have been likely. The main issue could be the low water content of the present-day atmosphere requiring low volcanic outgassing of any remaining water, but that could be explained by high surface pressure impairing water release into the atmosphere \citep{Gaillard2014}. Surface pressures would have remained high because of the CO$_2$-N$_2$ atmosphere and the lack of a mechanism to trap them at the surface in the absence of oceans \citep{sleep2001carbon}. No ocean would be possible because water was never outgassed early on and could not be released later on. The possibly dry nature of the mantle (with H trapped in the core, lost to space, or a low initial water abundance if Venus accreted dry material) in this evolution pathway would also be consistent with geoid-topography correlation data \citep{Kiefer1992} and the possible lack of a venusian asthenosphere. If Venus' interior is not dry, the impossibility to outgas water due to surface conditions would also fit this scenario, but not necessarily the geoid-topography correlation data. Further observation of the surface, possibly including estimates for the composition of lava flows may help discriminate between these scenarios.

As a consequence, the temperate Venus scenario requires a way to remove efficiently any surface/atmospheric water it did not lose during its primordial evolution, so that it can reach the present state of Venus (that has been in place since at least the mean age of the present-day surface). The longer the temperate phase lasted, the more difficult it is to remove water while building up the CO$_2$ atmosphere (unless the latter is primarily inherited from the magma ocean early outgassing) and meeting other observational markers. The dry evolution scenarios are more static, which has to be reconciled with Venus being an active planet. The present-day state of our understanding of mechanisms at work is still unable to distinguish between these evolutionary pathways.

\subsection{Discriminating between scenarios}
Considering the planet as a whole is a complex problem, but it is necessary to understand its evolution. Currently, our knowledge is limited by the lack of data about past (surface, mantle and atmospheric) conditions. This naturally reflects on how well we can assess interactions between interior and atmosphere, and in turn how they could have changed over the history of Venus. 
The further we step back in time, the wider the range of possible scenarios becomes.
Therefore, we search for ways to constrain evolution from observation and assess what could provide valuable landmarks in the evolution of Venus to anchor different scenarios.

Noble gases are powerful tracers of volatile element exchanges between the interior and exterior of terrestrial planets \citep[e.g.,][]{ozima_noble_2002,marty_origins_2020}. They are currently the most robust source of information on the distant past of Venus, despite a lack of data, measurement uncertainties and debates about their interpretation. See also \cite{Avice2022} for a review on noble gases in the present-day Venus atmosphere.
Many of the physical processes described above potentially altered the starting primordial elemental and isotopic composition of noble gases in the atmosphere of Venus. Depending on its accretion history, Venus could have, similarly to Earth, incorporated in its solid mantle noble gases which were originally present in the Solar nebula \citep{williams_capture_2018}. Gas partitioning between the melt in the magma ocean and the primitive atmosphere or early outgassing events would have led to the emergence of an early atmosphere of solar composition. However, atmospheric erosion by escape processes or by impacts could have modified the composition of this primordial atmosphere \citep[e.g.,][]{Pepin1991} or could have totally removed it (see sections \ref{imperosion} and \ref{hydrogen:oxygen:escape}).

Impacts of chondritic or cometary bodies at the end of Venus' accretion \citep{obrien_water_2014} would have delivered new noble gases (and other volatile elements) having distinct elemental and isotopic compositions depending on their origin \citep[e.g.,][]{busemann_primordial_2000-1,furi_nitrogen_2015,marty_xenon_2017,rubin_krypton_2018}. If there were remnants of a primordial solar atmosphere at the time of this late contribution, the mixing between solar and meteoritic and/or cometary gases could be detected in the resulting elemental and isotopic composition of atmospheric noble gases \citep[\textit{e.g.}][]{Marty2012,marty_meteoritic_2022}. Mixing between components could also result from the long-term degassing of noble gases into the atmosphere. Finally, noble gases could also track long-term escape processes. 

On Earth, the isotopic composition of atmospheric xenon evolved through time \citep[see][and refs. therein]{avice_evolution_2018}. This evolution could be due to the long-term escape of xenon ions from Earth's atmosphere to outer space in a photo-ionized hydrogen wind \citep{zahnle_strange_2019}. This escape process would progressively lead to the depletion and isotopic fractionation of atmospheric xenon. The fact that xenon on Earth and Mars share similar features (depletion and isotopic fractionation) means that the selective coupled H$^{+}$--Xe$^{+}$ escape mechanism could have been operating on both planets. The evolution of the isotopic composition of atmospheric xenon could thus be a tracer of hydrogen escape on terrestrial planets \citep{avice_perspectives_2020}. So far the elemental abundance and isotopic composition of atmospheric xenon on Venus remains unknown \citep[see][for a review on existing data]{Avice2022}. New measurements would certainly help to further understand how hydrogen escape shaped the atmosphere of Venus and may have influenced its geodynamical evolution \citep{baines_atmospheres_2013}.

Over the course of Venus' history, several isotopes of noble gases have been produced by extinct ($^{129}$I, $^{244}$Pu) and extant (e.g., $^{238}$U, $^{40}$K) radioactive nuclides present in the silicate portions of the planet. When a portion of Venus crust or mantle melted, these gaseous products migrated into the magmatic gas and were eventually degassed into the atmosphere inducing radiogenic/fissiogenic excesses on top of the primordial isotopic composition. Because the aforementioned radioactive nuclides have very different half-live times (e.g., $t_{1/2}$($^{129}$I)=16\,Myr and $t_{1/2}$($^{238}$U)=4.47\,Gyr), studying the abundances and relative proportions of their daughter products has the potential to provide important constraints on the amount and timing of exchanges of volatile elements between the interior and the atmosphere of Venus.

For example, atmospheric Xe on Mars and Earth have very distinct $^{129}$Xe/$^{130}$Xe ratios \citep{ozima_noble_2002}. The very high ratio for Mars atmospheric Xe could be due to intense and early episodes of atmospheric escape leaving an atmosphere depleted in primordial noble gases \citep{swindle_martian_2002}. Subsequent degassing of xenon from Mars' interior carrying a strong excess of radiogenic $^{129}$Xe would explain the high present-day $^{129}$Xe/$^{130}$Xe ratio of Mars atmospheric Xe. For Earth, recent studies demonstrated that, 3.3 Gyr ago, the atmospheric $^{129}$Xe/$^{130}$Xe ratio was significantly lower than the modern ratio \citep{avice_origin_2017,marty_geochemical_2019}. The difference has been used to estimate mantle degassing rates over the past 3.3 Gyr. 

Knowing the $^{3}$He/$^{4}$He ratio of Venus' atmospheric helium could also provide important information regarding the present and past interactions between the atmosphere and the interior. However, numerous events with changing magnitudes over time can modify this ratio such as: (i) deposition of $^{3}$He-rich extraterrestrial material, (ii) atmospheric escape causing isotopic fractionation of the remaining helium fraction and (iii) degassing of radiogenic $^{4}$He produced inside Venus. Only approaches coupling He and Ne-Ar-Xe isotope systematics would allow us to use the $^{3}$He/$^{4}$He ratio as an efficient probe of Venus' ancient geodynamics.

Another look at geodynamics can be obtained from the state of the core, although its composition is still debated, which can affect conclusions. A stagnant lid regime limits core-mantle heat flux (in the common view, all other things equal, the surface heat flux is smaller than for plate tectonics because of the lid, the mantle is hotter and the core cools more slowly), thus under these conditions the core of Venus would be expected to remain in a liquid state until present-day \citep{Nimmo2002,ONeill2014,orourke_thermal_2015,Smrekar2018}. The size and state of the core can be constrained by measurements of the moment of inertia factor and the tidal Love number $k_2$ \citep{Dumoulin2017}. Detection of remnant magnetisation of the crust caused by a past magnetic field could help shed light on the state and history of the core \citep{o2018prospects}. The absence of a Venusian solid inner core would therefore rule out a scenario involving long-term plate tectonics operating on Venus \citep{ONeill2021}. In case a solid inner core exists, even the presence of small amounts of hydrogen -- as suggested by the stifled outgassing scenario -- would increase considerably both the compressional and shear wave velocities in the core \citep{Caracas2015}.

Determining possible scenarios involving water could rely on the study of the very early history of Venus. Coupled models of magma ocean evolution and outgassing \citep[e.g.,][and \citealt{salvador2022} for a review]{Elkins-Tanton2008, Lebrun2013, Hamano2013, salvador2017relative, Nikolaou2019, Bower2019, Lichtenberg2021} and models of atmosphere-interior interactions can provide insights, however determining whether a temporarily habitable (e.g., Fig. \ref{Fig:VenusEvolution_KT21_v2.png}, right column) or never habitable (e.g., Fig. \ref{Fig:VenusEvolution_KT21_v2.png}, left column) evolutionary scenario was realized will require more observational constraints. 

However, the planet may still hold some clues for the links between the distant past and the present-day state on its surface. Water and its history are a key to understand how Venus reached its present-day state and diverged from Earth. A succession of increasingly difficult questions would ideally need to be answered to form a robust base for evolution scenarios. First, we need to look for any evidence of water (at the surface or in the mantle) over the course of the evolution of the planet, between the magma ocean freezing and the generation of the current state 200-1000 Ma ago. A second step would be to estimate the abundance and state of that water. Then, refining the timing of the observations would clarify whether the changes were catastrophic or progressive. Finally, proof of liquid water would set a standard for Venus' past climate. However, with the caveat that negative results might indicate either a lack of water (either at the surface or in the mantle, depending on the nature of the observation) or that no traces of water were preserved.

It will also be extremely important to gain more knowledge on the mechanisms that define possible evolution scenarios on Venus. This chapter has highlighted the advances made in recent years as well as gaps in our current understanding of the effects of those processes. We hope that a closer examination of both the surface composition and volatile species exchanges will allow to better link the past of Venus to its present state. 

The composition of the surface could retain traces of the mechanisms of atmosphere/crust interaction needed to understand the late evolution of Venus. It is important to look for signs of water and chemical, mineralogical and isotopic markers. On one hand, this could mean measuring the product of surface oxidation, or the composition of volcanic gas plumes in the atmosphere of Venus, to understand where the water comes from and where and how it disappeared. That also includes any evidence for past pyroclastic activity.

On the other hand, it can be even much more direct, by searching for actual traces of water on the surface, from surface composition, such as the presence of granite, or aqueous alteration products in the surface mineralogy. A detection of rocks similar to Earth's banded iron formations, silica-rich deposits, salt-rich formations or currently solid salt flows (e.g., in Venus canali) may quantify any past water activity \citep{Zolotov2009,Zolotov2019}. That also means looking for any evidence of fluvial \citep[e.g.,][]{Khawja2020}, lacustrine and/or shoreline processes and deposits in the tesserae and plains, and for evidence in any layered rocks exposed in deformed regions such as tesserae.

In the same way, terrains such as the tesserae, that could correspond to older parts of the surface of Venus \citep[e.g.,][]{Ivanov_Head1996}, may inform us about the past mechanisms of their formation, both from the point of view of the dynamics regime and, through their composition, from that of the evolution of the mantle composition. Important information about the water content in the interior of the planet can be obtained from the phase lag of the tidal Love number. The latter depends strongly on the mantle viscosity, which in turn depends on the temperature and volatile content. Four upcoming space missions to Venus \citep{widemann2022}, DAVINCI, VERITAS, ENVISION and Shukrayaan-1, possibly complemented by the Chinese mission VOICE (VOlcano Imaging and Climate Explorer), will provide information to better discriminate between an aqueous and a dry evolution scenario of the Earth's sister planet.

\begin{acknowledgements}
The authors thank S. Mojzsis and J. Head for their comments, as well as R. Wordsworth and an anonymous reviewer for their help in improving the manuscript. CG acknowledges the support of Rice University and the CLEVER planets group (itself supported by NASA and part of NExSS) and ET-HoME Excellence of Science programme. GA acknowledges support from the french Centre National d'Etudes Spatiales (CNES) for support of Venus-related studies. MJW acknowledges support from the Goddard Space Flight Center Sellers Exoplanet Environments Collaboration (SEEC), which is funded by the NASA Planetary Science Division's Internal Scientist Funding Model. MJW acknowledges support from NASA's Nexus for Exoplanet System Science (NExSS), the GSFC Sellers Exoplanet Environments Collaboration (SEEC), which is funded by the NASA Planetary Science Division's Internal Scientist Funding Model (ISFM) and the ROCKE-3D Project ISFM jointly funded by a NASA Planetary and Earth Science Divisions. AS acknowledges support from NASA's Habitable Worlds Program (No.~80NSSC20K0226). MYZ acknowledges support from the NASA Solar System Workings program.
\end{acknowledgements}

%
%

\bibliographystyle{spbasic}      
\bibliography{references}   

%
%

\end{document}